\documentclass[a4paper, 12pt, oneside,scrreprt]{Thesis}  
\graphicspath{Figures/}  
\usepackage[ numbers, comma, sort&compress]{natbib}  
\usepackage{verbatim}  
\usepackage{vector}  
\hypersetup{urlcolor=red, colorlinks=true}  
\usepackage{indentfirst}
\usepackage{txfonts}
\usepackage{multibib}
\usepackage[explicit]{titlesec}
\usepackage{amsmath}
\usepackage{amsfonts}
\usepackage{amsthm}
\usepackage{graphicx}
\usepackage{hyperref}
\usepackage[T1]{fontenc}
\usepackage{float}
\usepackage{graphics}
\usepackage[font=small,labelfont=normal, figurename=Fig.]{caption}

\usepackage[labelsep=space]{caption}
\usepackage{fancyhdr}
\usepackage{amssymb,amscd,calc}
\usepackage{lineno,hyperref}

\usepackage{amsmath,amsthm,amssymb,amsbsy, amstext,amsfonts,amscd}
\usepackage{enumerate}
\usepackage{epsfig}
\usepackage{authblk}
\usepackage{multicol}
\usepackage{float}
\usepackage{placeins}
\usepackage{epstopdf}
\usepackage{float}
\usepackage{natbib}
\floatstyle{plaintop}
\restylefloat{table}
\RequirePackage{hyperref}
\theoremstyle{definition}

\numberwithin{lem}{section}

\newtheorem{defn}{Definition}
\numberwithin{defn}{section}
\let\olddefinition\defn
\renewcommand{\defn}{\olddefinition\normalfont}

\newcommand{\crest}[1]{\includegraphics[width=0.5\textwidth]{#1}}

\title{A STUDY OF BLACK HOLES AND BEYOND: SHADOWS AND RELATIVISTIC ORBITS}

\date       {\today}
\subject    {}

\frontmatter      

\begin{document}

\maketitle

\clearpage
\setstretch{1.5}
\addcontentsline{toc}{chapter}{Dedication}
\pagestyle{plain}
\begin{center}
    \thispagestyle{empty}
    \vspace*{\fill}
    This thesis is dedicated to my beloved Alpa Talreja and Mihir Ramanuj
    \vspace*{\fill}
\end{center}

\clearpage
\setstretch{1.5}
\addcontentsline{toc}{chapter}{Certificate}
\pagestyle{plain}
\begin{center}
\textbf {\huge Certificate}
\end{center}
This is to certify that the thesis entitled \textbf{“A Study of Black Holes and Beyond: Shadows and Relativistic Orbits”} submitted by \textbf{Parthraj G. Bambhaniya (18DRNST003)} to the \textbf{Charotar University of Science and Technology} for the award of the Degree of \textbf{Doctor of Philosophy} is a bonafide record of research work carried out by him under my/our supervision. The contents of this thesis, in full or in parts, have not been submitted to any other Institute or University for award of any degree, diploma or titles.
\\
\\
\\
\\
\makebox[0 in][r]{} \textbf{Research Supervisor} 
\\
\makebox[0 in][r]{} Signature with Date: 
\\
\makebox[0 in][r]{} Name\makebox[0.91 in][r]{} : Prof. Pankaj S. Joshi 
\\
\makebox[0 in][r]{} Designation \makebox[0.41 in][r]{} : Distinguished Professor, Founding Director of ICSC.
\\
\makebox[0 in][r]{} Organization \makebox[0.36 in][r]{} : International Center for Space and Cosmology (ICSC), 
\\
\makebox[1.48 in][r]{} Ahmedabad University, Ahmedabad. 
\\
\\
\\
\\
\\
\makebox[0 in][r]{} \textbf{Dean}\\
\makebox[0 in][r]{} Signature with Date:\\
\makebox[0 in][r]{} Name\makebox[0.91 in][r]{} : Dr. Gayatri Dave\\
\makebox[0 in][r]{} Organization\makebox[0.46 in][r]{} : Faculty of Applied Sciences,\\
\makebox[1.48 in][r]{} Charotar University of Science and Technology, Changa.

\clearpage

\clearpage
\setstretch{1.5}
\addcontentsline{toc}{chapter}{Declaration}
\pagestyle{plain}
\begin{center}
\textbf {\huge Declaration}
\end{center}
I hereby declare that the thesis \textbf{“A Study of Black Holes and Beyond: Shadows and Relativistic Orbits”} submitted by me to \textbf{Charotar University of Science and Technology} for the degree of Doctor of Philosophy is the record of work carried out by me during the period from December 2018 to September 2023 under the supervision of \textbf{Prof. Pankaj S. Joshi}.  I also declare that this work has not formed the basis for the award of any degree, diploma or titles by any other Institution or University.
\\
\\
\\
\\
\makebox[0 in][r]{} Signature\makebox[1.38 in][r]{} : \\
\makebox[0 in][r]{} Date\makebox[1.7 in][r]{} : 20th September 2023\\
\makebox[0 in][r]{} Name of the Research Scholar\makebox[0 in][r]{} : Parthraj G. Bambhaniya\\
\makebox[0 in][r]{} ID Number\makebox[1.26 in][r]{} : 18DRNST003

\clearpage

\setstretch{1.5}
\addcontentsline{toc}{chapter}{Acknowledgement}
\pagestyle{plain}
\begin{center}
\textbf {\huge Acknowledgement}
\end{center}
\indent The journey of my doctoral study would not have been possible without some incredible people who have encouraged, supported, and mentored me throughout. The foremost person to be grateful to is my supervisor, Prof. Pankaj S. Joshi for providing me with such a nice opportunity to work under his guidance. His immense knowledge and profuse experience has encouraged me throughout my academic research and daily life. His generosity and flexibility gave me the freedom to work on a variety of exciting projects. Additionally, I'm very much grateful for his invaluable advice, continuous support, and patience during my Ph.D. study. I am highly thankful for the lab facility, and academic and financial support received from the Charotar University of Science and Technology under the Charusat Ph.D. student Research Fellowship (CPSF) scheme.

I would like to give very special thanks to my ideals: Mihir Ramanuj, Alpa Talreja, Khushboo Dhameja and Tushar Kapdi who planted the seed of curiosity about physics in me. If they had not been there, I would have never taken a turn toward Astrophysics. I would also like to express my gratitude to Dr. Dipanjan Dey for his valuable guidance to keep me interested and for encouragement at various stages of my training period. I would like to thank all the members and colleagues of the International Center for Cosmology (ICC), Dr. Rucha Desai, Dr. Kanwarpreet Kaur, Dr. Tapobroto Bhanja, Ashok, Vishva, Shailee, Jay, Divyesh Vithani for healthy discussions and support. I would also like to thank Prof. A. Hasmani, Prof. C. K. Sumesh, Prof. R. M. Patel, Prof. R. Rangarajan, and Prof. R. V. Upadhyay who have been evaluating my research constantly.

More than four years is a very long duration away from home and can make any person out of one's mind. But friends are an amazing gift to us that can transform and chanelise any bad energy into the best energy. I would like to give my sincere thanks to my little world, Tushar Kapdi, Anand, Divyesh Solanki, Siddharth Madan, Saurabh, Aadarsh, Anurag, Vishva, Kauntey, Kshitij, Ankit who “were always there for discussions on anything that I was unsure of” and “who have offered invaluable advice that will benefit me throughout my life. It is their kind care and support that has made my study and life on the Charusat campus a wonderful time. I also convey my special thanks to Dixitbhai and Bhavnaben for their constant help and good food. A very special thanks to my badminton team, Dr. Sunny, Dr. Mrugesh, Samruddhi, Stuti, Deep, Dweep, Ravi, Simmy, and Richi for making my research stress free. I would also like to thank visitors of ICC and well-wishers, Meet, Viraj, Suresh, Rittick, Prashant, Rajibul, Saikat, Karim, Koushiki, Cherian, Divya, Ameya, Nupur, Abhijeet, Chinmay, Vijay, Hari, Ashwathi, Sagar Baraiya, Nazir, Sagar Jethva, Hima, Bittu, Bina, Debayan, Shaileshbhai, Sonal Desai, Vittali, Dr. Krunal, Dr. Palash, Dr. Gayatri, Dr. Kinnari, Dr. Anjana, Dr. Manan, Dr. Shweta, Mukesh Yadav, Yatin Talati, Rakesh Dave, Sid Ramanuj, Mahipal, Yavkrit, Aditya, Chirag, Dhruv, Milan, and other MSc students for always being so kind and supportive.

Science is never the result of an individual endeavor. My all publications are the result of collaborations with amazing people, whom I have been extremely fortunate to be able to work along with. 

Finally, I would like to express my gratitude to my parents, my little brother, my lovely sisters, Yogesh, Jagdish, Darshita and my family members. Without their tremendous understanding and encouragement over the past few years, it would be impossible for me to complete my study. At last, I would like to thank all my academic teachers: Prof. S. P. Bhatnagar, Prof. N. K. Bhatt, Prof. G. M. Sutariya, Prof. Dipika Pandya, Prof. Umesh Dodiya, Prof. Maniyar, Prof. Bharat, Prof. Shilpa, Minal ma'am, Dr. Kanti, Krunal sir, Vishal sir,  Dr. Virbhadrasinh, Dharmendra sir and others for believing in me and constant encouragement towards my goal.

\makebox[3.2 in][l]{} \textbf{Parthraj G. Bambhaniya}

\clearpage
\setstretch{1.5}
\addcontentsline{toc}{chapter}{Abstract}
\pagestyle{plain}
\begin{center}
\textbf {\huge Abstract}
\end{center}

\indent This doctoral thesis is organized into seven chapters. The first chapter introduces readers to the formation of black holes and naked singularities as an end state of continuous gravitational collapse. The physical and geometrical properties of the Schwarzschild black hole, Joshi-Malafarina-Narayan-1 (JMN-1) naked singularity, and Janis-Newman-Winicour (JNW) naked singularity are summarised. The motivation and objectives to be derived are based on the literature reviews. The second chapter deals with the shadows of the mentioned compact objects. The equations of motion are explicitly calculated for general spherically symmetric and static spacetimes using the ray-tracing formalism and the null geodesics. The third chapter focuses on the construction of rotating naked singularity using the Newman-Janis-Algorithm (NJA). The NJA is used without complexification method and obtain rotating JNW naked singularity spacetime. The general formalism of the shadow shape is derived for rotating spacetime and obtain the shadow shapes for the rotating JNW, Kerr and deformed Kerr spacetimes. In the fourth chapter, the precession of timelike-bound orbits is investigated in the Schwarzschild, JMN-1, and JNW spacetimes. The fully relativistic orbit equations are derived for the provided models. The approximate solutions of the orbit equations are used to characterize the nature of orbital precession. The next chapter is on the precession of timelike bound orbits in the rotating Kerr and JNW spacetimes. The sixth chapter deals with the relativistic orbits of S-stars and discusses the orbital parameters of the real and apparent orbits. Astrometric data of the S2 star has been adopted from the available literature and use numerical techniques to study the relativistic orbits of the S2 star in the presence of a scalar field. The final chapter aims at summarising the results followed by some futuristic scopes that probe the nature of Sagittarius A* (Sgr A*) with a possible black hole mimicker.

\clearpage  


\clearpage
\pagestyle{fancy}  

\lhead{\emph{Generalized Caputo fractional differential equations and its applications}}  
\tableofcontents  

\listoffigures
\listoftables
\titleformat{\chapter}[display]
{\normalfont\bfseries\filcenter}{\Large\MakeUppercase\chaptertitlename\ \thechapter}{16pt}{\Large\MakeUppercase{#1}}
\titlespacing*{\chapter}{0pt}{0pt}{25pt}
\titleformat{\section}{\normalfont\bfseries}{\MakeUppercase\thesection}{1em}{\MakeUppercase{#1}}

\titleformat{\subsection}{\normalfont\bfseries}{\thesubsection}{0em}{{\hspace{1em}#1}}
\mainmatter	  
\pagestyle{plain}  


\chapter{Introduction}
Ever since the dawn of human civilization, the quest to determine the origin of the universe, has been the most curious one for individuals. With the invention of
scientific logic to approach the physical realm of the universe, humanity has moved strides
into the unknown like no other time in history. To be more philosophical, in the words of
Carl Sagan, "we are a way for the universe to know itself".

From the current understanding, it is believed that people have descended from stardust, the remains of a star that has run out of its nuclear fuel. This stardust is orbiting a second-generation star, i.e. the sun. There are billions of stars in the Milky Way galaxy. Like any other living entity, these stars have a life cycle with a beginning and an end. These stars arose from the clouds of dust and gas known as molecular clouds, which collapsed under their own gravity to form protostars. Researchers have classified stars into three categories based on their initial mass: small stars, average stars, and massive stars. All of these stars undergo the process of nuclear fusion in which they use their nuclear fuel to survive during their life cycle. At first, the fusion of hydrogen atoms takes place, resulting in the formation of helium atoms and production of an enormous amount of energy in the form of an exothermic process that balances the star's gravitational pull. When these stars exhaust hydrogen atoms in their core, the fusion of heavier elements takes place, such as carbon, oxygen, neon, silicon, and iron. These stars are also known as the main sequence
stars depending on the stages of their nuclear fusion. The fusion process in these main
sequence stars cannot fuse atoms heavier than iron, as the pressure from the thermal
energy produced in this process can no longer defy its gravitational pull, and the star starts
collapsing under its own gravity \citep{Joshi:2011rlc}. The small and average stars behave stable under their own quantum pressures, i.e. electron and neutron degeneracy pressures to form white dwarfs and neutron stars, respectively.

On the other hand, more massive stars cannot balance their gravitational pull by any
known quantum pressures. As Einstein's general theory of relativity predicts,
the continuous gravitational collapse of a massive star leads to a spacetime singularity
where the matter density and gravitational attraction become infinite \cite{penrose}. At this point, all
physical laws break down. This ultra-dense compact region can be hypothesized by
describing various astrophysical compact objects, e.g., the black hole, naked singularity, boson star, and other types of objects \cite{Oppenheimer:1939ue,psj1,psj2,jnw,bst,Visinelli:2021uve,Schunck:2003kk}. These compact objects have fascinating physical and geometrical properties, and it would be observationally significant to
distinguish them.  

 \section{Black hole}
The term "Black hole" is used for a compact astrophysical object from which
nothing can escape to a faraway observer. Even light cannot get away from its boundary which is an event horizon, due to the high gravitational field attraction. Therefore, one cannot see that dark region within event horizon where things can fall, but nothing will come out. The distinct
signature of a black hole is an "Event Horizon" that indicates presence of trapped surfaces. Within the general theory of relativity, the spacetime singularity can form at the center of such a black hole which is the actual boundary of the spacetime manifold \cite{penrose}. The spacetime singularity is the caustic point where all the geodesics are incomplete. All the physical quantities, such as matter densities, pressures, and spacetime curvatures, take arbitrarily large and diverging values near the singularity. Generally, when an event horizon covers the singularity, it is known as a black hole.

In the general theory of relativity, there are mainly two well-known non-rotating black
holes, namely (i) the Schwarzschild black hole and (ii) the Reissner-Nordstrom black hole. The
Schwarzschild black hole is a vacuum solution of the Einstein equations, which has
no charge and no rotation. Whereas the Reissner-Nordstrom black hole is a static
solution of the Einstein-Maxwell field equations with electric charge but no rotation.
The rotating generalizations of these static black holes are given as the Kerr (with
zero charge) and Kerr-Newman (with non-zero charge) black holes, respectively \cite{Hartle:2021pel}. In
this thesis, the focus is on the Schwarzschild and Kerr black holes for the comparative studies with the naked singularities.

\section{Schwarzschild black hole}
In 1916, Karl Schwarzschild found the first exact vacuum solution of Einstein's field equations \cite{Schwarzschild:1916uq}. This solution gives the spacetime metric outside a spherically symmetric and static massive body that does not contain an electric charge and angular momentum. The Schwarzschild black hole spacetime in spherical coordinates reads,
\begin{equation}
ds^2_{SCH} = -\left(1-\frac{2M}{r}\right)dt^2 + \left(1-\frac{2M}{r}\right)^{-1} dr^2 +r^2d\Omega^2\,\, , 
\label{schext}
\end{equation}
where $d\Omega^2=d\theta^2+\sin^2\theta d\phi^2$ represents the metric of the two spheres, and $M$ is the total mass of the black hole. The signature convention $(-, +, +, +)$ is used for spacetime metric throughout the thesis. In the above Schwarzschild spacetime, the singularity occurs at $r=0$, which is covered by a null surface, an event horizon which occurs at $r=2M$, also known as the Schwarzschild radius. As it is a vacuum solution of Einstein field equations, the energy density and pressures of this spacetime are,
\begin{equation}
    \rho=0, \; \; \;
    p_r=p_\theta=p_\phi=0.
\end{equation}
where $\rho$ is the energy density of the matter, $p_r$ is the radial pressure while $p_\theta$ and $p_\phi$ are tangential pressures.
The following conditions can easily verify the weak, strong and null energy conditions of this spacetime, respectively as,
\begin{equation}
    \rho\geq 0, \; \; \;
    \rho+p_r\geq 0, \; \; \;
    \rho+p_\theta\geq 0, \; \; \;
    \rho+p_\phi\geq 0, \; \; \;
    \rho+p_r+2p_\theta\geq 0. \; \; \;
\end{equation}

In 1939, the first dynamical collapse solution of a spherically symmetric and homogeneous dust cloud was derived by Oppenheimer, Snyder, and Datt \cite{Oppenheimer:1939ue,datt}, which was identified as OSD collapse later. It has been shown that the final state of a spherically symmetric and homogeneous dust collapse will always be a Schwarzschild black hole. Moreover, Birkhoff's theorem suggests that the Schwarzschild spacetime is the most general vacuum solution of Einstein's equations \cite{Hartle:2021pel}. However, the OSD collapse model depends on the ideal assumption, as the density should be homogeneous within the massive collapsing star. They also neglected gas pressure and took it to be zero to avoid dealing with the relatively complex Einstein equations \cite{Joshi:2011rlc}. In this work they have shown that an event horizon develops as collapse progresses in a way in which no material, particle, or light rays, can escape to a faraway observer. So, one must think beyond this idealistic scenario and consider more physical situations for the collapsing body.
 \section{Naked singularity}
In general, an astrophysical star is not supposed to have a homogeneous density profile within the star's surface. It can have high density at the core area, which could decrease as a function of radius as one go from the core area to the star's surface. Under a relatively more physically realistic situation, if inhomogeneity is introduced into the matter distribution profile and non-zero pressures are considered, then the end state of the dynamical collapse could result in a strong curvature singularity within the general theory of relativity \cite{Joshi:2011rlc,psj1,psj2,mosani1,mosani2,mosani3,Joshi:2013dva}. However, the end state of the dynamical gravitational collapse depends on the classes of initial conditions of the collapsing matter. 
In principle, theory of general relativity predicts that the spacetime singularity forms necessarily when large enough masses collapse under their own gravity. Nevertheless, it does not necessarily simultaneously enforce the formation (or otherwise) of an event horizon. If a singularity is formed without an event horizon, it is known as a naked or visible singularity \cite{Joshi:2008zz,Joshi:2018exh}.

Similar to a black hole, this is a region of an ultra-high-density where all the physical quantities will diverge arbitrarily. Naked singularity is an intriguing horizonless compact object. Although the cosmic censorship conjecture (CCC) prohibits the formation of strong horizonless singularities \cite{penrose}, several research studies reveal that the naked singularities can develop during the continuous gravitational collapse of an inhomogeneous matter cloud. In \cite{psj2}, Joshi, Malafarina and Narayan (JMN) have shown that a non-zero tangential pressure can prevent the formation of trapped surfaces around the core of high-density region of a collapsing matter cloud, resulting in a central naked singularity in large co-moving time. In this thesis, the focus is on the first type of Joshi-Malafarina-Narayan (JMN-1) and Janis-Newman-Winicour (JNW) naked singularity spacetimes \cite{psj2,jnw}. The details of these spacetimes are given as follows.

 \section{JMN-1 naked singularity}
 The equilibrium configurations of the collapsing fluid within the general theory of relativity are extensively studied in \cite{psj2}. The JMN-1 naked singularity can be formed as an end state of gravitational collapse with zero radial and non-zero tangential pressures and the corresponding spacetime is given as,
 \begin{equation}
     ds^2_{JMN-1} = -(1- M_0) \left(\frac{r}{R_b}\right)^\frac{M_0}{(1- M_0)}dt^2 + \frac{dr^2}{(1 - M_0)} + r^2d\Omega^2\,\, , 
\label{JMN1metric} 
\end{equation} 
where, $M_0$ and $R_b$ are positive constants. Here, $R_b$ represents the boundary radius of the distributed matter around the central singularity, and $M_0$ should be within the range $0<M_0<4/5$. Note that the upper limit of $M_0$ is defined by the fact that the sound speed cannot exceed unity. The stress-energy tensor of the JMN-1 spacetime gives the energy density $\rho$ and pressures $p$ as,
\begin{equation}
    \rho=\frac{M_0}{r^2}, \; \; \;
    p_r=0, \; \; \;
    p_{\theta}=\frac{M_0}{4(1-M_0)}\rho.
\end{equation}
This collapsing fluid is supported only by tangential pressure, and it can be verified that all energy conditions are satisfied by JMN-1 spacetime. 

The spacetime metric is modelled by considering a high-density compact region in a vacuum, which means that the spacetime configuration should be asymptotically flat. However, if it is not asymptotically flat, the interior spacetime must be matched to asymptotically flat exterior spacetime with a particular radius. On the other hand, asymptotically flat spacetimes need not be matched with any external spacetime. Two spacetimes can be matched at a particular space-like or timelike hypersurface. For that, one need to follow two junction conditions \cite{parth1}:
\begin{enumerate}

    \item The induced metrics of internal and external spacetimes should be identical on the matching hypersurface. 
    
    \item The extrinsic curvatures ($K_{ab}$) of internal and external spacetimes should be matched at the hypersurface, where the extrinsic curvatures are expressed in terms of the covariant derivative of normal vectors on the hypersurface:
    \begin{equation}
        K_{ab}=e^{\alpha}_ae^{\beta}_{b}\nabla_{\alpha}\eta_{\beta},
    \end{equation}
    where $e^{\alpha}_a$ and $e^{\beta}_b$ are the tangent vectors on the hypersurface and $\eta^{\beta}$ is the normal to that hypersurface. 
\end{enumerate}
The JMN-1 naked singularity spacetime is not asymptotically flat. The interior JMN-1 spacetime to exterior Schwarzschild spacetime can be smoothly matched at $r=R_b$ as,
\begin{equation}
    ds^2_{SCH} = -\left(1-\frac{M_0 R_b}{r}\right) dt^2+\left(1-\frac{M_0 R_b}{r}\right)^{-1}dr^2+r^2d\Omega^2,
    \label{matchSCH}
\end{equation}
where $M=\frac{1}{2}M_0 R_b$ is the Schwarzschild mass of the compact object. The extrinsic curvatures of JMN-1 and Schwarzschild spacetimes are automatically smoothly matched at $r=R_b$ \cite{parth1}. Since the JMN-1 spacetime has zero radial pressure.

\section{JNW naked singularity}
Janis, Newman, and Winicour obtained a minimally coupled mass-less scalar field solution of the Einstein field equations \cite{jnw}, and independently by Wyman found identical solution \cite{Wyman}. The Lagrangian density of the minimally coupled scalar field is given as,
\begin{equation}
\mathcal{L}=\sqrt{-g}\left(\frac12\partial^{\mu}\Phi\partial_{\mu}\Phi-V(\Phi)\right)\,\, ,
\end{equation}
where $\Phi$ is the scalar field and $V(\Phi)$ is the corresponding scalar field potential.
The minimal coupling conditions are defined as, 
\begin{equation}
    R_{\mu\nu}-\frac12 R g_{\mu\nu}=\kappa T_{\mu\nu}, 
   \end{equation}
\begin{equation}
    \Box\Phi(r)=V^{\prime}(\Phi(r)), 
\end{equation}
where $R$ is the Ricci scalar, $R_{\mu\nu}$ is the Ricci tensor, $T_{\mu\nu}$ is the energy-momentum tensor, and $g_{\mu\nu}$ is the metric tensor. $\kappa$ is a constant parameter, defined by $8\pi G/c^4$, where $G=c=1$. The Energy-momentum tensor ($T_{\mu\nu}$) for the minimally coupled scalar field can be written as,
\begin{equation}
    T_{\mu\nu}=\partial_{\mu}\Phi\partial_{\nu}\Phi-g_{\mu\nu}\mathcal{L}.
\end{equation} 
and the line element for the JNW spacetime is,
\begin{equation}
     ds^2_{JNW} = -\left(1-\frac{b}{r}\right)^n dt^2 + \left(1-\frac{b}{r}\right)^{-n}dr^2 + r^2\left(1-\frac{b}{r}\right)^{1-n}d\Omega^2\,\, ,
 \label{JNWmetric}
\end{equation}
where, $b=2\sqrt{M^2+q^2}$ and $ n=\frac{2M}{b}$. Here, $M$ and $q$ represent ADM mass and scalar field charge, respectively. As $b$ is positive and greater than $2M$, one can write $0<n<1$. The massless scalar field can be written as,
\begin{equation}
    \Phi=\frac{q}{b\sqrt{4\pi}}ln\left(1-\frac{b}{r}\right)\,\, .
\end{equation}
Further as, the JNW spacetime is asymptotically flat, it is not required to match with any other external spacetime. However, for $n=1$ or $q=0$, there are no effects of the scalar field ($\Phi$) on the spacetime, and JNW spacetime becomes Schwarzschild spacetime. Therefore, the JNW spacetime considers an extension of the Schwarzschild spacetime when the minimally coupled massless scalar field is included. 

Now, the question arises what could be the end state of the massive collapsing star? This question remains an open challenge. According to the Cosmic Censorship Conjecture (CCC), the end state of the massive collapsing body must be a black hole \cite{penrose}. However, as divulged before, several studies have shown that naked singularities can form due to the continual gravitational collapse of an inhomogeneous matter cloud. As pointed above, naked singularities form under physically realistic astrophysical situations and scenarios, their observational consequences become significant in the understanding of the universe such as the shadows cast by these objects, the strong gravitational lensing around them, the orbital precession of the stars in their vicinity and many others. In this thesis, the emphasis is mainly on the formation of a shadow by compact objects and the periastron precession of the star's orbits. From these studies, it can be predicted or probed that the possible observational signatures which would reveal the nature of the causal structure of any compact object. 

\section{Shadows of compact objects}
\begin{figure*}
\centering
\subfigure[]
{\includegraphics[width=120mm]{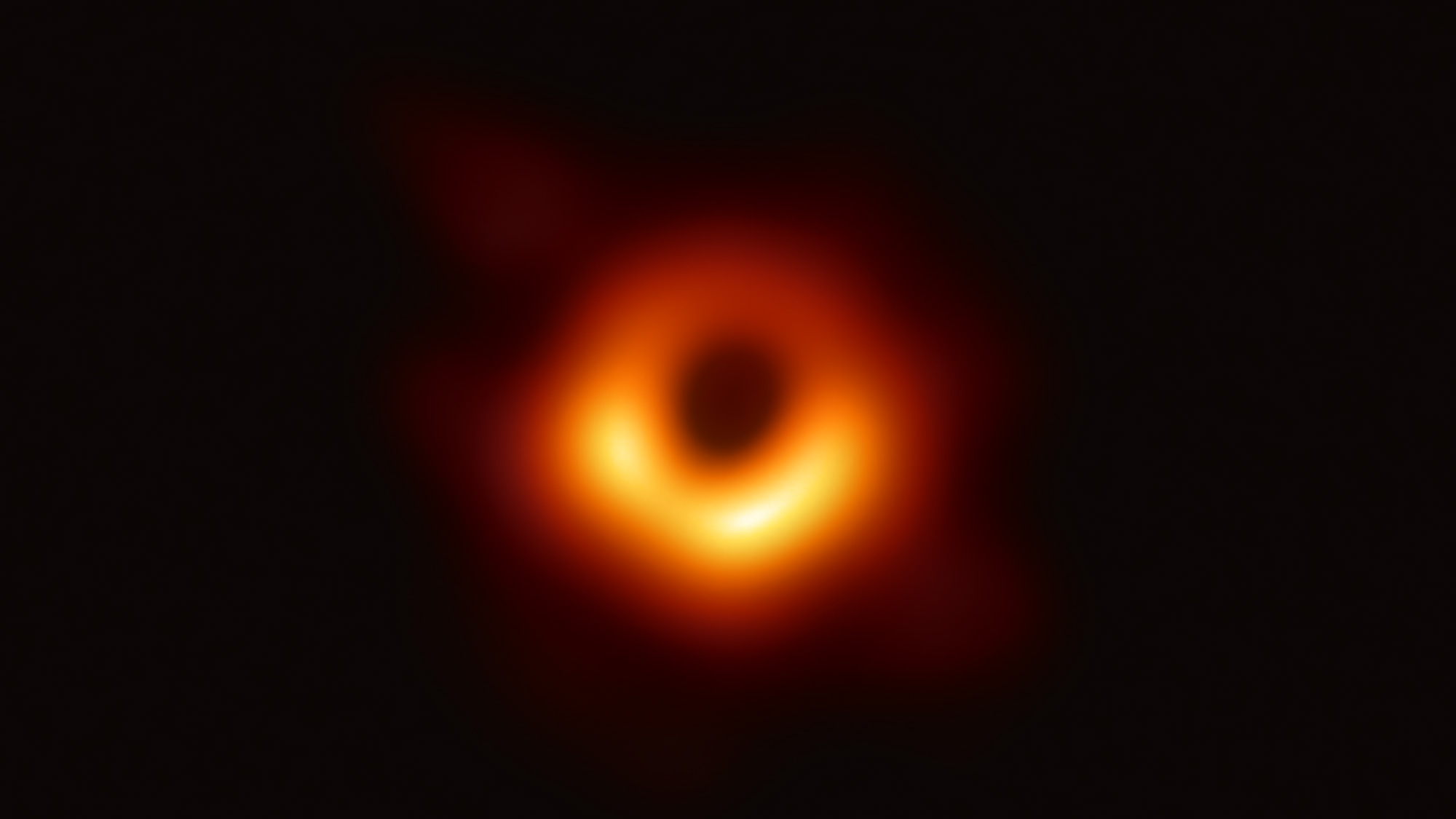}}
 \caption{Shadow image of M87*}
\label{M87}
\end{figure*}

\begin{figure*}
\centering
\subfigure[]
{\includegraphics[width=120mm]{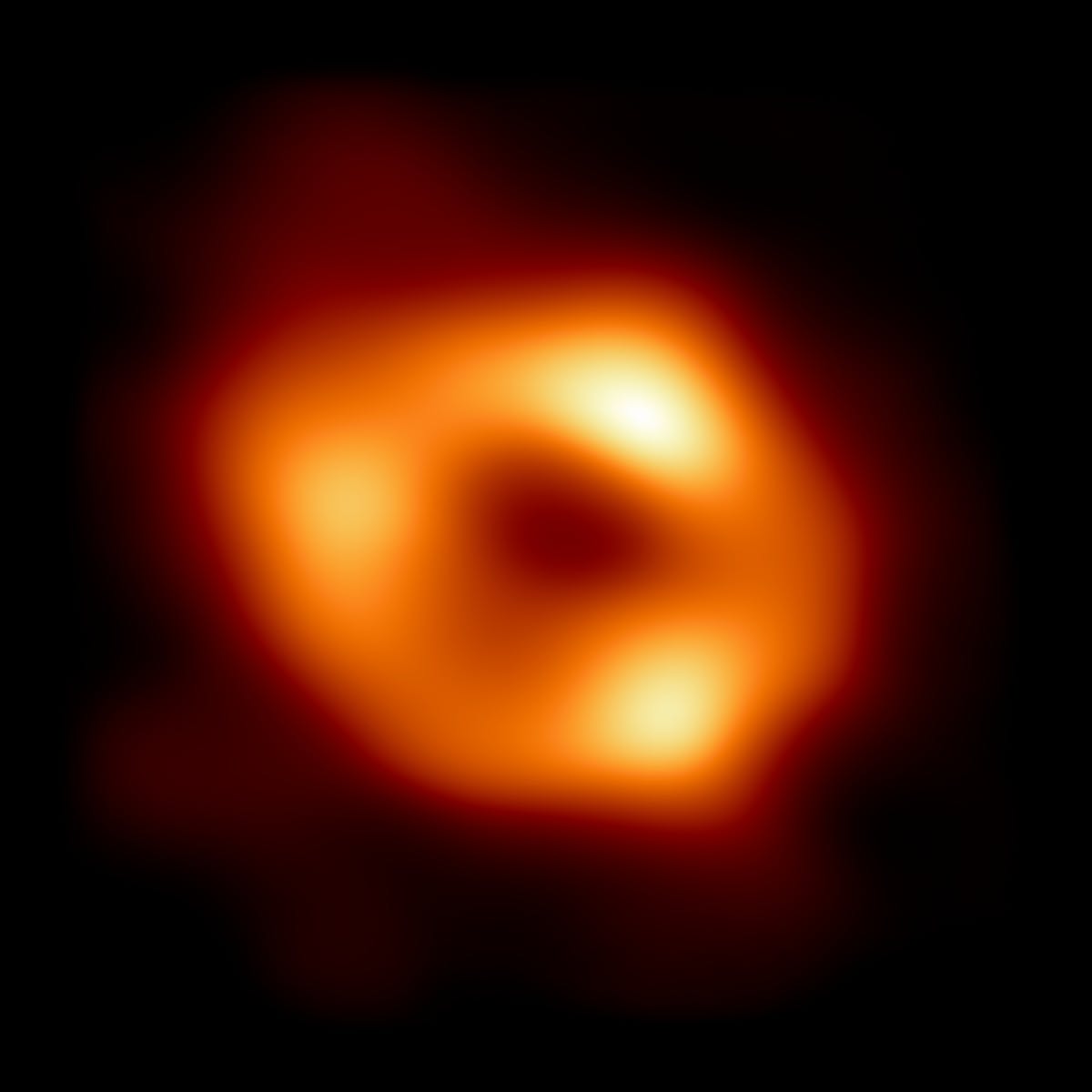}}
 \caption{Shadow image of Sgr A*}
\label{SgrA}
\end{figure*}

A supermassive black hole is believed to exist at the center of almost every galaxy \cite{Kormendy:2013dxa}. However, there is no conclusive evidence at this time about the nature of the compact object. In light of this, in April of 2019, the Event Horizon Telescope (EHT) acquired the first shadow image of a supermassive black hole at the heart of the M87 galaxy (see fig. \ref{M87}) \cite{M87}. Recently, the EHT collaboration has announced a major breakthrough in the imaging of an ultra-compact object at the centre of the Milky way galaxy (see fig. \ref{SgrA}) \cite{EventHorizonTelescope:2022xnr,EventHorizonTelescope:2022vjs,EventHorizonTelescope:2022wok,EventHorizonTelescope:2022exc,EventHorizonTelescope:2022urf,EventHorizonTelescope:2022xqj}. A bright emission ring around a core brightness depression in VLBI horizon-scale images of Sgr A*, with the latter linked to the shadow of black hole. The shadow boundary of the Sgr A* marks the visual image of the photon region and differentiates capture orbits from scattering orbits on the plane of a distant observer. The radius of the bright ring can be used as an approximation for the black hole shadow radius under some specific conditions and after proper calibration, with little reliance on the details of the surrounding accretion flux. While there is strong evidence that there is a massive concentration of mass in the center of the Milky Way galaxy, the question of whether or not it is a black hole is still open. The EHT shadow images of the M87 and Sgr A* galactic centers do not prove the notion of an event horizon and hence a supermassive black hole. Other compact objects can cast similar shadows \cite{Vagnozzi:2022moj}. The shadows created by compact objects, including black holes, naked singularities, gravastars, and wormholes, have received considerable attention \cite{shaikh1,gyulchev,null,Abdikamalov:2019ztb,Dey:2020haf,ohgami_2015,stuchlik_2019,Kaur,Sakai,Shaikh:2021cvl,Bambhaniya:2021ugr}. The Event Horizon Telescope (EHT) is now being upgraded to the next-generation Event Horizon Telescope (ngEHT) in order to monitor the galactic center of the Milky Way (Sgr A*) and other prospective radio sources. 
As discussed earlier, the gravitational collapse with zero radial and non-zero tangential pressures can form the JMN-1 naked singularity. It could be achieved as a quasi-equilibrium configuration of the collapsing fluid employing the general theory of relativity. The shadows of JMN-1 naked singularity spacetimes are explored and compared to a Schwarzschild black hole shadow \cite{shaikh1}. It is generally speculated that the existence of a photon sphere causes shadows. Given that there is no photon sphere and hence no shadow in JMN-1 naked singularity for the range of characteristic parameter $0<M_0<2/3$, an exciting full moon image is produced. JMN-1 naked singularity, on the other hand, can create a similar shadow for $2/3<M_0<1$, just as the Schwarzschild black hole produces. Furthermore, another naked singularity model introduced in 1968 by Janis-Newman-Winicour (JNW) can create a substantially similar shadow for the characteristic parameter value $0.5<n<1$ as the Schwarzschild black hole does \cite{gyulchev,Saurabh:2022jjv}. 

In the recent work, it is illustrated that a new spherically symmetric naked singularity solution of the Einstein field equation, despite the absence of a photon sphere, can cast a shadow \cite{null}. The general prerequisites for a shadow to arise without a photon sphere are then defined for null and timelike naked singularities, where each satisfies all of the energy conditions \cite{Dey:2020bgo}. Thus, when two distinct sorts of equally massive compact objects create shadows of the same size, it becomes challenging to identify them. However, because their spacetime geometries are distinct, the nature of the null and timelike geodesics around the two compact objects would be different. The existence of an upper bound of the effective potential of null geodesics induces the formation of a shadow, according to these studies \cite{null,Dey:2020bgo}, implying that if a spacetimes' effective potential of null geodesics has an upper bound, that spacetime can cast a shadow, such that the existence of a shadow does not require the presence of an event horizon and a photon sphere around a singularity. Consequently, these models can provide viable alternatives for testing or investigating the nature of the galactic center's supermassive compact object (e.g. Milky Way).

\section{Precession of timelike bound orbits}

Half of the Nobel prize in physics 2020 was awarded to Prof. Roger Penrose for discovering that black hole formation is a robust prediction of the general theory of relativity. At the same time, the half prize was bestowed to Prof. Reinhard Genzel and Prof. Andrea Ghez for discovering a supermassive compact object (Sgr A*) at the center of the Milky Way galaxy \cite{Do,GRAVITY:2018ofz,Hees:2017aal,GRAVITY:2020gka}. The Sgr A* is assumed to be a supermassive black hole (SMBH) that exists at the core of the Milky Way galaxy. There is, however, no concrete evidence for the same, i.e. the Sgr A* is indeed a SMBH. The nature of Sgr A* is still unknown. It is expected that the mass of Sgr A* is about four million times the mass of the sun, which is located at a distance of 8.2 kpc (1 kpc = $3*10^{16}km$) from the Earth. In recent days, the analysis of the nature of Sgr A* has become a subject of great interest. To envisage the nature of the Sgr A*, it is vital to investigate the different possible observational features, such as the shadow properties, the stellar orbits of the S-stars around the Sgr A*, accretion matter properties, etc. These observational features would be helpful in verifying the nature of the Sgr A*, whether it is a supermassive black hole or a naked singularity.

In the second chapter, the study of observational appearance at $230$ GHz is covered with probing the nature of Sagittarius-A* (Sgr A*) as a naked singularity \cite{Saurabh:2022jjv}. The JMN-1 and JNW naked singularity spacetimes which are anisotropic fluid solutions of the Einstein field equations are considered. Motivated by radiatively inefficient accretion flows (RIAF) \cite{Pu:2018ute,Pandya:2016qfh}, An analytical model for emission and absorption coefficients is used to solve the general relativistic radiative transfer equation. The resulting emission is then utilized to generate images to predict the nature of the Sgr A* with synthetic Very Long Baseline Interferometry (VLBI) images from current and future Event Horizon Telescope (EHT) arrays. Three different EHT array configurations are being used to simulate the models of naked singularities and a black hole. This may have little effect on the baseline, but it would increase the u-v plane gridding, making it feasible to capture a better-resolved image. Therefore, it is pretty exciting and valuable for probing the shadow image of the Sgr A* to predict whether it is a supermassive black hole or a naked singularity.

 In the third chapter, the rotating Janis-Newman-Winicour (JNW) naked singularity spacetime is constructed using Newman-Janis Algorithm (NJA) as the rotating spacetimes are more physically realistic for investigating observational properties \cite{solanki}. There is an analysis of NJA with and without complexification methods and found that the energy conditions that were satisfied when the complexification step was skipped. The shadows cast by rotating JNW were studied deeply along with deformed Kerr spacetimes then compared them with those cast by the Kerr black hole \cite{deformed}.

In fourth chapter, the fully relativistic general orbit equation is derived for any spherically, symmetric and static spacetime. The precession of timelike bound orbits is studied in the JMN-1 and JNW naked singularity spacetimes and compared them with the Schwarzschild black hole case \cite{parth1,Dey:2019fpv,Joshi:2019rdo}. The approximate solutions are find for the given models to analyze the nature of orbital precession. Several exciting differences emerge from those results, including the conclusion that particle-bound orbits in naked singularity spacetimes can precess in the opposite direction of particle motion, which is impossible in Schwarzschild spacetime. Such reverse precession and other distinctions may help distinguish a naked singularity from a black hole observationally.

The fifth chapter focuses on, precession of the timelike bound orbits in the rotating Kerr, and JNW spacetimes \cite{solanki,Bambhaniya:2020zno}. Using approximation, it is shown that the negative precession is not possible in Kerr spacetime. While rotating, JNW spacetime can admit negative precession for the particular range of scalar field charge $q$ and spin parameter $a$. This distinct feature of the timelike bound orbits in rotating JNW spacetime can be observationally significant to distinguish from the Kerr and other rotating compact objects.  

\begin{figure*}
\centering
\subfigure[]
{\includegraphics[width=130mm]{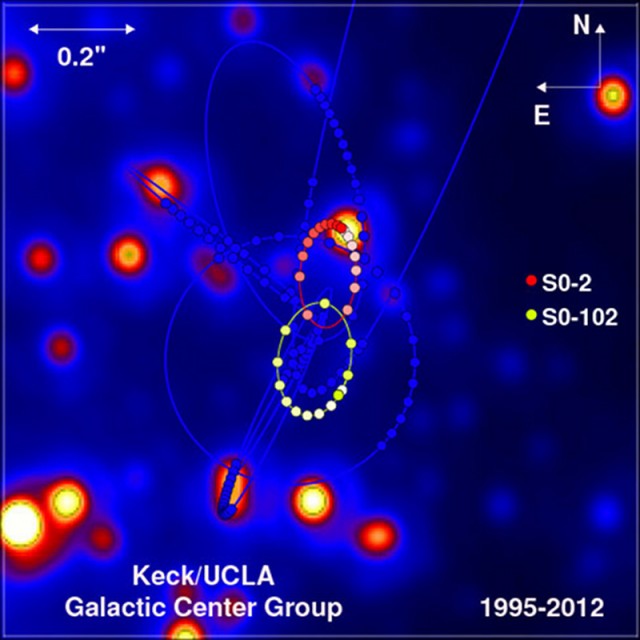}}
 \caption{S-stars near Milky Way galactic center}
\label{Stars}
\end{figure*}

Several S-stars are orbiting around the Milky Way galactic center (See fig. \ref{Stars}). Since they are very close to the Sgr A*, a general relativistic effect can be seen in their dynamics. Hence, their dynamics may distinguish a naked singularity from a black hole. GRAVITY, SINFONI, and UCLA galactic center groups shared their most recent data on the stellar motions of the S2 star around the Sgr A*, which is one of the crucial stars in the S-stars family \cite{Do,GRAVITY:2018ofz,Hees:2017aal,GRAVITY:2020gka}. The S2 star's orbit can give the essential information about the nature of the Sgr A* and its dynamics. Therefore, one can verify the precession of the S star's orbit within the background of a black hole and naked singularity spacetimes. 

Recently, the GRAVITY collaboration published a work on the S2 star's orbital precession by considering the Sgr A* as a black hole, where they have found that the precession angle of the S2 star's orbit is about 12 arc-minute which is too small to detect whether the orbital shift is positive or negative \cite{GRAVITY:2020gka}. More orbital data points with precise locations are required to make a concrete conclusion on the nature of Sgr A*. Hence, any evidence of negative precession of any `S' star can raise a big question on the existence of a black hole at the Milky Way galaxy center. In this context, the best-fitting orbit of the S2 star in the static JNW spacetime is obtained and investigated the scalar field effect on the orbital dynamics of that star in the sixth chapter \cite{Bambhaniya:2022xbz}. The best fitting parameters are estimated for the JNW metric using the Monte-Carlo-Markov-Chain (MCMC) technique and obtain the lowest Chi-squared value is 4.71. Here, it is predicted that the nature of the Sgr A* using the available observed astrometric data of the S2 star. It can be inferred that JNW naked singularity could be a plausible candidate which might represent the spacetime structure of Sgr A* which mimics the Schwarzschild black hole spacetime.


\chapter{Shadows cast by non-rotating black hole and naked singularities}

The study of null geodesics is useful for describing the formation of the shadows cast by ultra-compact objects. The study is about an observational appearance of the shadow image of the Sgr A* to predict whether it is a supermassive black hole or a naked singularity. The shadow properties are examined in JMN-1 and JNW naked singularity spacetimes. The novel features of the results are discussed and compared with the non-rotating black hole case, i.e. Schwarzschild black hole.

\section{Null geodesics in spherically symmetric and static spacetime}

Null geodesics represent the light ray orbits. To describe the world line of light rays, one has to consider the coordinates $x^\alpha$ as a function of an affine parameter $\lambda$.
The general spherical symmetric and static spacetime can be written as,
\begin{equation}
    ds^2 = - g_{tt}(r)dt^2 + g_{rr}(r)dr^2 + g_{\theta\theta}(r)d\theta^2 + g_{\phi\phi}(r)\sin^2\theta d\phi^2\,\, ,
    \label{static}
\end{equation}
where $g_{tt}$, $g_{rr}$, $g_{\theta\theta}$ and $g_{\phi\phi}$ are functions of $r$ only.
The conserved quantities for spherically symmetric and static spacetime (\ref{static}) corresponding to light ray can be defined as,  
\begin{equation}
     \gamma = -\xi \cdot k= g_{tt}(r)\left(\frac{dt}{d\lambda}\right)\,\, ,\,\,\,
      \label{con}
\end{equation}
\begin{equation}
   h =\eta \cdot k= g_{\phi\phi}(r)\left(\frac{d\phi}{d\lambda}\right)\,\, ,
   \label{congen1}
\end{equation}
where $\lambda$ is an affine parameter and $\xi$ and $\eta$ are killing vector corresponding to temporal and spherical symmetries. Here, $\gamma$ and $h$ represent the conserved energy and angular momentum per photon. The normalization of null vectors defines as,
\begin{equation}
 k\cdot k=g_{\alpha\beta}\frac{dx^\alpha}{d\lambda}\frac{dx^\beta}{d\lambda}=0,  
\end{equation}
 now, using the conserved quantities and normalization condition into general spherically symmetric and static spacetime (\ref{static}) for null geodesics, the following expression can be taken into consideration, 
\begin{equation}
    -\frac{\gamma^2}{g_{tt}(r)}+g_{rr}(r)\left(\frac{dr}{d\lambda}\right)^2+\frac{h^2}{g_{\phi\phi}(r)}=0.
\end{equation}
For simplicity, the light ray geodesics are considered in an equatorial plane $\theta=\pi/2$. By multiplying $g_{tt}(r)/h^2$ term to the above equation,
\begin{equation}
    \frac{1}{b^2}=\frac{g_{tt}(r)g_{rr}(r)}{h^2}\left(\frac{dr}{d\lambda}\right)^2+W_{eff}(r),
    \label{energyintegral}
\end{equation}
where, $b=h/\gamma$ is the impact parameter and $W_{eff}(r)$ is the effective potential for photon orbits given as,
\begin{equation}
    W_{eff}(r)=\frac{g_{tt}(r)}{g_{\phi\phi}(r)}.
    \label{lightweff}
\end{equation}
The effective potential is useful to determine the stability of the light orbits. Unstable circular null orbits are observed when the following conditions are satisfied,
\begin{equation}
    W^{\prime}_{eff}(r_{ph})=0, \; \; \;
    W^{\prime\prime}_{eff}(r_{ph})<0,
    \label{unstableorbit}
\end{equation}
the above conditions imply that the effective potential has a maximum value at $r=r_{ph}$ for unstable circular photon orbits, where $r_{ph}$ is the radius of the photon sphere. The path of the light ray can change due to the spacetime curvature. Using the equations (\ref{con}), (\ref{congen1}) and (\ref{energyintegral}), one can derive an expression to define the shape of the light orbits as,  
\begin{equation}
    \left(\frac{dr}{d\phi}\right)^2=\pm\frac{g_{\phi\phi}^2(r)}{g_{tt}(r)g_{rr}(r)}\left[\frac{1}{b^2}-W_{eff}(r)\right].
    \label{photonshape}
\end{equation}
Here, the $\pm$ signs represent the outgoing and incoming light ray paths, respectively. 
An orbit equation of the light ray (photon) can be obtained by differentiating above eq.~(\ref{photonshape}) with respect to $\phi$, 
\begin{equation}
  \frac{d^2u}{d\phi^2}+\frac{g_{\phi\phi}^2(u)u^4}{2g_{tt}^2(u)g_{rr}(u)b^2}\left(\frac{dg_{tt}(u)}{du}\right)-D(u)\left(\frac{dg_{rr}(u)}{du}\right) 
  +E(u)\left(\frac{dg_{\phi\phi}(u)}{du}\right) - F(u) = 0\,\, ,
  \label{orbiteqlight}
    \end{equation}
    where $u=\frac1r$ and,
    \begin{equation}
        D(u)=\left[\frac{g_{\phi\phi}^2(u)u^4}{2g_{tt}(u)g_{rr}^2(u)b^2}-\frac{g_{\phi\phi}(u)u^4}{2g_{rr}^2(u)}\right],
    \end{equation}
\begin{equation}
    E(u)=\left[\frac{u^4}{2g_{rr}(u)}-\frac{g_{\phi\phi}(u)u^4}{g_{tt}(u)g_{rr}(u)b^2}\right]\,\, ,
\end{equation}
and
\begin{equation}
    F(u)=\left[\frac{2g_{\phi\phi}^2(u)u^3}{g_{tt}(u)g_{rr}(u)b^2}-\frac{2g_{\phi\phi}(u)u^3}{g_{rr}(u)}\right].
\end{equation}
The above equation (\ref{orbiteqlight}) represents the light ray orbit equation in the general spherically symmetric and static spacetime which can be solved numerically for a given spacetime.

\section{Observable intensity}
It is widely studied that the shadows of compact objects can be formed due to a photon sphere. The photon sphere corresponds to the unstable circular light ray orbits. In addition to the existence of a photon sphere, one more condition has to be satisfied: an effective potential of the spacetime for the null geodesic should have an upper bound at $r=r_{ph}$. The behavior of the light rays near the compact object can be understood by the condition of impact parameter $b$. From the equation (\ref{energyintegral}), one can write the impact parameter corresponding to the turning point of the light ray,
\begin{equation}
    b_{tp}=\frac{r_{tp}}{\sqrt{g_{tt}(r_{tp})}}.
    \label{impact}
\end{equation}
where $dr/d\lambda=0$ at the turning point and $r_{tp}$ is the radius of the turning point from the compact object. Note that, for null geodesics, if only one extremum value of effective potential exists and that extremum value is the unstable photon orbit or maximum value of potential, then the minimum impact parameter at the turning point of null geodesics is,
\begin{equation}
    b_{tp}=b_{ph},
\end{equation}
where $b_{ph}$ is the impact parameter corresponding to the photon sphere. The light rays coming from the background source with the impact parameter $b<b_{ph}$ could not reach an asymptotic observer, they would be trapped inside the photon sphere. Hence, the shadow with radius $b_{ph}$ would form in the observer's sky.
There are numbers of literature where authors show that the shadow can form without a photon sphere in particular types of geometries \cite{null,Dey:2020haf,Dey:2020bgo}. However, if the photon sphere is absent in a given spacetime and the effective potential diverges near the center, then the shadow will not occur in that spacetime \cite{shaikh1}.  

Now, it is crucial to visualize the shadow cast by any compact object in a physical situation. This can be done by calculating the intensity distribution of the light rays coming from the accretion matter around the compact object.
The radially freely falling thin accretion matter radiates monochromatic light, where the emission rate of that light per unit volume describe as,
\begin{equation}
    j(\nu_e) \propto \frac{\delta(\nu_e-\nu_*)}{r^2}\,\, ,
\end{equation} 
where $\nu_e$ is the emitted frequency of a photon as measured in the rest frame of the emitter. Note that, for simplicity, in this section, the case in which the thin accretion matter is in spherically symmetric form is considered. The observed intensity of the light coming from the accretion matter could be obtained in the asymptotic observer's sky with the coordinates $X, Y$ as,
\begin{equation}
    I_{\nu_o}(X,Y)= \int_{\gamma\prime}^{} g^3 j(\nu_e) dl_{prop}\,\, ,\label{intensity}
\end{equation}
where the infinitesimal proper length ($dl_{prop}$) of the emitter in the rest frame can be defined as,
\begin{equation}
    dl_{prop}=-k_\alpha u^\alpha_e d\lambda,
\end{equation}
where $k_{\alpha}$ is the null four velocities of photons emitted by emitter and $u^{\alpha}_e$ is the time-like 4-velocity of emitter. $\lambda$ is the affine parameter and $g=\nu_o/\nu_e,$ is the gravitational red-shift factor defined by,
\begin{equation}
    g=\frac{k_\alpha u^\alpha_o}{k_\beta u^\beta_e}.
\end{equation}
The subscripts $0$ and $e$ represent the observed and emitted frequencies, respectively. Note that in the above equation (\ref{intensity}), the integration is carried out along the null trajectory $\gamma\prime$. One can write the four-velocity components of the radially freely falling accretion matter for general spherically symmetric and static spacetime as,
\begin{equation}
    u^t_e = \frac{1}{g_{tt}(r)}, \; \; \; \; \; \;
    u^r_e = -\sqrt{\frac{(1-g_{tt}(r))}{g_{tt}(r)g_{rr}(r)}}, \; \; \; \; \; \;
    u^\theta_e = u^\phi_e = 0,
\end{equation}
using these four velocity components, one can define the gravitational red-shift factor as,
\begin{equation}
    g = \frac{1}{\frac{1}{g_{tt}(r)}-\frac{k_r}{k_t} \sqrt{\frac{(1-g_{tt}(r))}{g_{tt}(r)g_{rr}(r)}}}\,\, ,
\end{equation}
where
\begin{equation}
    \frac{k^r}{k^t} = \sqrt{\frac{g_{tt}(r)}{g_{rr}(r)} \left(1-\frac{g_{tt}(r) b^2}{r^2}\right)}\,\, .
\end{equation}
Now, the intensity distribution in the plane of observer's sky $(X, Y)$ can be obtained by integrating equation (\ref{intensity}) for all observed frequencies as \cite{shaikh1},
\begin{equation}
    I_o(X,Y) \propto - \int_{\gamma\prime} \frac{g^3 k_t dr}{k^r r^2 }\,\, ,
\label{intensity1}
\end{equation}
where $X^2+Y^2=b^2$. The above equation (\ref{intensity1}) represents the observed intensity of the light coming from the thin accretion matter around the compact object. Which can be used to simulate the shadow of a compact object in the observer's sky. This equation can also show how the intensity varies with the impact parameter $b$.

\section{Radiative Modelling} \label{sec:level1}
In this section, an analytic model of emissivity and absorptivity coefficients are considered for radiative transfer calculations. In what follows, the prescription for radiative transfer is described \cite{Gold:2020iql}. The BH mass (M) for Sgr A* to be $4.3$ x $10^6 M_\odot$ and source distance (D) from the Earth is $8200 pc$ \cite{Do} are adopted for radiative transfer calculations.
The covariant form of the general relativistic radiative transfer is expressed as  
\begin{equation}
\frac{d\mathcal{I}}{d\tau_\nu} = -\mathcal{I} + \frac{{\eta}}{\chi},
\end{equation}
where $\mathcal{I}$ is the Lorentz-invariant intensity and is related to the specific intensity via $\mathcal{I} = I_\nu/\nu^3 = I_{\nu_o}/\nu_0^3$ where the subscript `o' denotes quantities in the local rest frame. $\tau_\nu$ is defined as the optical depth. $\chi$ and $\eta$ are the invariant absorptions and emission coefficients at frequency $\nu$. The number density of the fluid is given by
\begin{eqnarray}
    N = n_0 \exp{\left\{-0.5\left[\left(\frac{r}{10}\right)^2 + z^2\right]\right\}},
\end{eqnarray}
where $z=h\cos\theta$. $n_0$ is the reference number density, and $h$ is the vertical scale height.
The Keplerian angular momentum profile $\Omega = 1/r^{3/2}$ is used and the fluid four-velocity is thus given by 
\begin{equation}
    u_\mu = \bar{u}(-1, 0, 0, \Omega),
\end{equation}
with $\bar{u} = \sqrt{-(g^{tt} + g^{\phi\phi}\Omega^2 - 2g^{t\phi}\Omega)}$ such that $u_\mu u^\mu = -1$. The specific emissivity and absorption coefficients are given by
\begin{eqnarray}
    j_\nu &=& \mathcal{C}N\left(\frac{v}{v_{obs}}\right)^{-\alpha},\\
    \alpha_\nu &=& \mathcal{A}\mathcal{C}N\left(\frac{v}{v_{obs}}\right)^{-(\alpha+\beta)},
\end{eqnarray}
where $\mathcal{C}$ and $\mathcal{A}$ are constants controlling the effect of absorption and emission coefficients.
The Lorentz-invariant coefficients are then
\begin{eqnarray}
    \eta &=& \frac{j_\nu}{\nu^2}, \\
    \chi &=& \nu\alpha_\nu.
\end{eqnarray}
Table (\ref{table:model}) shows the values of different parameters for the accretion and the models of naked singularities. Note that JMN-1 spacetime without a photon sphere is defined as model-I, JMN-1 spacetime with photon sphere is defined as model II, and JNW spacetime is defined as model III.
\begin{table}
	\centering
	\caption{Model details}
	\label{table:model}
	\begin{tabular}{lllll} 
		\hline 
		\hline 
		\textbf{Parameter} & \textbf{Model-I} &\textbf{Model-II} &\textbf{Model-III} &\textbf{Sch. BH} \\
		\hline
		$M (M_\odot)$              & $4.3$ x $10^6$              &  $4.3$ x $10^6$               &   $4.3$ x $10^6$   &  $4.3$ x $10^6$\\
		$M_0$            & 0.63             &  0.70             &   -   &  -\\	
        $R_b$            & 2.857            &  3.175            &   -   &  -\\
        $\gamma$         & -                &  -                & 0.51  &  -\\
        $\mathcal{A}$    & 100              &  100              & 100    & 100\\
        $\mathcal{C}n_0$ (x$10^{-17}$) & 2.5 &  2.0 & $10^{-3}$ & 2.0\\
        $\alpha$         & -2               &  -2               & -2      & -2\\
        $\beta$          & 2.5              &  2.5              &  2.5    & 2.5\\
        $h$              & 10/3             &  10/3             &  10/3     & 10/3\\
	    \hline
	\end{tabular}
\end{table}

\section{Ray-Tracing Formalism}
\label{sec:rtf}

In order to calculate the shadow image of a black hole or naked singularity, one must first
solve the geodesic equations in the background spacetime under
consideration. The approach for ray-traced images of the models described in the previous section is considered. A system of six differential equations ($\dot{t}, \dot{r}, \dot{\theta}, \dot{\phi}, \dot{p_r}, \dot{p_\theta})$ is solved.

\subsection{Geodesic equations of motion}

For a given metric $g_{\alpha \beta}$, the Lagrangian can be written as
\begin{equation}
2\mathcal{L} = g_{\alpha \beta}\,\dot{x}^{\alpha}\dot{x}^{\beta}, 
\end{equation}
where an overdot denotes differentiation with respect to the affine
parameter, $\lambda$. 
From the Lagrangian, the covariant four-momenta of geodesics are:
\begin{eqnarray}
p_\alpha = \frac{\partial \mathcal{L}}{\partial \dot{x}^\alpha}.
\end{eqnarray}
Using the conservation of energy and angular momentum, this can be broken down to write $p_t = -E$ and $p_\phi=L$ where $E$ is the particle's total energy and $L$ is the angular momentum in the direction of $\phi$. The final set of geodesic equations that will  be solved is then given by
\begin{eqnarray}
\dot{t} &=& \frac{E}{g_{tt}} \\
\dot{r} &=& \frac{p_r}{g_{rr}}\\
\dot{\theta} &=&   \frac{p_\theta}{g_{\theta\theta}}\\
\end{eqnarray}
\begin{eqnarray}
\dot{\phi} &=& \frac{L}{g_{\phi\phi}} \\
\dot{p_r} &=& \frac{1}{2g_{tt}}\left[ -{p_r}^2\left(\frac{g_{tt}}{g_{rr}}\right)'- (Q +  L^2)\left(\frac{g_{tt}}{g_{rr}}\right)' \right]  \\
\dot{p_\theta}  &=& \frac{L^2}{g_{\theta\theta}}\frac{\cos{\theta}}{{\sin^3{\theta}}}
\end{eqnarray}
where $(')$ denotes differentiation with respect to $r$ and $Q$ is Carter's constant, which is the third constant of motion.

\subsection{Initial Conditions}
The initial conditions which are required to solve the system of an ordinary differential equation (ODE) are described in the previous section. In literature, different authors have assumed different initial conditions based on the type of work and ODEs which are needed to be solved. The formalism described in \cite{Younsi:2016azx} is used. An observer is at some distance from the source. The observer is placed far away from the Black hole   ($r_{\mathrm{obs}} = 10^{3}\,M$), where the spacetime is assumed to be flat (asymptotic flatness). The observer's position is specified in Boyer-Lindquist (oblate spheroidal)
coordinates as $(r_{\mathrm{obs}}, \theta_{\mathrm{obs}},  \phi_{\mathrm{obs}})$ where $\theta_{obs}$ and $\phi_{obs}$ are the inclination and azimuthal angles of the observer. The transformation from Cartesian coordinates to Boyer-Lindquist coordinates is given by
\begin{eqnarray}
\ \! r^{2} &=& \sigma+\sqrt{\sigma^{2}+a^{2}Z^{2}} \,, \label{CartBLr} \\
\cos \theta &=& Z/r \,, \label{CartBLtheta} \\
\tan \phi &=& Y/X \,, \label{CartBLphi}
\end{eqnarray}
where
\begin{eqnarray}
X &\equiv& \mathcal{D}\cos\phi_{\mathrm{obs}}-x\sin\phi_{\mathrm{obs}} \,, \\
Y &\equiv& \mathcal{D}\sin\phi_{\mathrm{obs}}+x\cos\phi_{\mathrm{obs}} \,, \\
Z &\equiv& r_{\mathrm{obs}}\cos\theta_{\mathrm{obs}}+y\sin\theta_{\mathrm{obs}} \,,
\end{eqnarray}
and
\begin{eqnarray}
\sigma &\equiv& \left(X^{2}+Y^{2}+Z^{2}-a^{2}\right)/2 \,, \\
\mathcal{D}
&\equiv& \sin\theta_{\mathrm{obs}} \sqrt{r_{\mathrm{obs}}^{2}+a^{2}} -
y \cos\theta_{\mathrm{obs}}\,.
\end{eqnarray}
Finally, to obtain the ray’s velocity components in Boyer- Lindquist coordinates one has to differentiate Eqs.(\ref{CartBLr})--(\ref{CartBLphi}), yielding:
\begin{eqnarray}
-\Sigma \ \dot{x}^{r} &=& r \mathcal{R} \sin\theta \sin\theta_{\mathrm{obs}} \cos \Phi+\mathcal{R}^{2} \cos\theta \cos\theta_{\mathrm{obs}} \,, \label{rdot_initial} \\
-\Sigma \ \dot{x}^{\theta} &=& \mathcal{R} \cos\theta \sin\theta_{\mathrm{obs}} \cos \Phi - r \sin\theta \cos \theta_{\mathrm{obs}} \,, \label{thetadot_initial}\\
\mathcal{R} \ \dot{x}^{\phi} &=& \sin\theta_{\mathrm{obs}} \sin \Phi \ \!  \mathrm{cosec}\theta \,, \label{phidot_initial}
\end{eqnarray}
where
\begin{eqnarray}
\Sigma &\equiv& r^{2} +a^{2}\cos^{2}\theta \,, \\
\mathcal{R} &\equiv& \sqrt{r^{2}+a^{2}} \,, \\
\Phi &\equiv& \phi-\phi_{\mathrm{obs}} \,.
\end{eqnarray}

\begin{figure*}
\centering

{\includegraphics[scale=0.095]{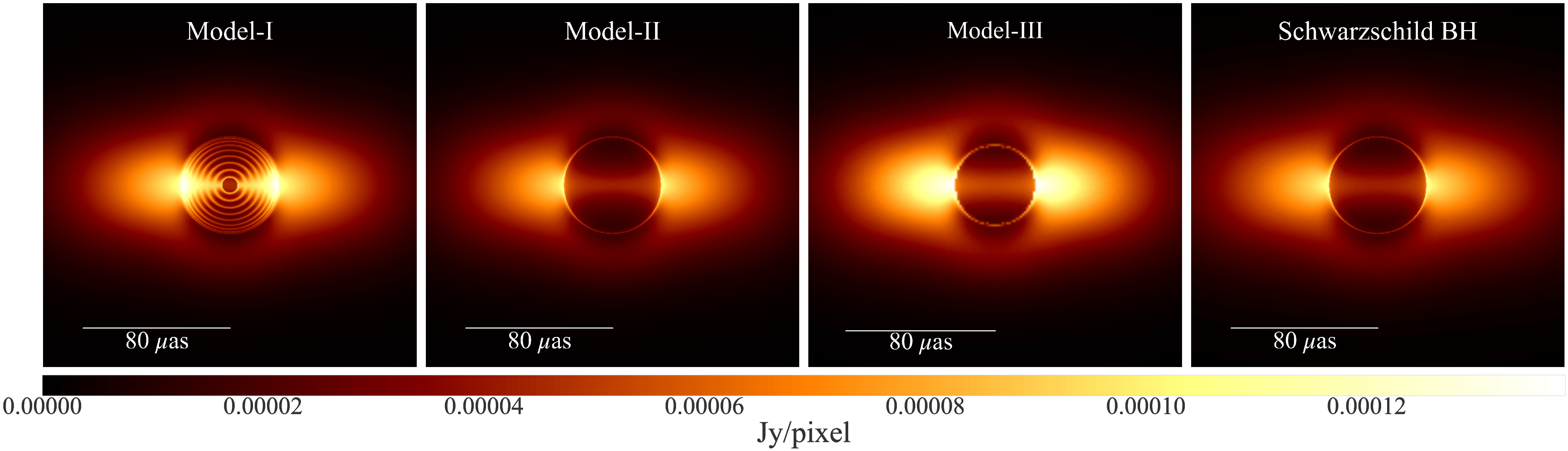}}
\hspace{0.2cm}
\caption
{Model images for Sgr A*, parameters details are given in the Table  (\ref{fig:model}).}\label{fig:model}
\end{figure*}

Using the above formalism, the images for various models described in the previous section and parameter values in Table~\ref{table:model} can be generated. Different columns in Figure~\ref{fig:model} depict these ray-traced models. Contrary to Model II and Schwarzschild BH, Model-I produces a particular image with multiple Einstein rings without casting a shadow. The authors previously produced a similar full-moon type image in \cite{shaikh1}. On the other hand, Model III casts a relatively smaller shadow than Model-II and Schwarzschild BH.

\hspace{0.4cm}

\section{EHT VLBI Observations}
\label{VLBI}
Continuing the analysis of the previous section, the next step of visualizing these images with VLBI is applied.
For this purpose, Three different array configurations are utilised, including current and future observations. The label EHT-I, EHT-II and EHT-III for 2017, 2021 and 2025 observations are given, respectively, as mentioned in Table ~(\ref{table:sites}). It should be noted that EHT-III is one of the possible configurations for future (2025) observations. Meanwhile, EHT-II completed the observations for 2021 in early April. The images produced in the preceding section are infinite-resolution in that they are produced by a pristine source that encounters no interstellar matter on its way to a perfect detector that can detect every photon that passes through it. In reality, interstellar scattering will occur, and the baselines will only cover a portion of the observing region.

\begin{table}
	\centering
	\caption{Locations of Existing/Future Sites in the Event Horizon Telescope Array.}
	\label{table:sites}
	\begin{tabular}{lccr} 
		\hline
		\hline
		\textbf{EHT-I}    &    \textbf{Latitude}    &    \textbf{Longitude}    &    \textbf{SEFD.}$^{\dagger}$ \\
		\textbf{}    &    \textbf{($\deg$)}    &    \textbf{($\deg$)}    &    \textbf{(Jy)} \\
		\hline
		\hline
		ALMA & -23.03 & -67.75 & 90    \\
        APEX & -23.01 & -67.76 & 3500  \\
        JCMT & 19.82 & -155.48 & 6000  \\
        LMT & 18.98 & -97.31 & 600   \\
        IRAM$^{\star}$    & 36.88  & -3.39   & 1400  \\
        SMA & 19.82 & -155.48 & 4900  \\
        SMT & 32.70 & -109.89 & 5000  \\
        SPT & -90.00 & 45.00 & 5000  \\
        \hline
        \hline
        \textbf{EHT-II, (Additional)}    &    \textbf{Latitude}    &    \textbf{Longitude}    &    \textbf{SEFD.}$^{\dagger}$ \\
        \textbf{}    &    \textbf{($\deg$)}    &    \textbf{($\deg$)}    &    \textbf{(Jy)}\\
        \hline
        \hline
        GLT & 76.54 & -68.69 & 10000 \\
        KP & 31.96 & -111.61 & 10000 \\
        NOEMA & 44.63 & 5.91 & 700 \\
        \hline
        \hline
        \textbf{EHT-III, (Additional)}    &    \textbf{Latitude}    &    \textbf{Longitude}    &    \textbf{SEFD.}$^{\dagger}$ \\
        \textbf{}    &    \textbf{($\deg$)}    &    \textbf{($\deg$)}    &    \textbf{(Jy)} \\
        \hline 
        \hline
        BAJA  & 30.87  & -115.46 & 10000 \\
        BOL   & -16.25 & -68.13  & 10000 \\
        CARMA & 37.1   & -118.14 & 10000 \\
        DRAK  & -29.3  & 29.27   & 10000 \\
        GAM   & 23.25  & 16.17   & 10000 \\
        HAY   & 42.43  & -71.49  & 2500  \\
        KAUAI & 21.79  & -159.51 & 10000 \\
        KEN   & -0.15  & 37.31   & 10000 \\
        PDB{}   & 44.44  & 5.91    & 1500  \\
        PIKES & 38.65  & -105.04 & 10000 \\
        VLT   & -24.48 & -70.4   & 10000 \\
		\hline
	\end{tabular}
	
     $\star$ sites used for the observation only in EHT-I, III. \\
     $^{\dagger}$System Equivalent Flux Density
\end{table}

\begin{figure*}
\centering

{\includegraphics[scale=0.20]{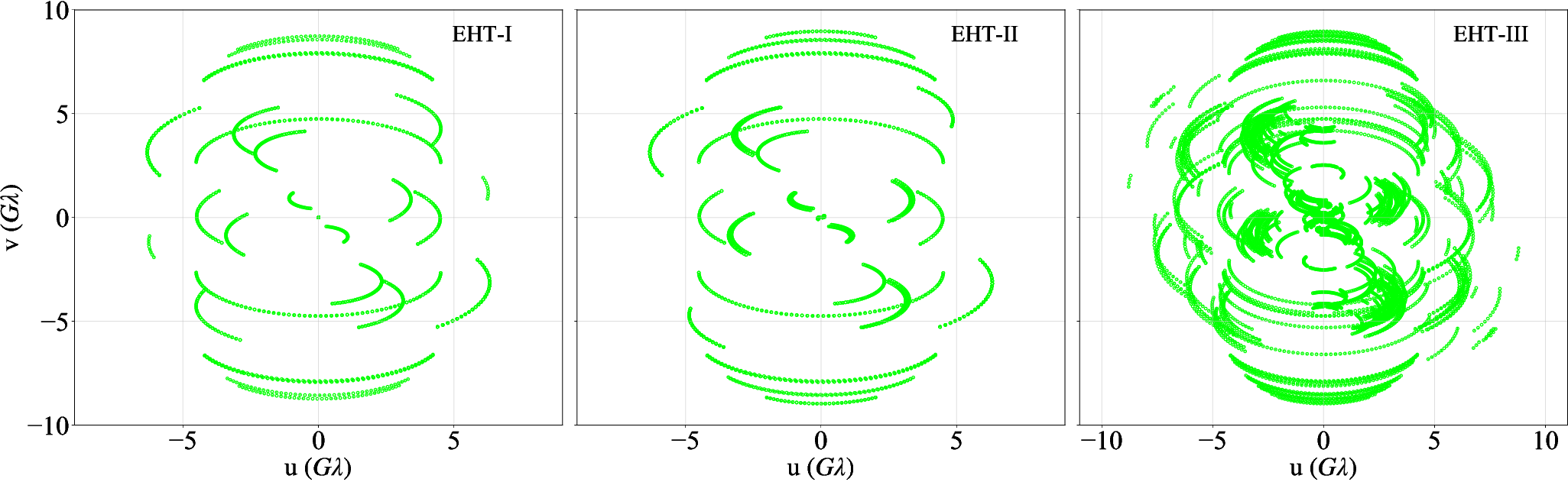}}
\hspace{0.2cm}
\caption
{Baseline coverage of Sgr A$^\star$ for the array configuration of EHT mentioned in Table~\ref{table:sites}.}\label{fig:uv}
\end{figure*}

\begin{figure*}
\centering

{\includegraphics[scale=0.1]{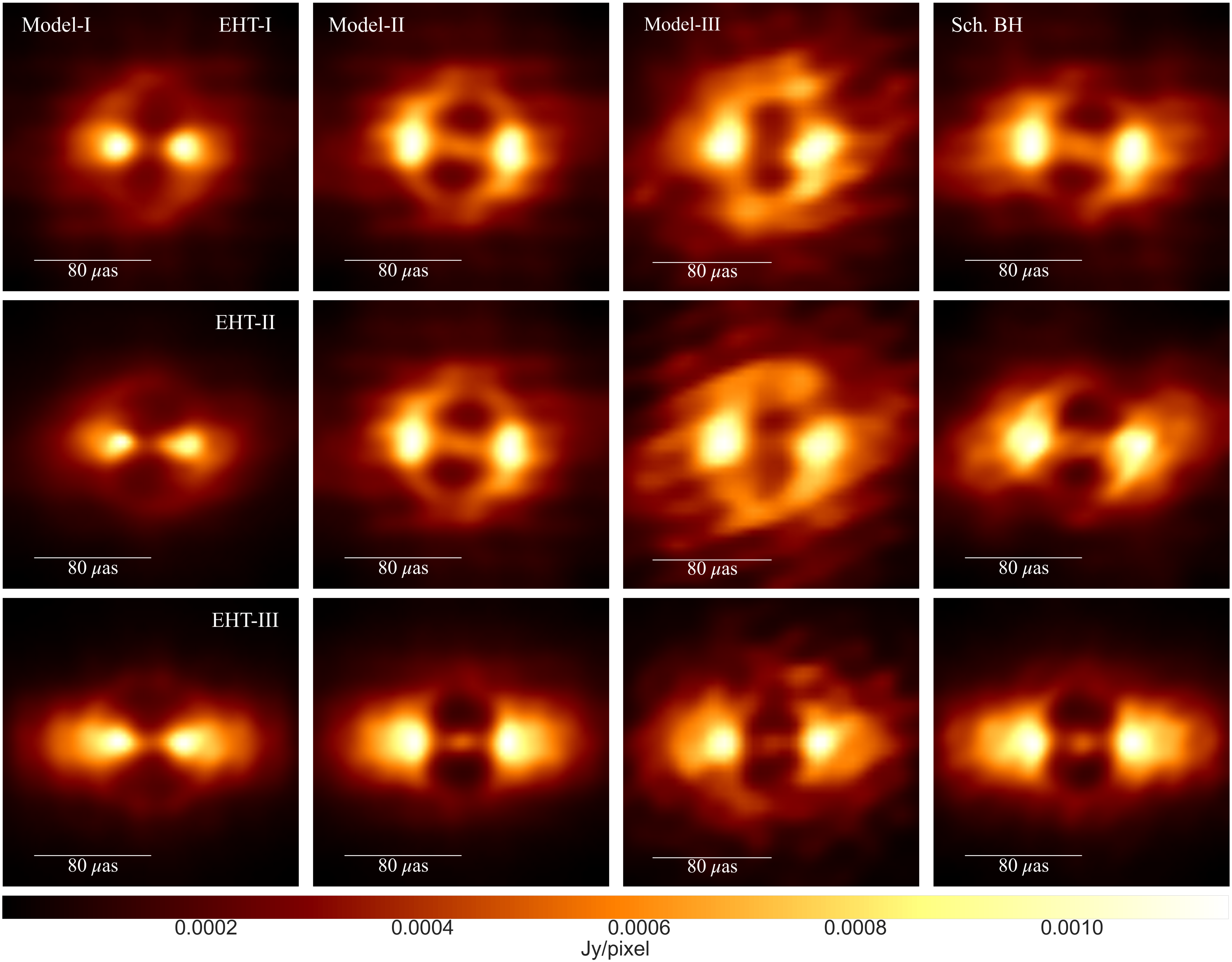}}
\hspace{0.2cm}
\caption
{Reconstructed images of Sgr A* for models described in section-I with different array configurations of EHT (top to bottom: EHT-I, EHT-II, EHT-III).}\label{fig:rec-images}
\end{figure*}

The synthetic radio images  are generated with $\texttt{ehtim}$ \cite{Chael:2018oym} using the array configurations mentioned in Table~(\ref{table:sites}). The following parameters were used in the simulations: $\Delta\nu = 4$ GHz bandwidth, $t = 24$ hours, corresponding to a full day, at a central frequency of $230$ GHz. The total observation time is one of the most important parameters in the imaging process, and this unusually long observation time allows to present the ideal-case scenario. The (u, v) coverage grows larger as the observation time increases, and reconstruct a better image. The Maximum Entropy Method (MEM) is used to reconstruct static images from synthetic VLBI data.
Following this procedure, the synthetic observations at the Galactic center are performed. The visibility amplitudes are calculated by Fourier transforming the images and sampling them over the projected baselines of different arrays (EHT-I, EHT-II and EHT-III). The effects of thermal noise and phase errors are included to mimic realistic observations during the simulations. As it can be seen in Fig. (\ref{fig:uv}), the (u,v) coverage is subsequently increasing as more antennas are added to the configuration. Images reconstructed for different models and arrays are presented in Fig. (\ref{fig:rec-images}). It should also be noted that the same imaging script was used for all the reconstructions used here. In the next sections, a qualitative and quantitative analysis of the results produced in this section would discuss.

\subsection{Qualitative analysis}

Overall visual inspection of the images clearly shows the difference in the images of different models with different arrays. Starting from the left-hand side, as discussed in the previous section Model-I produces an interesting concentric rings type image. So the corresponding synthetic image for different arrays does not give a shadow-type structure. Due to the spherically symmetric nature of the model, the image in the third row for EHT-III shows a nearly circular structure. Those multiple rings are not visible in the reconstruction. This aberration is due to the resolution limit, making it very difficult to resolve them.

Model II casts a shadow and is clearly visible in all the arrays. Due to additional emissions in the line of sight, the shadow structure appears to be divided. Although, it should be noted that it is a consequence of the radiative transfer modelling and raytracing that can be seen such emission along with the particular inclination. There are differences between Model-I and Model II in the image structure. Comparing Model-II with Schwarzschild BH is evident that significant difference does not appear due to the same shadow size and similar emission regions. Hence, it can be stated that Model II and Schwarzschild BH mimic each other, and one cannot tell the difference between them from the VLBI images.

For Model III, the shadow size is smaller than the other models prescribed here, which can be seen from the ray-traced images. So, in the EHT-I image, the shadow region is not as noticeable. The same may be said for EHT-II. The situation improves substantially with EHT-III, and the shadow region becomes more apparent. Model III appears to be distinct from the Black Hole case. Even so, the change is not all that substantial. This coincides with the previous case as well for Model II. On a relative scale, Model III is more distinct than Model II.

\begin{table}
	\centering
	\caption{Normalized Cross-Correlation coefficient for the comparison between the reconstructed of various models with the Schwarzschild BH.}
	\vspace{0.35cm}
	\label{table:comp}
	\begin{tabular}{llll} 
		\hline 
		\hline 
		\textbf{Sch. BH}    &   \textbf{Model-I}    &    \textbf{Model-II}     &     \textbf{Model-III}\\
		\hline
		\textbf{EHT-I}    & 0.88 &  0.96 & 0.74 \\	
        \textbf{EHT-II}   & 0.90 &  0.98 & 0.73 \\
        \textbf{EHT-III}  & 0.92 &  0.99 & 0.75 \\

	    \hline
	\end{tabular}
\end{table}

\subsection{Quantitative Analysis}

Image-comparison measures, such as the Normalized Cross-correlation coefficient (NCC), structural dissimilarity index etc., can be used to make a more quantitative evaluation of the degree of similarity among the various images under consideration. For this purpose, NCC has been utilised to compare the resultant images.
If two images are similar to each other, the maximum value NCC can have 1, and the maximum dissimilarity will give NCC value to be around 0.
This can be incorporated with three different arrays: EHT-I, EHT-II and EHT-III. NCC is then computed between the corresponding reconstructed Schwarzschild BH image and the other models, as seen in Figure~\ref{fig:rec-images}. Lower NCC values suggest that these models can be differentiated from the Schwarzschild BH case. Given that, Model-I and Model-III produce distinct features (as explained in the previous section), it is expected to deduce a similar result from NCC.
Table~\ref{table:comp} summaries these results. Model-I and Model III give relatively lower values than Model II, suggesting that the images are distinguishable.
Model-I having no shadow, and Model-III having a smaller shadow size as shown in  Figure~\ref{fig:model}, are responsible for various structures in the reconstructed images resulting in relatively lower values of NCC than Model-II, which has the same shadow size as Schwarzschild BH.

\section{Discussions and Conclusions}
In this chapter, Sgr A* as the JMN-1 or JNW naked singularity has been modelled with the possibility of having either a shadow-like structure or a pristine source (concentric ring-like structure) emitting from all the regions. The raytracing algorithm is applied to solve the geodesic equations and then solve the general relativistic radiative transfer equation. An analytic model for emission and absorption coefficients is utilised for generating emission maps. Afterwards, the synthetic observations have been produced for the prescribed models of the source considering three different array configurations, namely EHT-I, EHT-II and EHT-III and reconstruct the images.

From the overall analysis, it is seen that the naked singularity models are distinguishable from a Black hole with VLBI imaging. Both qualitative and quantitative analyses of the images provide a rough estimate of the differences between the models. Model-I and Model-II exhibit different accretion structures and imaging artifacts, which can be seen in the image reconstructions. Meanwhile, Model II remains a Black hole mimicker and is almost similar to a Black hole in size and emission region. The NCC quantifies this inference and has values near $1$ for only Model II. However, for other models, the values are relatively smaller, differentiating the models quantitatively. The NCC can be used to explore these potential differences to some extent.

The synthetic VLBI images for the EHT-I array for Model-II, Model-III and Schwarzschild BH give shadow-like features. However, on the other hand, Model-I does not hint at that since it does not theoretically produce a shadow. As observed further the number of arrays increase in the configuration, the Fourier domain fills up significantly, generating a clear image. So, when the models are compared based on increased arrays, they retain their original state in shadow. 

The above analysis concludes that when Sgr A* is observed on these baselines and an image is reconstructed, if the image showcase a shadow-like structure, one cannot distinguish between a Black hole and a naked singularity. Moreover, if Sgr A* does not contain a BH but a naked singularity (Model-I), then one can expect an image without shadow as seen in Fig ~(\ref{fig:rec-images}). No matter how much the baseline is increased or the number of arrays are added, there will not be any image having a shadow feature. Indeed, the absence of the shadow in the JMN-1 spacetime introduces essential and fundamental differences in the flow dynamics and images. It can act as a possible observable to test the presence of a naked singularity. This can thus allow to probe the nature of Sgr A* with shadows and VLBI imaging. 

Although the analysis presented in this work is based on the images, there are a few caveats that require specific attention. Firstly, the radiative modelling used for the accretion region is a simple analytical model. The number density can be considered analogous to the electron number density in RIAF modelling. RIAF spatial distributions of the electron temperature $T_e$ and thermal electron density $n_e$ are usually described by a hybrid combination \cite{Pu:2018ute}. Different eDF's (electron distribution functions) are also incorporated in the emission and absorption coefficients to model the emission regions more accurately \cite{Pandya:2016qfh}. The radiative cooling time-frame for RIAF is significantly longer than the accretion timescale for a cold, geometrically thin, Keplerian revolving disk, resulting in a hot, geometrically thick flow with a sub-Keplerian rotation \cite{Narayan:1994is,Narayan:1996wu}. Introducing different flow dynamics can produce some interesting and distinguishable features in the resulting images \cite{Pu:2016qak} as well.
To conclude, these following caveats would be interesting topics to follow up the current work in the future. Additionally, they would provide a more robust way to probe the nature of Sgr A* for several observable. Recent results from the 2017 and 2021 observational campaigns of EHT may provide more insight into the argument. Concerning the general relativity (and not the other modified gravity theories), these naked singularity solutions present an excellent alternative to the conventionally used Black hole. Since, rotation is pervasive in astrophysical systems, the rotating solution of the compact body is believed to be the most realistic for studying any physical process that occurs in the vicinity of a black hole. Therefore, the next chapter, would cover the shadow properties in the rotating compact objects. 

\chapter{Shadows cast by rotating black holes and naked singularities}

The inclusion of rotation in the non-rotating spacetime geometry makes it physically more realistic spacetime.
In 1965, Newman and Janis showed that the Kerr metric could be obtained from the Schwarzschild metric using a complex coordinate transformation \cite{Newman:1965tw}. Using the same method for the static Reissner-Nordström metric, one can derive the Kerr-Newman metric, which represents a spacetime geometry of the electrically charged and rotating black hole \cite{Newman:1965my}. The application of the NJA for generating interior solutions which match smoothly to the external Kerr metric is studied in \cite{Drake1997}. In this algorithm, authors have used two different types of complexification with no apparent reason. There are many ways to complexify the coordinates. However, in the literature, as mentioned above, the only reason given for choosing those particular complexifications is that it is successful in generating the Kerr and Kerr-Newman metric. However, the success of the NJA is limited to the spacetimes which at least satisfy the reciprocal condition, $g_{tt}g_{rr} = -1$. Later, it was proved by S.P. Drake, and P. Szekeres that the only perfect fluid solution generated by the NJA is the Kerr metric, and the only Petrov typed D solution to the Einstein-Maxwell equation is the Kerr-Newman metric \cite{Drake2000}.

In this chapter, the rotating JNW naked singularity spacetime is constructed using NJA. The NJA is analyzed with and without complexification methods and find that the energy conditions do satisfied when the complexification step is skipped. Then the shadows cast by rotating JNW naked singularity is studied and compared with those cast by the Kerr black hole. The shadow of the deformed Kerr metric is also studied and discussed.

\section{Construction of the rotating JNW spacetime}
\label{Sec_Rotating_JNW_NJA}

In this section, the rotating JNW naked singularity spacetime is determined using the NJA. The final result of the NJA is obtained in the null coordinates, which must be transformed into the Boyer-Lindquist coordinates (BLC). The BLCs are more convenient to work with the rotating spacetime metric. It is analysed that it is not possible to perform the BLC transformation when considering the JNW spacetime metric (\ref{JNW_metric}) as a seed metric. The complexification step in NJA \cite{Mustapha3} is dropped to resolve this problem.

\subsection{JNW metric and the Newman-Janis Algorithm}
The JNW spacetime metric is given by,
\begin{eqnarray}
    ds^2 = -\left(1-\frac{2M}{r\nu} \right)^{\nu} dt^2 + \left(1-\frac{2M}{r\nu} \right)^{-\nu} dr^2 + r^2 \left(1-\frac{2M}{r\nu} \right)^{1-\nu} d\Omega^2
    \label{JNW_metric}
\end{eqnarray}
where, $d\Omega^2 = d\theta^2 + \sin^2\theta d\phi^2$. It is straightforward to show that the following expression is the final form of NJA, considering the JNW metric (\ref{JNW_metric}) as a seed metric.
\begin{eqnarray}
    ds^2 = - \left(1-\frac{2Mr}{\nu \rho^2} \right)^{\nu} du^2 - 2 du dr -2a\sin^2\theta \left[1 - \left(1 - \frac{2Mr}{\nu \rho^2} \right)^{\nu} \right] du d\Phi \nonumber \\ +\, \rho^2 \left(1 - \frac{2Mr}{\nu \rho^2} \right)^{1-\nu} d\Omega^2 + 2a\sin^2\theta dr d\Phi + a^2\sin^4\theta \left[2 - \left(1-\frac{2Mr}{\nu \rho^2} \right)^{\nu} \right] d\Phi^2
    \label{JNW_metric_EFC}
\end{eqnarray}
The above form of the spacetime metric seems rather complicated. It can be made simpler and symmetrical by transforming it into the Boyer-Lindquist coordinates (BLC). The BLC $(t, r, \theta, \phi)$ represents the rotating black hole spacetime metrics because all the off-diagonal terms of the metric, except $dt d\Phi$, vanish; and its axial symmetry becomes apparent. The transformation into the BLC requires the following:
\begin{eqnarray}
    && du = dt - \xi(r) dr
    \label{du},\\
    && d\Phi = d\phi - \chi(r) dr.
    \label{dphi}
\end{eqnarray}
It is noted that the transformation functions, i.e., $\xi(r)$ and $\chi(r)$, strictly depend on coordinate $r$ only; otherwise eq. (\ref{du}) and (\ref{dphi}) will not be integrable. To write eq. (\ref{JNW_metric_EFC}) into the BLC like form, the transformation functions must take the following form,
\begin{eqnarray}
    && \xi(r,\theta) = \frac{1}{\Delta} \left[\rho^2 \left(1 - \frac{2Mr}{\nu \rho^2} \right)^{1-\nu} + a^2\sin^2\theta \right],\\
    && \chi(r) = \frac{a}{\Delta},\\
    && \Delta = r^2 - \frac{2Mr}{\nu} + a^2.
\end{eqnarray}
It can be seen that the transformation function $\xi$ also depends on coordinate $\theta$, and therefore the eq. (\ref{du}) is not integrable. In these coordinates, eq. (\ref{JNW_metric_EFC}) is written as

\begin{eqnarray}
    ds^2 = -\left(1-\frac{2Mr}{\nu \rho^2} \right)^{\nu} dt^2 + \rho^2 \left(1 - \frac{2Mr}{\nu \rho^2}\right)^{1-\nu} \left(\frac{dr^2}{\Delta} + d\Omega^2 \right)\nonumber \\-\, 2a\sin^2\theta \left[1 - \left(1 - \frac{2Mr}{\nu \rho^2} \right)^{\nu} \right] dt d\phi +  a^2\sin^4\theta \left[2 - \left(1 - \frac{2Mr}{\nu \rho^2} \right)^{\nu} \right] d\phi^2.
    \label{Rotating_JNW_NJA}
\end{eqnarray}
It can be seen that the above metric reduces to the JNW spacetime metric (\ref{JNW_metric}) when $a=0$. However, the transformation into the BLC was improper in obtaining eq. (\ref{Rotating_JNW_NJA}), one can check the energy conditions because the NJA is just an algorithm that successfully works for Kerr and Kerr-Newman metric. Thus, it cannot be commented whether the final result eq. (\ref{Rotating_JNW_NJA}) is a valid solution of the Einstein field equation. It can only be answered once it is known whether or not the energy conditions for the metric (\ref{Rotating_JNW_NJA}) are satisfied. The energy conditions for the above spacetime metric are discussed in Appendix A. It can be concluded that the rotating JNW spacetime metric (\ref{Rotating_JNW_NJA}) was obtained using NJA with complexification violates the energy conditions.
Now, the $\theta$ dependency of the transformation function $\xi$ is dropped by considering a slow rotation approximation in which the second and higher-order terms of rotation parameter ($a$) are ignored. Thus, eq. (\ref{Rotating_JNW_NJA}) reduces to the following form,
\begin{eqnarray}
    ds^2 = -\left(1-\frac{2M}{\nu r} \right)^{\nu} dt^2 + \left(1-\frac{2M}{\nu r} \right)^{-\nu} dr^2 - 2a\sin^2\theta \left[1 - \left(1-\frac{2M}{\nu r}\right)^{\nu} \right] dt d\phi \nonumber \\ +\, r^2 \left(1-\frac{2M}{\nu r} \right)^{1-\nu} d\Omega^2
    \label{Slow_rotation_JNW}
\end{eqnarray}
And, the transformation functions $\chi$ and $\xi$ become
\begin{eqnarray}
     && \xi(r) = \left(1-\frac{2M}{\nu r} \right)^{-\nu},  \\
    && \chi(r) = \frac{a}{r^2} \left(1-\frac{2M}{\nu r} \right)^{-1}. 
\end{eqnarray}
Eq. (\ref{Slow_rotation_JNW}) is the slow rotation form of the JNW spacetime metric. The rotating JNW spacetime metric (\ref{Rotating_JNW_NJA}) obtained using the NJA with complexification is not properly transformed into the BLC because of the complexification of the coordinates. This issue can be solved by skipping the complexification step, which is discussed in the next section.

\subsection{Newman-Janis Algorithm without Complexification}

\noindent Consider a general form of the static and spherically symmetric spacetime metric as,
\begin{equation}
ds^2=-G(r)dt^2+\frac{dr^2}{F(r)}+H(r)d\Omega^2.
\label{general_metric}
\end{equation}
To apply the algorithm, transform it into the null coordinates $\{u,r,\theta,\phi\}$ using the coordinate transformation,
\begin{equation}
du=dt-\frac{dr}{\sqrt{F(r)G(r)}}
\end{equation}
In null coordinates, the spacetime metric (\ref{general_metric}) takes the form,
\begin{equation}
ds^2=-G(r)du^2-2\sqrt{\frac{G(r)}{F(r)}}dudr+H(r)d\Omega^2.
\label{general_metric_null}
\end{equation}
The null tetrads of the above spacetime metric are,
\begin{eqnarray}
&& l^\mu=\delta^\mu_1\,, \label{l_mu} \\
&& n^\mu=\sqrt{\frac{F(r)}{G(r)}}\delta^\mu_0-\frac{1}{2}F(r)\delta^\mu_1\,, \label{n_mu} \\
&& m^\mu=\frac{1}{\sqrt{2H(r)}} \left(\delta^\mu_2+\frac{i}{\sin\theta}\delta^\mu_3 \right). \label{m_mu}
\end{eqnarray}
The next step of the NJA is the complexification, which generalize the function (i.e. $G(r), F(r), H(r)$) as the real function of radial coordinate $r$ and it's complex conjugate $\bar{r}$. The trick is to skip this step and proceed directly to the next step, which is the complex coordinate transformation given as,
\begin{eqnarray}
&& r^\prime=r+ia\cos\theta \label{r_prime}, \\
&& u^\prime=u-ia\cos\theta \label{u_prime}.
\end{eqnarray}
Under the coordinate transformation, by considering that the components of the metric tensor (\ref{general_metric}) are transformed as $$\{G(r),F(r),H(r)\}\to\{A(r,\theta,a),B(r,\theta,a),\psi(r,\theta,a)\},$$ where the following conditions are satisfied.
\begin{eqnarray}
&& \lim\limits_{a\to 0}A(r,\theta,a)=G(r)\,, \\
&& \lim\limits_{a\to 0}B(r,\theta,a)=F(r)\,, \\
&& \lim\limits_{a\to 0}\psi(r,\theta,a)=H(r).
\end{eqnarray}
Now, perform the complex coordinate transformation (\ref{r_prime}), (\ref{u_prime}) on the null tetrad given in equations (\ref{l_mu}), (\ref{n_mu}), (\ref{m_mu}),
\begin{eqnarray}
&& l^\mu=\delta^\mu_1\,, \\
&& n^\mu=\sqrt{\frac{B}{A}}\delta^\mu_0-\frac{1}{2}B\delta^{\mu}_1\,, \\
&& m^\mu=\frac{1}{\sqrt{2\psi}} \left(\delta^\mu_2+ia\sin\theta (\delta^\mu_0-\delta^\mu_1) +\frac{i}{\sin\theta}\delta^\mu_3 \right).
\end{eqnarray}
The fourth null vector $\bar{m}^\mu$ is just a complex conjugate of $m^\mu$. These are the components of the null tetrad of the final metric tensor in null coordinates $(u, r, \theta, \Phi)$.
\begin{eqnarray}
    ds^2 = A du^2+2\sqrt{\frac{A}{B}}dudr+2a\sin^2\theta\left(\sqrt{\frac{A}{B}}-A \right)dud\Phi - 2a\sin^2\theta\sqrt{\frac{A}{B}}drd\Phi \nonumber \\-\, \psi d\theta^2  - \sin^2\theta\left[\psi+a^2\sin^2\theta\left(2\sqrt{\frac{A}{B}}-A \right)\right]d\Phi^2. \,\,\,\,\,\,\,\,\,\,\,
    \label{general_rotating_null}
\end{eqnarray}
The spacetime metric is determined in the null coordinates $\{u,r,\theta,\Phi\}$, which can be transformed into the BLC $\{t,r,\theta,\phi\}$ as,
\begin{eqnarray}
&& du=dt-\lambda(r)dr,
\label{du new}\\
&& d\Phi=d\phi-\chi(r)dr.
\label{dphi new}
\end{eqnarray}
The transformation functions $\chi$ and $\lambda$ should strictly depend on the radial coordinate ($r$) only for eq. (\ref{du new}), (\ref{dphi new}) to be integrable, which is not possible in NJA with complexification. In the Boyer-Lindquist form, the spacetime metric contains only one off-diagonal term, which is $dtd\phi$. Therefore, using the above coordinate transformation (\ref{du new}) and (\ref{dphi new}), and dropping all off-diagonal terms except $dtd\phi$, the following two expressions are derived as,

    \begin{eqnarray}
        && A\lambda(r)-\sqrt{\frac{A}{B}}+a\sin^2\theta\left(\sqrt{\frac{A}{B}}-A \right)\chi(r)=0
        \label{dtdr},\\
        && a\left(\sqrt{\frac{A}{B}}-A \right)\lambda(r)+a\sqrt{\frac{A}{B}}+\left[\psi-a^2\sin^2\theta\left(2\sqrt{\frac{A}{B}}-A \right)\chi(r) \right]=0.
        \label{drdphi}
    \end{eqnarray}
Solving these expressions for $\lambda$ and $\chi$, 
\begin{eqnarray}
&& \chi(r,\theta)=\frac{a}{\psi B+a^2\sin^2\theta}\,, \\
&& \lambda(r,\theta)=\frac{\psi\sqrt{\frac{B}{A}}+a^2\sin^2\theta}{\psi B+a^2\sin^2\theta}.
\end{eqnarray}
The BLC transformation functions $\chi$ and $\lambda$ depends on $\theta$, but this time there are three unknown functions $A(r,\theta,a), B(r,\theta,a)$ and $\psi(r,\theta,a)$ which play an important role in dropping the $\theta$ dependency of $\chi$ and $\lambda$. The following form of the transformation functions is chosen because it successfully works in Kerr and Kerr-Newman case. Therefore above equations reduces into,
\begin{eqnarray}
&& \chi(r)=\frac{a}{FH+a^2},
\label{chi}\\
&& \lambda(r)=\frac{K+a^2}{FH+a^2}
\label{lambda},
\end{eqnarray}
where, $K=H\sqrt{\frac{F}{G}}$.
Substitute the above expression of $\chi$ and $\lambda$ into the equations ($\ref{dtdr}$) and ($\ref{drdphi}$), and solve them for the unknown functions $A(r,\theta,a)$ and $B(r,\theta,a)$,
\begin{eqnarray}
&& A(r,\theta,a)=\frac{(FH+a^2\cos^2\theta)\psi}{(K+a^2\cos^2\theta)^2},\\
&& B(r,\theta,a)=\frac{FH+a^2\cos^2\theta}{\psi}.
\end{eqnarray}
The function can be verified by $A(r,\theta,a)$ and $B(r,\theta,a)$ respectively, reduce to the function $G(r)$ and $F(r)$ when $a\to 0$ as,
\begin{eqnarray}
&& \lim\limits_{a\to 0}A(r,\theta,a)=G(r),\\
&& \lim\limits_{a\to 0}B(r,\theta,a)=F(r).
\end{eqnarray}
Therefore, under the BLC coordinate transformation, the metric (\ref{general_rotating_null}) becomes,

    \begin{eqnarray}
         ds^2=-\frac{(FH+a^2\cos^2\theta)\psi}{(K+a^2\cos^2\theta)^2}dt^2+\frac{\psi}{FH+a^2}dr^2-2a\sin^2\theta\left(\frac{K-FH}{(K+a^2\cos^2\theta)^2} \right)\psi dtd\phi \nonumber \\+\,\psi d\theta^2 + \psi sin^2\theta\left[1+a^2\sin^2\theta\left(\frac{2K-FH+a^2\cos^2\theta}{(K+a^2\cos^2\theta)^2} \right) \right]d\phi^2,
        \label{general_rotating_BLC}
    \end{eqnarray}
where, $\psi(r,\theta,a)$ is still unknown. It must satisfy the following constraint, which corresponds to the vanishing component of the Einstein tensor $G_{r\theta}=0$,
\begin{equation}
    3a^2\sin{2\theta}\psi^2 K_{,r} + (K+a^2\cos^2\theta)^2 (3 \psi_{,\theta}\psi_{,r} - 2 \psi \psi_{,r\theta}) = 0.
\end{equation}
This non-linear partial differential equation can be solved for $\psi$. One can verify that a solution of the above partial differential equation is,
\begin{equation}
    \psi(r,\theta,a) = K(r) + a^2\cos^2\theta.
    \label{psi}
\end{equation}
For the Kerr and Kerr-Newman metric, $K(r)=r^2$, and the above expression reduces to $\psi=r^2+a^2\cos^2\theta$. After substituting equation (\ref{psi}) into the general form of the rotating spacetime metric (\ref{general_rotating_BLC}), if the static spacetime metric (\ref{general_metric}) can be obtained in the limit $a\to 0$, the rotating solution is known as normal fluid. Otherwise, it is known as conformal fluid. The normal fluid solution requires,
\begin{equation}
    \lim\limits_{a\to 0}\psi(r,\theta,a)=H(r),
\end{equation}
which implies $F(r)=G(r)$.
The metric (\ref{general_rotating_BLC}) can be written in Kerr-like form as,
\begin{eqnarray}
    ds^2 = -\left(1-\frac{2f}{\rho^2} \right)dt^2 + \frac{\rho^2}{\Delta} dr^2 + \rho^2 d\theta^2 + \frac{\Sigma\sin^2\theta}{\rho^2} d\phi^2 -  \frac{4af\sin^2\theta}{\rho^2} dt d\phi, \,\,\,\,\,\,\,\,\,\,
    \label{general_rotating_metric}
\end{eqnarray}
where,
\begin{eqnarray}
    && \rho^2 = K(r) + a^2\cos^2\theta\,, \\
    && f = \frac{K(r)-F(r)H(r)}{2}\,, \\
    && \Delta = F(r) H(r) + a^2\,, \\
    && \Sigma = \big(K(r)+a^2 \big)^2 - a^2\Delta\sin^2\theta.
\end{eqnarray}
When $a\to 0$, the rotating solution (\ref{general_rotating_metric}) reduces to the static, and spherically symmetric spacetime metric if $F(r) = G(r)$.

\subsection{Rotating JNW metric}
A new rotating spacetime metric is formed that reduces to the static JNW naked singularity spacetime metric (\ref{JNW_metric}) in the limit $a\to 0$. The components of the JNW spacetime metric is substituted into eq. (\ref{general_rotating_metric}) and the following expression is obtained,
\begin{eqnarray}
    ds^2 = -\left(1-\frac{2f}{\rho^2} \right)dt^2 + \frac{\rho^2}{\Delta} dr^2 + \rho^2 d\theta^2 + \frac{\Sigma\sin^2\theta}{\rho^2} d\phi^2 - \frac{4af\sin^2\theta}{\rho^2} dt d\phi, \,\,\,\,\,\,\,\,\,\,\,\,
    \label{JNW_rotating}
\end{eqnarray}
where,
\begin{eqnarray}
    && 2f = r^2 \left(1 - \frac{2M}{r\nu} \right) \left[-1 + \left(1 - \frac{2M}{r\nu} \right)^{-\nu} \right]\,, \\
    && \rho^2 = r^2 \left(1 - \frac{2M}{r\nu} \right)^{1-\nu} + a^2\cos^2\theta\,, \\
    && \Delta = r^2 - \frac{2Mr}{\nu} + a^2\,, \\
    && \Sigma = (\rho^2 + a^2\sin^2\theta)^2 - a^2\Delta\sin^2\theta.
\end{eqnarray}
One can verify that it reduces to the Kerr metric when $\nu = 1$. Also, it reduces to eq. (\ref{Slow_rotation_JNW}) under the slow rotation approximation. To check whether the spacetime metric is physically valid, one must proceed to check the energy conditions. Consider the inverse form of the rotating spacetime metric (\ref{JNW_rotating}),
\begin{eqnarray}
    (\partial_s)^2 = - \frac{(K+a^2)^2-a^2\Delta\sin^2\theta}{\rho^2\Delta} (\partial_t)^2 + \frac{\Delta}{\rho^2} (\partial_r)^2 + \frac{1}{\rho^2} (\partial_\theta)^2 + \frac{\Delta-a^2\sin^2\theta}{\rho^2\Delta\sin^2\theta} (\partial_\phi)^2 \nonumber\\-\, \frac{4 a M K}{r\rho^2\Delta} \partial_t\partial_\phi. \,\,\,
\end{eqnarray}
It can also be written as,
\begin{eqnarray}
   (\partial_s)^2 = -\left(\frac{K+a^2}{\rho\sqrt{\Delta}}\partial_t + \frac{a}{\rho\sqrt{\Delta}}\partial_{\phi} \right)^2 + \frac{\Delta}{\rho^2} (\partial_r)^2 
+ \frac{1}{\rho^2} (\partial_{\theta})^2 + \left(\frac{a\sin\theta}{\rho}\partial_t + \frac{1}{\rho\sin\theta}\partial_{\phi} \right)^2. \,\,\,
\label{Inverse_general_rotating_metric}
\end{eqnarray}
Now, choose a set of orthonormal basis $\{u^{\mu}, e_{r}^{\mu}, e_{\theta}^{\mu}, e_{\phi}^{\mu} \}$, where $u^\mu$ can be considered as a four velocity of the fluid.
\begin{eqnarray}
    && u^\mu = \left(\frac{K+a^2}{\rho\sqrt{\Delta}}, 0, 0, \frac{a}{\rho\sqrt{\Delta}} \right)\,, \\
    && e_r^\mu = \left(0, \frac{\sqrt{\Delta}}{\rho}, 0, 0 \right)\,, \\
    && e_\theta^\mu = \left(0, 0, \frac{1}{\rho}, 0 \right)\,, \\
    && e_\phi^\mu = \left(\frac{a\sin\theta}{\rho}, 0, 0, \frac{1}{\rho\sin\theta} \right).
\end{eqnarray}
One can verify that $u^\mu u_\mu = -1$, $e_{i}^{\mu} (e_i)_\mu = 1$, $u^\mu (e_i)_\mu = 0$; where, $(i \to r,\theta,\phi)$. In terms of these orthonormal bases, eq. (\ref{Inverse_general_rotating_metric}) is written as,
\begin{equation}
    (\partial_s)^2 = (-u^\mu u^\nu + e_r^\mu e_r^\nu + e_\theta^\mu e_\theta^\nu + e_\phi^\mu e_\phi^\nu) \partial_\mu \partial_\nu,
\end{equation}
where the metric tensor is,
\begin{equation}
    g^{\mu\nu} = -u^\mu u^\nu + e_r^\mu e_r^\nu + e_\theta^\mu e_\theta^\nu + e_\phi^\mu e_\phi^\nu.
\end{equation}
Similarly, the energy-momentum tensor can be decomposed as,
\begin{equation}
    T^{\mu\nu} = \rho_e u^\mu u^\nu + P_r e_r^\mu e_r^\nu + P_\theta e_\theta^\mu e_\theta^\nu + P_\phi e_\phi^\mu e_\phi^\nu,
\end{equation}
where, $\rho_e$ is the energy density, and $P_i$ $(i\to r,\theta,\phi)$ are the principal pressure components. The Einstein field equation in the form $G_{\mu\nu}=T_{\mu\nu}$ requires,
\begin{eqnarray}
    && \rho_e = u^\mu u^\nu G_{\mu\nu} \label{rho general}\,, \\
    && P_r = e_r^\mu e_r^\nu G_{\mu\nu} = g^{rr} G_{rr} \label{Pr general}\,, \\
    && P_\theta = e_\theta^\mu e_\theta^\nu G_{\mu\nu} = g^{\theta\theta} G_{\theta\theta} \label{Ptheta general}\,, \\
    && P_\phi = e_\phi^\mu e_\phi^\nu G_{\mu\nu} \label{Pphi general}.
\end{eqnarray}
The components of the energy-momentum tensor of the rotating JNW spacetime are as follows,
\begin{eqnarray}
    && \rho_e = \frac{M^2 (1 - \nu^2) (\Delta+a^2)}{r^6 \nu^2} \left(1 - \frac{2M}{r\nu} \right)^{-3+\nu} \label{rho}, \\
    && P_r = - P_\theta = \frac{M^2 (1-\nu^2)}{r^4 \nu^2} \left(1 - \frac{2M}{r\nu} \right)^{-2+\nu}, \label{Pr_Ptheta}
\end{eqnarray}
\begin{eqnarray}
    P_\phi = P_\theta + \frac{2a^2}{r^4} \left(1-\frac{2M}{r\nu} \right)^{-3+\nu} \bigg[\left(1-\frac{2M}{r\nu} \right)^{1+\nu} - \left(1-\frac{M(1+\nu)}{r\nu}\right)^2 \bigg].\,
    \label{Pphi}
\end{eqnarray}
It can be verified that the rotating JNW spacetime metric (\ref{JNW_rotating}) satisfies the weak energy condition ($\rho_e \geq 0$, $\rho_e+P_i \geq 0$), null energy condition ($\rho_e+P_i \geq 0$), strong energy condition ($\rho_e+\sum P_i \geq 0$, $\rho_e+P_i \geq 0$), and the dominant energy condition ($\rho_e > |P_i|$). It is noted that when $\nu=1$, all the components of (\ref{rho}), (\ref{Pr_Ptheta}), (\ref{Pphi}) vanish, which is evident as it is already seen that the rotating JNW metric reduces to the Kerr metric when $\nu=1$. Also, $P_r \neq P_\theta \neq P_\phi$, therefore, it is an anisotropic fluid solution.

\section{The Kerr spacetime} \label{Kerr metric}
Using the above NJA, one can get the stationary, axisymmetric and rotating Kerr spacetime from the Schwarzschild spacetime in Boyer-Lindquist coordinates as,
\begin{eqnarray}
    ds^2 = -\left(1-\frac{r_s r}{\Sigma}\right)  dt^2 + \frac{\Sigma}{\Delta} dr^2 + \Sigma d\theta^2 &+& \left(r^2 + a^2 + \frac{r_s r a^2 \sin^2\theta}{\Sigma}\right) \sin^2\theta d\phi^2 \nonumber\\ &-& \frac{2 r_s r a \sin^2\theta}{\Sigma}  dt d\phi\,\, , 
    \label{kerr}
\end{eqnarray}
where $r_s= 2 M$ and
\begin{equation}
    \Sigma = r^2 + a^2 \cos^2\theta,
\end{equation}
\begin{equation}
    \Delta = r^2 + a^2 - r_s r.
\end{equation}
The spin parameter $a$ in length dimension is defined corresponding to the total angular momentum $J$ as $a = J/M $. Although, the dimensionless spin parameter $\tilde{a}$ can be written as $\tilde{a}=J/M^2$, where $G=c=1$. The horizons are determined by the relation $g_{rr}\rightarrow\infty$, which implies that the solutions of $\Delta = 0$ can give us the positions of the horizons as,
    \begin{equation}
        r_\pm = M \pm \sqrt{M^2 - a^2}\,\,.
        \label{horizon}
    \end{equation}
where $\pm$ indicates the event horizon and Cauchy horizon, respectively. These horizons coincide at $r=M$ for $a = M$, which is known as an extreme Kerr black hole. On the other hand, $a > M$ is considered, the horizons disappear, and hence the Kerr black hole becomes a timelike Kerr naked singularity.

\section{General formalism of the shadow shapes for rotating spacetimes}
\label{Sec_Shadow}

In this section, the general formalism of the null geodesics is reviewed using the Hamilton-Jacobi separation method. The general expressions for finding the shape and nature of a shadow is obtained. In general relativity, the Hamilton-Jacobi equation is given by,
\begin{equation}
   \frac{\partial S}{\partial \lambda}+\mathcal{H} = 0, \;
   \mathcal{H}=\frac{1}{2} g^{\mu \nu}p_{\mu}p_{\nu}
   \label{Hamilton},
\end{equation}
where $\lambda$ is an affine parameter, S is the Jacobi action, $\mathcal{H}$ is the Hamiltonian, and $p_
{\mu}$ is the momentum
defined by,
\begin{equation}
   p_
{\mu}=\frac{\partial S}{\partial x^{\mu}}=g_{\mu \nu}\frac{d x^{\nu}}{d\lambda}.
\end{equation}
In the above equation (\ref{Hamilton}), the Hamiltonian does not depend explicitly on $t$ and $\phi$ coordinates. Therefore there are two constants of motion, the conserved energy $E=-p_t$ and the conserved angular momentum
$L=p_\phi$ (about the axis of symmetry).
The separable solution of the above differential equation is given (\ref{Hamilton}), the Jacobi action can be written in terms of
already known constants of the motion as,
\begin{equation}
   S = \frac{1}{2}\mu^2 \lambda - E t + L\phi + S_{r}(r) + S_{\theta}(\theta) \label{action},
\end{equation}
 where $\mu$ is the rest mass of the test particle. Therefore, for a photon or null geodesic, $\mu=0$ is considered. The metric tensor components of general rotating spacetime (\ref{general_rotating_BLC}) are given as, 

 \begin{equation}
   g^{tt}=-\frac{H-a^2 G \sin^2\theta+2 a^2 \sqrt{\frac{G}{F}}\sin^2\theta}{P}, 
 \end{equation}
  \begin{equation}
   g^{rr}=\frac{FH+a^2\sin^2\theta}{H},
 \end{equation}
 \begin{equation}
   g^{\theta\theta}=\frac{1}{H},
 \end{equation}
 \begin{equation}
   g^{\phi\phi}=-\frac{G\csc^4\theta}{2a^2G^2-4a^2G\sqrt{\frac{G}{F}}+a^2\left(\frac{G}{F}\right)-GH\csc^2\theta},
 \end{equation}
 \begin{equation}
   g^{t\phi}=g^{\phi t}=\frac{a\left( \sqrt{\frac{G}{F}}-G\right)}{P}, 
 \end{equation}
where,

\begin{equation}
    P=-GH+2a^2G^2\sin^2\theta-4a^2G\sqrt{\frac{G}{F}}\sin^2\theta+a^2 \left(\frac{G}{F}\right)\sin^2\theta.
\end{equation}
Substituting eq. (\ref{action}) into eq. (\ref{Hamilton}), furthermore, using the components of the metric tensor, the following expression can be obtain:

\begin{eqnarray}
     -\left(FH+a^2\sin^2\theta\right)\left(\frac{dS_r}{dr}\right)^2+\frac{\left(\left(\sqrt{\frac{F}{G}}H+a^2\sin^2\theta\right)E-aL\right)^2}{FH+a^2\sin^2\theta}-\left(L-aE\right)^2 \nonumber \\=\,\left(\frac{dS_{\theta}}{d\theta}\right)^2+L^2\cot^2\theta-a^2E^2\cos^2\theta.
    \label{p1}
\end{eqnarray}
Note that the left and right-hand sides of the above equation (\ref{p1}) are only functions of $r$ and $\theta$, respectively. Therefore one can write each side equal to a separation constant since the Jacobi action principle is the separable solution,

\begin{equation}
  -\left(FH+a^2\sin^2\theta\right)\left(\frac{dS_r}{dr}\right)^2+\frac{\left(\left(\sqrt{\frac{F}{G}}H+a^2\sin^2\theta\right)E-aL\right)^2}{FH+a^2\sin^2\theta}-\left(L-aE\right)^2=\mathcal{K},  
\end{equation}

\begin{equation}
  \left(\frac{dS_{\theta}}{d\theta}\right)^2+L^2\cot^2\theta-a^2E^2\cos^2\theta=\mathcal{K},  
\end{equation}
where the separation constant $\mathcal{K}$ is known as the Carter constant. Using the above differential equations, one can write down the following separated null geodesic equations for radial ($r$) and tangential ($\theta$) parts as,
\begin{equation}
    H^2\left( \frac{dr}{d\lambda}\right)^2 -R(r) = 0\,\, ,
    \label{rmotion}
\end{equation}
\begin{equation}
    H^2\left( \frac{d\theta}{d\lambda}\right)^2 -\Theta(\theta) = 0\,\, ,
    \label{thetamotion}
\end{equation}
where $R(r)$ and $\Theta(\theta)$ can be considered as radial and tangential parts of effective potentials, respectively, which can be defined as,
\begin{equation}
   R(r) = \left[X(r) - a \xi \right]^2 - \Delta(r) \left[\eta + (\xi - a)^2 \right]\,\, ,
   \label{rpoten}
\end{equation}
\begin{equation}
\Theta(\theta)=\eta+(\xi-a)^2-\left(\frac{\xi}{\sin\theta}-a\sin\theta\right)^2\,\,
\label{thetapoten}
\end{equation}
where, $\xi = \frac{L}{E}$ and $\eta = \frac{\mathcal{K}}{E^2}$. The functions $X(r)$ and $\Delta(r)$ can be define as,
\begin{equation}
    X(r)=\sqrt{\frac{F}{G}}H+a^2\sin^2\theta,
\end{equation}
\begin{equation}
    \Delta(r)=FH+a^2\sin^2\theta.
\end{equation}
Note that, $R(r)$ and $\Theta(\theta)$ must be non-negative for photon motion. In a general rotating spacetime, unstable circular orbits exist when the following conditions hold,
\begin{equation}
     R(r_{ph}) = 0, \;   \frac{d R(r_{ph})}{dr}=0, \;    \frac{d^2 R(r_{ph})}{dr^2} \leq 0\,\,. \label{eq3}
\end{equation}
where $r=r_{ph}$ is the radius of the unstable photon orbit.
Using the above conditions (\ref{eq3}), the critical impact parameters can be obtain corresponding to the maxima of the $R(r)$ as,
\begin{equation}
     \xi = \frac{X_{ph}\Delta'_{ph}-2\Delta_{ph}X'_{ph}}{a\Delta'_{ph}},
     \label{xi}
\end{equation}
\begin{equation}
    \eta = \frac{4a^2X'_{ph}\Delta_{ph}-\left[(X_{ph} - a^2)\Delta'_{ph}-2X'_{ph}\Delta_{ph}\right]^2}{a^2\Delta'^2_{ph}},
    \label{eta}
\end{equation}
where prime denoted for derivative concerning $r$ and subscript ``ph" denoted for the quantities which are evaluated at $r=r_{ph}$. The above expressions are the general forms of the critical impact parameters $\xi$ and $\eta$ for the unstable photon orbits, which would be important to find out the shape of the shadows. However, one can obtain the apparent shape of the shadow by using the celestial coordinates $\alpha$ and $\beta$, which lie in the line of sight of the observer direction. The expressions of the celestial coordinates $\alpha$ and $\beta$ can be written in the following way,
\begin{eqnarray}
    \alpha &=& \lim_{r_0\to\infty} \left(-r_0^2 \sin{\theta_{0}} \frac{d \phi}{dr}\right)\,\, ,\\
    \beta &=& \lim_{r_0\to\infty} \left(r_0^2 \frac{d \theta}{dr}\right)\,\, ,
\end{eqnarray}
where $r_0$ and $\theta_0$ are the coordinates of the asymptotic observer. Note that, if the general rotating metric to be an asymptotically flat in the limit of $r\to \infty$, then one can consider $G\to 1$, $F\to 1$, $H\to r^2$, $X\to r^2$ and $\Delta\to r^2$. Therefore, the above expressions of the celestial coordinates $\alpha$ and $\beta$ becomes,
\begin{eqnarray}
    \alpha &=& -\frac{\xi}{\sin{\theta_{0}}}\,\, ,\\
    \beta &=& \pm \sqrt{\eta + a^2 \cos^2{\theta_{0}} - \xi^2\cot^2{\theta_{0}}}.\,\, 
\end{eqnarray}

\subsection{Shadows in the rotating JNW spacetime}
From the above expressions (\ref{xi}) and (\ref{eta}), one can obtain the celestial coordinates $\alpha$ and $\beta$ for the rotating JNW spacetime by using its metric tensor components. The JNW metric tensor components are used without adopting any complexification method since the general rotating metric is obtained by skipping the complexification method. The metric tensor components of the JNW spacetime are given as,
\begin{equation}
    G = F = \left(1-\frac{2M}{r\nu}\right)^{\nu}\,\, ;\
    H = r^2\left(1-\frac{2M}{r\nu}\right)^{1-\nu}
\end{equation}
using the above equations, one can find that the expressions for $X(r)$ and $\Delta(r)$,
\begin{equation}
    X(r)=r^2\left(1-\frac{2M}{r\nu}\right)^{1-\nu}+a^2\sin^2\theta,
\end{equation}
\begin{equation}
    \Delta(r)=r^2\left(1-\frac{2M}{r\nu}\right)+a^2\sin^2\theta.
\end{equation}
Therefore, using the above expressions, one can calculate the critical impact parameters and celestial coordinates for the rotating JNW spacetime. The shape of the shadows are constructed using the expressions of celestial coordinates for rotating JNW spacetime, where the inclination angle of the observer is $\theta_{0}=\pi/2$.

\begin{figure*}
\centering
\subfigure[Shadow for $q =0.4$ and $a = 0.3$.]
{\includegraphics[width=44mm]{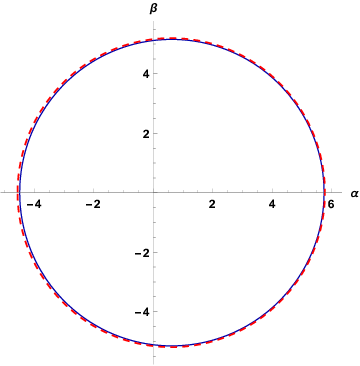}\label{re1n}}
\hspace{0.2cm}
\subfigure[Shadow for $q =0.4$ and $a = 0.4$.]
{\includegraphics[width=44mm]{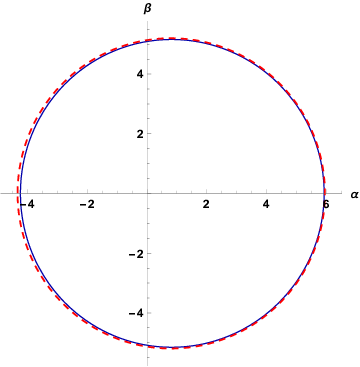}\label{re2n}}
\subfigure[Shadow for $q =0.4$ and $a = 0.5$.]
{\includegraphics[width=44mm]{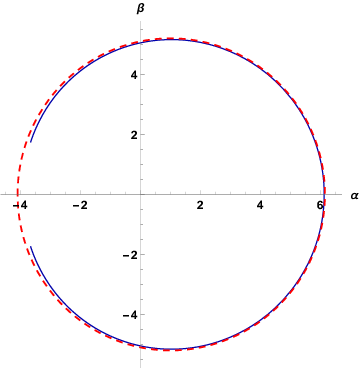}\label{re3n}}\\
\subfigure[Shadow for $q =0.6$ and $a = 0.3$.]
{\includegraphics[width=44mm]{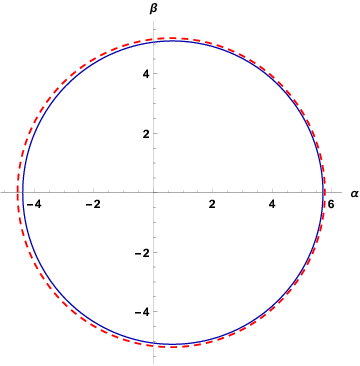}\label{re1p}}
\hspace{0.2cm}
\subfigure[Shadow for $q =0.6$ and $a = 0.4$.]
{\includegraphics[width=44mm]{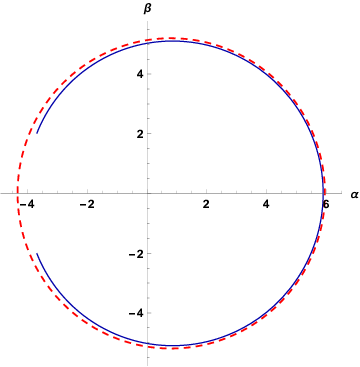}\label{re2p}}
\subfigure[Shadow for $q =0.6$ and $a = 0.5$.]
{\includegraphics[width=44mm]{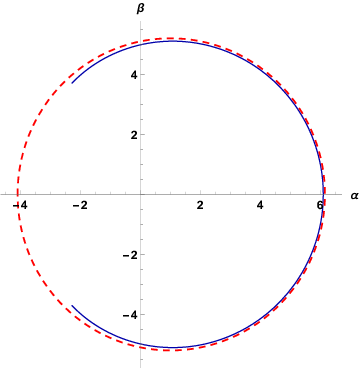}\label{re3p}}\\
\subfigure[Shadow for $q =0.8$ and $a = 0.3$.]
{\includegraphics[width=44mm]{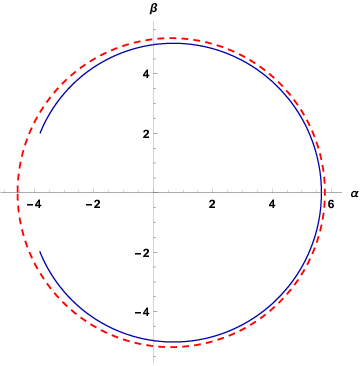}\label{re11p}}
\hspace{0.2cm}
\subfigure[Shadow for $q =0.8$ and $a = 0.4$.]
{\includegraphics[width=44mm]{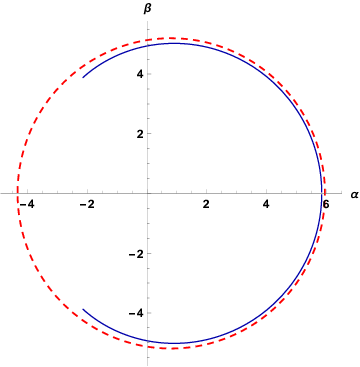}\label{re22p}}
\subfigure[Shadow for $q =0.8$ and $a = 0.5$.]
{\includegraphics[width=44mm]{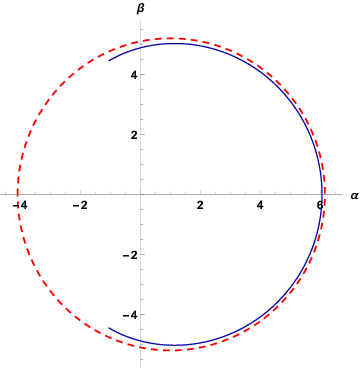}\label{re33p}}\\
 \caption{Figure shows the shape of the shadow cast by the rotating JNW naked singularity for various values of scalar field charge $q$ and spin parameter $a$. The values of $M=1$ and $\theta=\pi/2$ are considered for the shape of shadows in celestial coordinates. The details of this figure are discussed in the text.}
\label{figshadow}
\end{figure*}

In fig.(\ref{figshadow}), it is shown that the shape of the shadows for different values of spin parameter $a$ and scalar field charge $q$, where the blue line indicates the shadows in rotating JNW spacetime and the red dotted line indicate the shadows in the Kerr spacetime. It is well known that the shadow shapes in the Kerr black hole spacetime always be in contour shape, and as the spin parameter increasing, it will become more and more prolate. One can see in fig.(\ref{figshadow}), that the horizontal radius of the red dotted line will decrease from the left side as the value of a spin parameter increasing. On the other hand, it is found that the shadows in rotating JNW spacetime change from contour shape to arc shape as the values of the spin parameter and scalar field charge or one of them are increasing. In the next section, the shape of the shadow cast by deformed Kerr spacetime is shown. 

\subsection{Shadows in the deformed Kerr spacetime}
The deformed Schwarzschild spacetime is considered a solution in the modified Quadratic gravity \cite{Yunes:2011we}. In that paper, the authors show that parametric deformation to the Schwarzschild metric solves the modified field equations in the small coupling approximations. The deformed Schwarzschild spacetime is given as,
\begin{equation}
    ds^2 = -\left(1-\frac{2M}{r}\right)\left(1+h(r)\right) dt^2 +\frac{ \left(1+h(r)\right)}{ \left(1-\frac{2M}{r}\right)}dr^2+ r^2 d\Omega^2 \,\, ,
    \label{DSch metric}
\end{equation}
where $M$ is the mass of the deformed Schwarzschild black hole. $d\Omega^2 = d\theta^2 + \sin^2\theta d\phi^2$ is the metric of two sphere, and the parametric deformation
\begin{eqnarray}
    h(r) = \frac{\epsilon M^3}{r^3}.
\end{eqnarray}
The above deformation represents the deviation from the Schwarzschild spacetime, where $\epsilon$ is a deformation parameter. When $\epsilon=0$, the above metric reduces to the original Schwarzschild metric. 
The specific deformation is considered here, and the reasons are described as follows: The general form of the deformation function is given as, 
\begin{equation}
    h(r) =  \sum_{k=0}^{\infty} \epsilon_k \left(\frac{M}{r}\right)^k.
    \label{generaldeform}
\end{equation}
  In general relativity, it can be shown that the stationary and asymptotically flat spacetime should be Schwarzschild-like \cite{Beig:1980be}. Therefore, in the above equation (\ref{generaldeform}), one need to consider $k\geq 2$. One can show that asymptotically flat spacetimes with slower fall off ($0\leq k<1$) can generate gravitational radiation, and hence, the spacetime cannot be stationary \cite{Kennefick:1993zr}. From the Lunar Laser Ranging experiment, it is shown that the value of $\epsilon_2$ is $|\epsilon_2| \leq 4.6\times 10^{-4}$ \cite{Johannsen:2011dh,Williams:2004qba}. So that, one can consider $\epsilon_2 =0$. In \cite{Psaltis:2008bb}, it is shown that the value of $|\epsilon_3|$ is unconstrained. Hence, in this work, $\epsilon_3\neq 0$ is considered and neglected the higher order contributions (i.e., $k>3$) \cite{Johannsen:2011dh}. 
\begin{figure*}
\centering
\subfigure[Spin $a=0.6$]
{\includegraphics[scale=0.5]{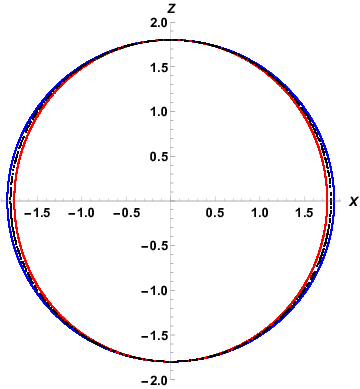}}
\hspace{0.2cm}
\subfigure[Spin $a=0.7$]
{\includegraphics[scale=0.5]{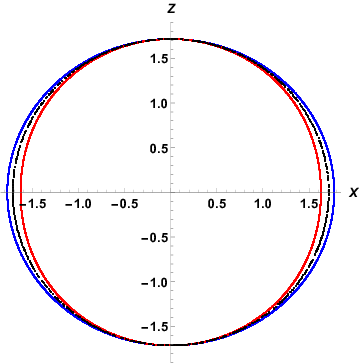}}
\hspace{0.2cm}
\subfigure[Spin $a=0.8$]
{\includegraphics[scale=0.4]{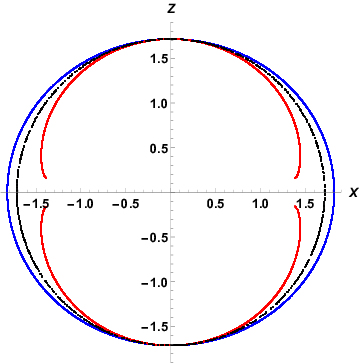}}
\hspace{0.2cm}
\subfigure[Spin $a=0.8$]
{\includegraphics[scale=0.4]{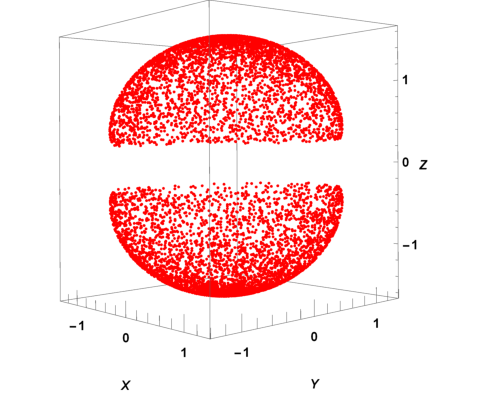}}
\hspace{0.2cm}
\caption
{Event horizon position for $\epsilon = 0$ (black), $\epsilon = 1$ (Red), $\epsilon = -1$ (Blue) with different spin and 3D view of directional naked singularity for $\epsilon = 1$,  $a = 0.8$.}\label{Pos1}
\end{figure*}

One can determine the Kerr-like deformed spacetime using the NJA, 
\begin{eqnarray}
    ds^2 = -\bigg(1-\frac{2Mr}{\rho^2} \bigg) (1+h) dt^2 &+& \frac{\rho^2 (1+h)}{\Delta + a^2 h \sin^2\theta} dr^2 + \rho^2 d\theta^2 - \frac{4Mar}{\rho^2} (1+h) \sin^2\theta dt d\phi\nonumber \\ &+& \bigg[\rho^2 + a^2\sin^2\theta (1+h) \bigg(1+\frac{2Mr}{\rho^2} \bigg) \bigg] \sin^2\theta d\phi^2\,.
    \label{DK metric}
\end{eqnarray}
In \cite{Johannsen:2011dh}, the above metric is defined as deformed Kerr spacetime. One can see that the above metric reduces to the original Kerr metric when $\epsilon=0$. Kerr geometry has an oblate shape. However, the deformed Kerr spacetime is more oblate for $\epsilon < 0$ and more prolate for $\epsilon > 0$. One can derive the position of the inner and outer horizon of deformed Kerr spacetime by solving the following equation,
\begin{equation}
g_{t\phi}^2+g_{tt}g_{\phi\phi}=0\,\, .
\end{equation}
The position and shape of an outer event horizon of the deformed Kerr spacetime is shown for various values of the deformation parameter ($\epsilon$) and spin parameter ($a$). Figs.~(\ref{Pos1}) show the shapes and position of outer event horizon for spin parameter $a=0.6, 0.7, 0.8$ respectively. It can be seen from the above figures that for $ \epsilon=1$ and $a=0.8$, the outer event horizon is not closed, and it has a hole near the equatorial plane of deformed Kerr spacetime. Therefore, the singularity at the centre becomes naked for the particular values of $\epsilon$ and $a$.

\begin{figure*}
\centering
{\includegraphics[width=80mm]{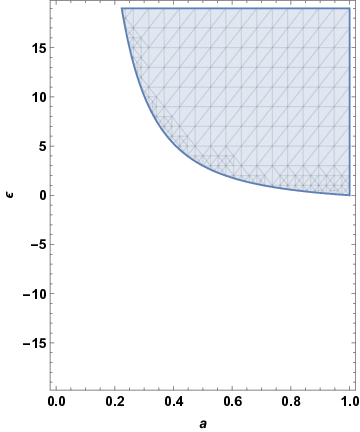}}
\caption{The shaded region in the figure shows the existence of a naked singularity in deformed Kerr spacetime.}
\label{nakedcond}
\end{figure*}
In Fig.~(\ref{nakedcond}), the parameters' space in $\epsilon$ and $a$ frame is shown for which the central singularity of deformed Kerr spacetime becomes naked. The shaded region in that figure shows the possible values of $\epsilon$ and $a$ for the existence of a naked singularity in the deformed Kerr spacetime. On the other hand, the unshaded region shows the existence of a deformed Kerr black hole. From Fig.~(\ref{nakedcond}), it can be seen that for $0\leq a\leq 1$, the central singularity can be naked when the deformation parameter $\epsilon >0$. However, when $\epsilon<0$ and $0\leq a\leq 1$, the central singularity of the deformed Kerr spacetime cannot be naked.

\begin{figure*}
\centering
\subfigure[Shadow for $\epsilon = -1$ and $a = 0.6$.]
{\includegraphics[width=44mm]{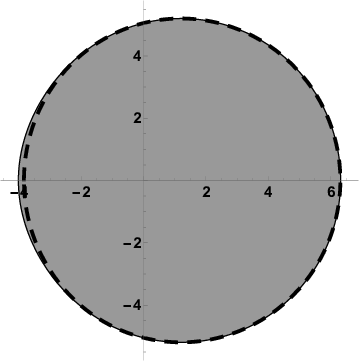}\label{re1n}}
\hspace{0.2cm}
\subfigure[Shadow for $\epsilon = -1$ and $a = 0.7$.]
{\includegraphics[width=44mm]{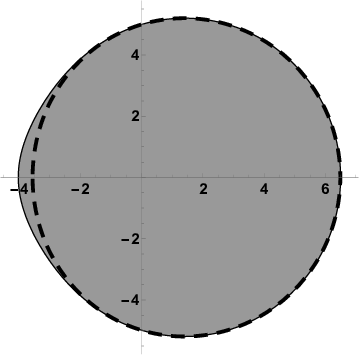}\label{re2n}}
\subfigure[Shadow for $\epsilon = -1$ and $a = 0.8$.]
{\includegraphics[width=44mm]{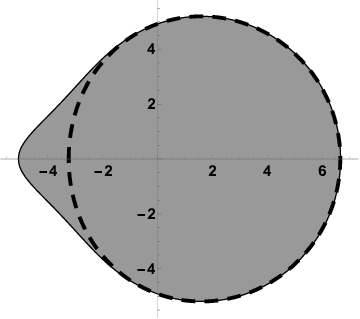}\label{re3n}}\\
\subfigure[Shadow for $\epsilon = 1$ and $a = 0.6$.]
{\includegraphics[width=44mm]{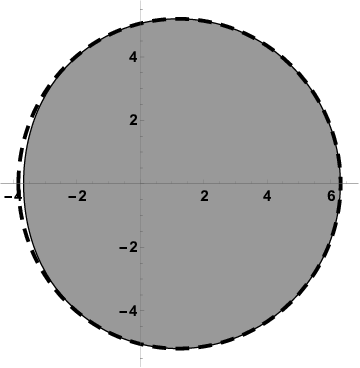}\label{re1p}}
\hspace{0.2cm}
\subfigure[Shadow for $\epsilon = 1$ and $a = 0.7$.]
{\includegraphics[width=44mm]{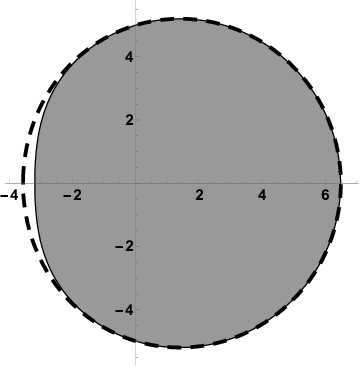}\label{re2p}}
\subfigure[Shadow for $\epsilon = 1$ and $a = 0.8$.]
{\includegraphics[width=44mm]{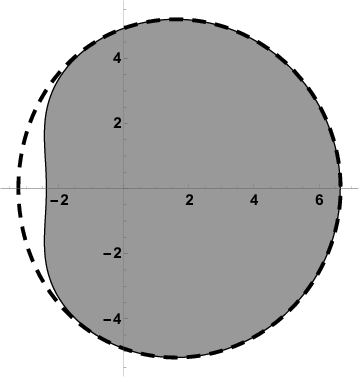}\label{re3p}}\\
\subfigure[Shadow for $\epsilon = 18$ and $a = 0.6$.]
{\includegraphics[width=44mm]{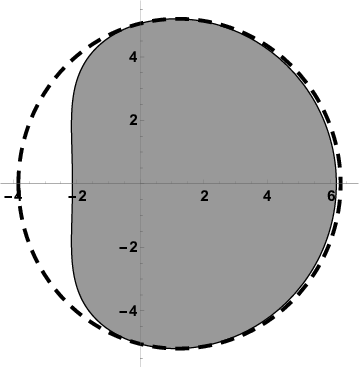}\label{re11p}}
\hspace{0.2cm}
\subfigure[Shadow for $\epsilon = 18$ and $a = 0.7$.]
{\includegraphics[width=44mm]{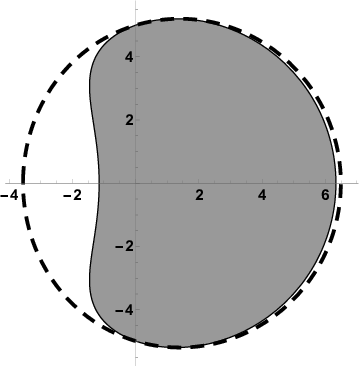}\label{re22p}}
\subfigure[Shadow for $\epsilon = 18$ and $a = 0.8$.]
{\includegraphics[width=44mm]{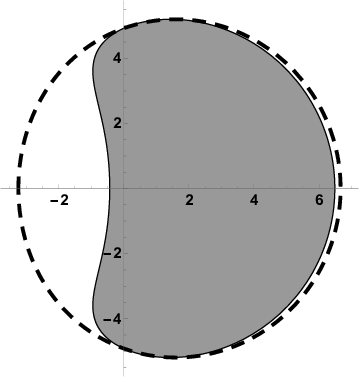}\label{re33p}}\\
 \caption{Figure shows the shape of the shadow cast by the deformed Kerr spacetime for various values of deformation parameter $\epsilon$ and spin parameter $a$. The details of this figure are discussed in the text.}
\label{figshadow}
\end{figure*}
Now from the definition of the celestial coordinates, one can write down the following expressions of celestial coordinates for deformed Kerr spacetime, 
\begin{eqnarray}
    \alpha &=& -  \frac{ - \Delta \xi}{(\Delta + a^2 h \sin^2{\theta_{0}})\sin{\theta_{0}}}\,\, ,\\
    \beta &=& \sqrt{\eta - a^2 \sin^2{\theta_{0}} - \xi^2 \csc^2{\theta_{0}}\,\, .}
\end{eqnarray}
Using the above expressions of celestial coordinates, it can be verified that they are not independent of the deformation parameter $\epsilon$. Therefore, though the behaviour of $R(r)$ is similar for both Kerr and deformed Kerr spacetime, the shadow shapes of both spacetimes are distinguishable. In Fig.~(\ref{figshadow}), the shapes of the shadows cast by deformed Kerr spacetime are shown for various values of $\epsilon$ and $a$, where the shaded region shows the shape of the shadow cast by deformed Kerr spacetime and the dashed lines show the shape of the shadow cast by Kerr spacetime. Figs.~(\ref{re1n}),(\ref{re2n}) and (\ref{re3n}) show the shadow shape for $a=0.6, 0.7$ and $0.8$ respectively, where the $\epsilon=-1$. On the other hand, in Figs.~(\ref{re1p}),(\ref{re2p}) and (\ref{re3p}), the shadow shape for $a=0.6, 0.7$ and $0.8$ are shown respectively, where the $\epsilon=1$. Those figures show how the shadow shape changes when $\epsilon$ changes its signature. Figs.~(\ref{re11p}),(\ref{re22p}) and (\ref{re33p})also show the shadow shape of deformed Kerr spacetime for the same values of spin parameter, where $\epsilon=18$ is taken. As discussed above, the central singularity in deformed Kerr spacetime cannot be naked when $\epsilon<0$ where $0\leq a\leq1$. Therefore, all the shadow shapes shown in Figs.~(\ref{re1n}),(\ref{re2n}) and (\ref{re3n}) are corresponding to the deformed Kerr black hole. However, though the Figs.~(\ref{re1p}),(\ref{re2p}) are showing the shape of the shadow cast by deformed Kerr black hole, Figs.~(\ref{re3p}),(\ref{re11p}),(\ref{re22p}) and (\ref{re33p}) are showing the shadow shape of the deformed Kerr naked singularity, which can be verified from Fig.~(\ref{nakedcond}). Therefore, the deformed Kerr naked singularity can cast a shadow, unlike Kerr naked singularity.

The amount of the deformation ($D$) of the shadow shape from the shadow shape of the Kerr black hole can be defined as,
\begin{equation}
    \alpha_{dk} - \alpha_{k} = D\, ,
\end{equation}
where the $\beta=0$ axis to measure the deformation of the shadow shape and $\alpha_{dk}$ and $\alpha_{k}$ are celestial coordinates for deformed Kerr and Kerr spacetimes, respectively. One can derive the following equation of $D$ for deformed Kerr spacetime,
\begin{equation}
     D = \frac{a \epsilon M^3 (r^2 (r - 3M) + a^2 (M + r))}{(M - r)(r^4 (r - 2M) + a^2 (\epsilon M^3 + r^3))}\,\, ,
\end{equation}
where
\begin{equation}
     r = \frac{\left(M + (2a^2 M - M^3 + 2 a M \sqrt{a^2 - M^2})^{1/3}\right)^2}{(2a^2 M - M^3 + 2 a M\sqrt{a^2 - M^2})^{1/3}}\,\, .
\end{equation}
\begin{figure*}
\centering
{\includegraphics[width=135mm]{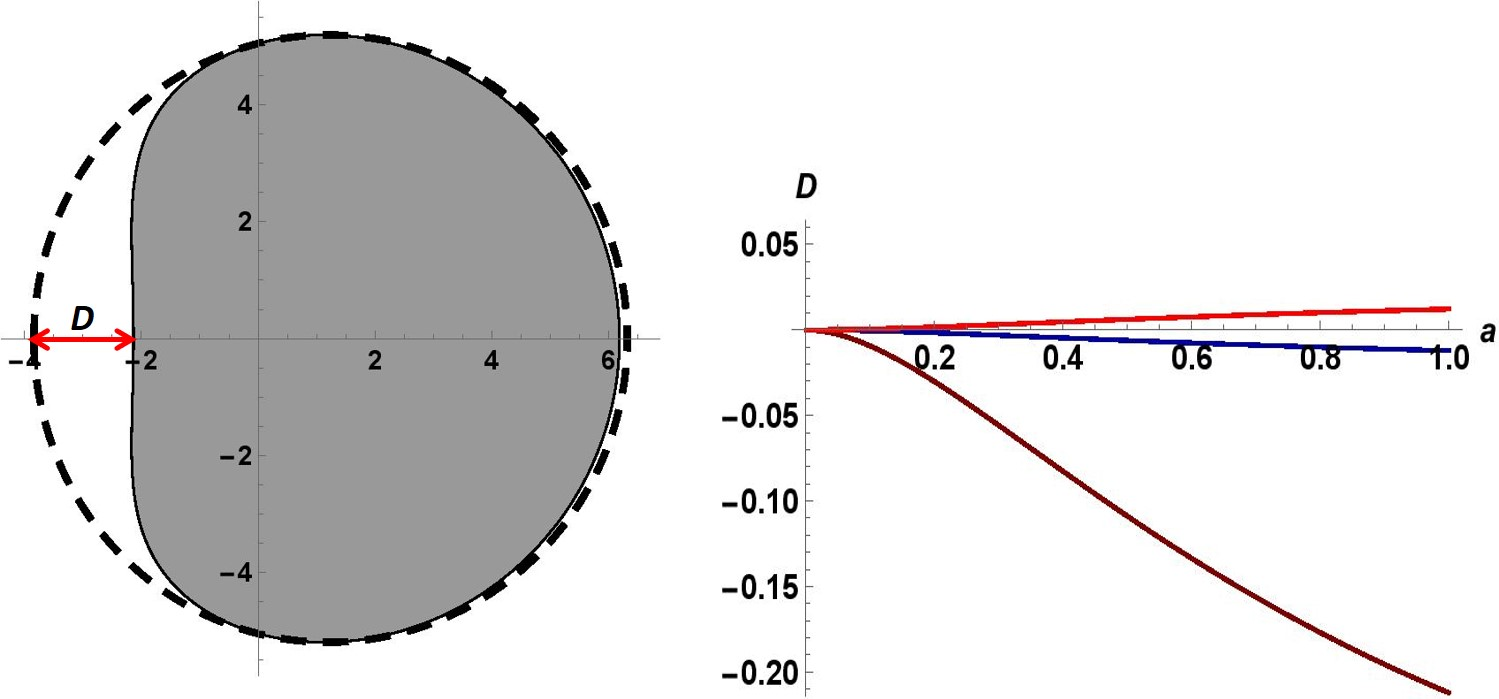}}
 \caption{Figure shows how the deformation of shadow shape ($D$) changes with the value of spin parameter $a$, where the red, blue and brown lines correspond to the deformed Kerr spacetime having $\epsilon=-1,1$ and $18$ respectively.}\label{D}
\end{figure*}
Fig.~(\ref{D}) shows how the deformation of the shadow shape $D$ changes with spin parameter $a$ for various values of $\epsilon$. The left side figure in Fig.~(\ref{D}) shows how one can define the deformation in the shadow shape. On the other hand, in the right side figure in Fig.~(\ref{D}), the red, blue and brown curves show how the deformation of the shape of the shadow changes with spin parameter $a$, when the $\epsilon= -1, 1$ and $18$ respectively.

\section{Discussions and conclusions}
 \label{sec_discussion}
 The conclusions of this chapter are in the following way: the rotating JNW spacetime is consructed using the NJA without employing the complexification of coordinates. The rotating JNW spacetime obtained using the NJA, considering the complexification step, cannot be written into the BLC. If that metric is written into the BLC-like form by performing an improper coordinate transformation, it does not satisfy the energy conditions. To resolve this problem, the complexification step is dropped in the NJA to properly transform the rotating solution into the BLC. As a result, the rotating JNW spacetime is obtained into the BLC, which does satisfy the energy conditions.
 
The rotating Kerr black hole can cast a prolate contour shape shadow (see dotted red line in fig.(\ref{figshadow})). While the spin parameter increase up to $M\ge a$, it will become more and more prolate contour, where $M$ is the mass of the black hole. Nevertheless, as it reach the $M<a$ limit, the Kerr black hole becomes the Kerr naked singularity, which may not cast contour shaped shadow. To get more insight, in this paper, the shape of the shadows cast by rotating JNW naked singularity is obtained and compared with the shape of shadows cast by the Kerr black hole. It is found that the shape of the shadows in rotating JNW spacetime becomes prolate arc shape from the prolate contour shape as the scalar field charge $q$ and/or spin parameter $a$ is increases (see blue line in fig. (\ref{figshadow})). 

In addition, it is shown that the deformed Kerr naked singularity can cast a shadow, unlike the Kerr naked singularity. If there exists a finite amount of deformation of Kerr spacetime, then the effect of the deformation may have become very significant near the singularity of deformed Kerr spacetime. In the context of the observations by the EHT/ngEHT group, these studies on shadows are beneficial. This novel signature of the shadow shapes in rotating JNW spacetime (prolate arc) and the shadow shape of the deformed Kerr naked singularity could be observationally significant to differentiate the naked singularities from the black holes. However, the existence and observation of a shadows are not sufficient, or may not ensure the existence of an event horizon necessarily in the galactic center. This necessitates studying different observables such as accretion spectra, light curves, etc. Since the main focus is on Sgr A*, the motion of various stars around the central object are helpful via astrometric observations which can show precession in their orbits.  Therefore, the precession of relativistic orbits in the vicinity of black hole and its mimickers are discussed.

\chapter{Precession of timelike bound orbits in static spacetimes}
The term `geodesic' notion represents the extremum distance between two points on a spherical surface or any other curved surface. The motion of the freely falling test particle described the shortest possible route. Thus, it always follows the timelike or null geodesics in a given curvature of spacetime. The nature of the causal structures and spacetime curvatures can be scrutinized by analyzing the geodesic motions of the test particles. Timelike geodesics are essential to describe the orbital motion of a massive test particle. Therefore, in this chapter, the timelike bound orbits of a test particle is analysed and discuss the orbital precession (periastron precession) in naked singularity and black hole spacetimes.
\section{Timelike geodesics in spherically symmetric and static spacetime}

The spherical and temporal symmetry of the general spacetime (\ref{static}) can be illustrated by the conservation of angular momentum and energy of a freely falling particle. Therefore the angular momentum and energy of a particle per unit rest mass are always conserved. All possible types of orbits are classified under the most conserved angular momentum and energy of a test particle. The conserved angular momentum ($h$) per unit rest mass and conserved energy ($\gamma$) per unit rest mass of a freely falling particle in the spherically symmetric and static spacetime can be written as,
\begin{equation}
     h = g_{\phi\phi}(r)\frac{d\phi}{d\tau}\,\, ,\,\,\,
   \gamma =  g_{tt}(r)\frac{dt}{d\tau}\,\, ,
   \label{congen}
\end{equation}
where the proper time of the particle is given by  $\tau$. the sign convention of the metric components $g_{\mu\nu}$ is considered as $(-,+,+,+)$, and employing this sign convention to the normalization of four velocities for timelike geodesics, one can have the relation $u^{\mu}u_{\mu}=-1$, where $u^{\mu}$ is the four velocity of a particle. Now utilizing the conservation laws given in (\ref{congen}) and normalization of four velocity relation, one can derive the general expression of the total energy of a freely falling test particle as,
\begin{equation}
    E=\frac{g_{rr}(r)g_{tt}(r)}{2}\left(\frac{dr}{d\tau}\right)^2+V_{eff}(r)\,\, ,
    \label{totalE}
\end{equation}
where $E$ is given by defining the constant as $E=\frac{\gamma^2-1}{2}$ and radial effective potential can be written as,
\begin{equation}
     V_{eff}(r) = \frac{1}{2}\left[g_{tt}(r)\left(1+\frac{h^2}{g_{\phi\phi}(r)}\right)-1\right]
    \label{veffgen}\,\, ,
\end{equation}
note that an equatorial plane ($\theta=\frac{\pi}{2}$) is considered for the particle's trajectories since the spacetime configuration is spherically symmetric. The above equation of effective potential plays an important role in the motion of the particles in a given general form of spacetime (\ref{static}). By defining some conditions in between the total energy and effective potential as,
\begin{equation}
    V_{eff}(r)=E, \; \; \;
    V_{eff}^{\prime}(r)=0,
\end{equation}
  one can get types of bound orbital shapes, for example, circular orbits and elliptical orbits. For circular orbits, the required expressions for conserved angular momentum ($h$) and energy ($\gamma$) are calculated by using the above conditions,
\begin{equation}
    h^2=\frac{g_{\phi\phi}(r)}{\left(\frac{2g_{tt}(r)}{\sqrt{g_{\phi\phi}(r)} g_{tt}^{\prime}(r)}-1\right)}\, ,\,\,\, \gamma^2=\frac{2g_{tt}^2(r)}{\sqrt{g_{\phi\phi}(r)} g_{tt}^{\prime}(r)}\frac{1}{\left(\frac{2g_{tt}(r)}{\sqrt{g_{\phi\phi}(r)} g_{tt}^{\prime}(r)}-1\right)}\,\, ,
    \label{he}
\end{equation}
where prime denoted for derivative with respect to $r$ and the above conditions $V_{eff}(r)=E$, $V_{eff}^{\prime}(r)=0$, are valid for both stable and unstable circular orbits. The stability of the circular orbits defined by the second derivatives of effective potential with respect to $r$ as,
\begin{equation}
   V_{eff}^{\prime\prime}(r)>0 \, ,\,\,\, 
    V_{eff}^{\prime\prime}(r)<0
    \,\,,
\end{equation}
for stable and unstable circular orbits, respectively. One can get bound elliptical orbits by employing the condition $V_{eff}(r) = E$, in which the minimum ($r_{min}$) and maximum ($r_{max}$) radial approaches of the freely falling test particle from the center are established. Therefore, the necessary conditions to get the stable bound orbits can be define for freely falling test particles in the following manner,
\begin{equation}
   V_{eff}(r_{min})=V_{eff}(r_{max})=E\, , \,\, 
   \label{bound1}
   \end{equation}
   \begin{equation}
   E-V_{eff}(r)>0\, ,\,\,\,\forall r\in (r_{min},r_{max}).
   \label{bound2}
\end{equation}
The shape of the timelike bound orbits can describe using eq.~(\ref{totalE}) with the value of conserved energy $\gamma$ and angular momentum $h$,

\begin{equation}
    \frac{dr}{d\phi}=\frac{g_{\phi\phi}(r)}{h}\frac{\sqrt{2(E-V_{eff})}}{\sqrt{g_{rr}(r)g_{tt}(r)}}.
    \label{shapeorbit}
\end{equation}
The above equation (\ref{shapeorbit}) represents how the particle changes its radial coordinate $r$ with azimuth coordinate $\phi$ in an equatorial plane. One can derive an orbit equation of a freely falling test particle by differentiating above eq.~(\ref{shapeorbit}) with respect to $\phi$, 
\begin{equation}
  \frac{d^2u}{d\phi^2}+\frac{\gamma^2g_{\phi\phi}^2(u)u^4}{2g_{tt}^2(u)g_{rr}(u)h^2}\left(\frac{dg_{tt}(u)}{du}\right)-A(u) \left(\frac{dg_{rr}(u)}{du}\right)
  +B(u)\left(\frac{dg_{\phi\phi}(u)}{du}\right) - C(u) = 0\,\, ,
  \label{orbiteqgen}
    \end{equation}
    where $u=\frac1r$ and,
    \begin{equation}
        A(u)=\left[\frac{\gamma^2g_{\phi\phi}^2(u)u^4}{2g_{tt}(u)g_{rr}^2(u)h^2}-\frac{g_{\phi\phi}(u)u^4}{2g_{rr}^2(u)}-\frac{g_{\phi\phi}^2(u)u^4}{2g_{rr}^2(u)h^2}\right],
    \end{equation}
\begin{equation}
    B(u)=\left[\frac{\gamma^2g_{\phi\phi}(u)u^4}{g_{tt}(u)g_{rr}(u)h^2}-\frac{u^4}{2g_{rr}(u)}-\frac{g_{\phi\phi}(u)u^4}{g_{rr}(u)h^2}\right]\,\, ,
\end{equation}
and
\begin{equation}
    C(u)=\left[\frac{2\gamma^2g_{\phi\phi}^2(u)u^3}{g_{tt}(u)g_{rr}(u)h^2}+\frac{2g_{\phi\phi}(u)u^3}{g_{rr}(u)}+\frac{2g_{\phi\phi}^2(u)u^3}{g_{rr}(u)h^2}\right].
\end{equation}
\\
It is very complicated to derive the exact analytical solution of the above orbit equation (\ref{orbiteqgen}) since the derived orbit equation is a second-order non-linear differential equation. However, one can solve it numerically, and one can get the nature and shape of the timelike bound orbits for the given conserved energy $\gamma$ and angular momentum $h$. Note that the nature and shape of the timelike bound orbits depend on the causal structure of spacetime geometry.

\section{Fully relativistic orbit equations} \label{relativisticorbiteqs}
The fully general relativistic orbit equation of a freely falling particle can be derived using the timelike geodesics. In the above section, the general procedure to derive an orbit equation of a test particle is described in spherically symmetric and static spacetime. This section analyzes the orbit equations for the Schwarzschild, JMN-1, and JNW models, which are spherically symmetric and static spacetimes. 

\subsection{Schwarzschild orbit equation}
The Schwarzschild spacetime is defined as given in equation (\ref{schext}); it has a strong singularity at $r=0$, which is covered by an event horizon. One can see that the metric tensor components of the Schwarzschild spacetime are independent of $t$ and $\phi$. Therefore, due to the temporal and spherical symmetries of the Schwarzschild spacetime, the conserved energy and angular momentum per unit rest mass can be written of the freely falling test particle in this spacetime as,
\begin{equation}
     \gamma_{SCH} = \left(1-\frac{2M}{r}\right) \left(\frac{dt}{d\tau} \right)\,\, ,
\end{equation}
\begin{equation}
     h_{SCH}= r^2  \left(\frac{d\phi}{d\tau}\right)\,\,,
\end{equation}
\\
the effective potential for the Schwarzschild spacetime can be derived using the general expression given in (\ref{veffgen}),

\begin{equation}
    (V_{eff})_{SCH}= \frac{1}{2}\left[\left(1 -\frac{M_0R_b}{r}\right)\left(1 + \frac{h_{SCH}^2}{r^2}\right) - 1\right].\,\,
    \label{veffsch}
\end{equation}
\\
where $2M=M_0 R_b$. The expressions of the conserved energy and angular momentum for circular geodesics are defined as from the conditions $V_{eff}(r)=E$ and $V_{eff}^{\prime}(r)=0$,
\begin{equation}
    \gamma_{SCH}^2=\frac{2\left(r-M_0R_b\right)^2}{r\left(2r-3M_0R_b\right)}\,,\, ,\label{esch}
\end{equation}
\begin{equation}
    h_{SCH}^2=\frac{M_0R_br^2}{\left(2r-3M_0R_b\right)}\,\,
    \label{hsch}
\end{equation}
\\
from the above expressions, one can see that no circular orbit is possible within the range $0\leq r\leq\frac{3M_0 R_b}{2}$, or in terms of the total mass of the black hole, one can write $0\leq r\leq 3M$, where $M=\frac{3M_0 R_b}{2}$. Note that the above range is valid for stable and unstable circular orbits. Therefore, for stable circular orbits, one has to use the 
$ (V_{eff}^{\prime\prime})_{SCH}>0$ condition as,
\\
\begin{equation}
   (V_{eff}^{\prime\prime})_{SCH}= \frac{2M}{r}\left(\frac{6M-r}{3M-r}\right)\,\, .
   \label{isco}
\end{equation}
\\
from the above expression, one can see that no stable circular orbits are possible below $r=6M$ in the Schwarzschild spacetime. Thus, $r=6M$ is the innermost stable circular orbit (ISCO) in the Schwarzschild spacetime. Note that for bound orbits, the conditions given in equations (\ref{bound1}) and (\ref{bound2}) should be fulfilled. Now, an orbit equation of the test particle can be written using the general expression of the orbit equation given in (\ref{orbiteqgen}) for the Schwarzschild spacetime as, 
\begin{equation}
   \frac{d^2u}{d\phi^2} + u = \frac{3M_0R_b}{2}u^2 + \frac{M_0R_b}{2h^2}\,\, .
   \label{orbiteqsch}
\end{equation}
which is the second-order non-linear differential equation.

\subsection{JMN-1 orbit equation}
The JMN-1 naked singularity is the non-vacuum solution of Einstein equations. As discussed above, JMN-1 is formed as an end state of quasi-static gravitational collapse. It has a strong curvature singularity at the center without an event horizon. As it is a spherically symmetric and static solution, the conserved energy and angular momentum per unit rest mass is obtained as,
\begin{equation}
     \gamma_{JMN-1} =(1-M_0) \left(\frac{r}{R_b}\right)^{\frac{M_0}{1-M_0}} \left(\frac{dt}{d\tau} \right)\,\, ,
\end{equation}
\begin{equation}
     h_{JMN-1}= r^2  \left(\frac{d\phi}{d\tau}\right)\,\,,
\end{equation}
now, employing the general formalism of the orbit equation, the effective potential for the JMN-1 spacetime is derived as,
\begin{equation}
    (V_{eff})_{JMN-1} = \frac{1}{2}\left[(1- M_0)\left(\frac{r}{R_b}\right)^\frac{M_0}{(1- M_0)}\left(1 + \frac{h_{JMN-1}^2}{r^2}\right) - 1\right]\,\, ,\\
\end{equation}
where $h_{JMN-1}$ is the conserved angular momentum per unit rest mass of the test particles in JMN-1. The conserved quantities for circular orbits are defined as,
\begin{equation}
    \gamma^2_{JMN-1}=\frac{2(1-M_0)^2\left(\frac{r}{R_b}\right)^{\frac{M_0}{1-M_0}}}{\left(2-3M_0\right)}\, , \, \\
\end{equation}
\begin{equation}
     h^2_{JMN-1}=\frac{r^2M_0}{2-3M_0}\,\, ,\\
\end{equation}
the stable circular orbit can be obtained by using the condition,
\begin{equation}
    (V_{eff}^{\prime\prime})_{JMN-1}=\frac{M_0}{R_b^2}\left(\frac{r}{R_b}\right)^{\frac{3M_0-2}{1-M_0}}>0\,\, 
\end{equation}
from the above equations, it is seen that for $M_0<2/3$, JMN-1 spacetime can have stable circular orbits of any radius. In JMN-1 spacetime, the unstable circular orbits do not exist within the region $0<M_0<2/3$. Since, $h_{JMN-1}$ and $\gamma_{JMN-1}$ become imaginary for $M_0>2/3$, the stable circular orbits are not possible at any value of $r$ in JMN-1 spacetime. Now, an orbit equation of a freely falling test particle in JMN-1 spacetime can write as,
\begin{equation}
     \frac{d^2u}{d\phi^2} + (1 - M_o) u - \frac{\gamma^2}{2h^2}\frac{M_0}{(1- M_0)}\left(\frac{1}{u}\right)\left(\frac{1}{uR_b}\right)^\frac{-M_0}{(1- M_0)}=0\,\, ,\label{orbiteqJMN-1}\\
\end{equation}

\subsection{JNW orbit equation}
The JNW spacetime is a minimally coupled mass-less scalar field solution of the Einstein equations. Unlike JMN-1  naked singularity spacetime, JNW naked singularity spacetime is asymptotically flat. Therefore, this spacetime need not match with the external Schwarzschild spacetime. From the equation (\ref{JNWmetric}), the metric tensor components of this spacetime are independent of $t$ and $\phi$. Thus, one can define the conserved energy and angular momentum of a test particle per unit rest mass as, 
\begin{equation}
    h_{JNW} = r^2\left(1-\frac{b}{r}\right)^{1-n}\frac{d\phi}{d\tau}\, ,\,\,
\label{conservedJNW}
\end{equation}
\begin{equation}
    \gamma_{JNW} =  \left(1-\frac{b}{r}\right)^{n}\frac{dt}{d\tau}\,\,.
\end{equation}
Similarly, using the general formalism, the effective potential of a freely falling test particle can be obtain for JNW spacetime as,
\begin{equation}
    (V_{eff})_{JNW} = \frac{1}{2}\left[\left(1-\frac{b}{r}\right)^n\left(1+\frac{h_{JNW}^2}{r^2}\left(1-\frac{b}{r}\right)^{n-1}\right)-1\right]\,\, ,
\end{equation}
the conserved quantities for circular geodesics can be defined as,
\begin{equation}
    \gamma_{JNW}^2 = \left(1-\frac{b}{r}\right)^n\left[\frac{2r-b(n+1)}{2r-b(2n+1)}\right]\,\, ,\\
\end{equation}
\begin{equation}
    h_{JNW}^2 = r^2\left[\frac{bn\left(1-\frac{b}{r}\right)^{1-n}}{2r-b(2n+1)}\right]\,\, ,
\end{equation}
where the stable circular orbits would be excess by applying the condition $(V^{\prime\prime}_{eff})>0$. Although, In JNW spacetime, $(V^{\prime\prime}_{eff})$ is not always positive \cite{parth1}. One can find $(V^{\prime\prime}_{eff})>0$ at any point in JNW spacetime for $n<0.447$. But, for $0.447<n<0.5$, one certain radial interval ($r_{-},r_{+}$) is defined, in which no stable circular orbits are possible, where $r_{-}$ and $r_{+}$ can be written as,
\begin{equation}
    r_{-}=\frac14\left(b~(2+6n)-4.472~b\sqrt{n^2-0.2}\right)\,\, ,\label{oscojnw}\\
\end{equation}
\begin{equation}
    r_{+}=\frac14\left(b~(2+6n)+4.472~b\sqrt{n^2-0.2}\right)\,\, .
    \label{iscojnw}
\end{equation}
Note that stable circular orbits are not possible in the certain radial interval $r_{-}<r<r_{+}$. However, one can have stable circular orbits outside this radial interval as there is $(V^{\prime\prime}_{eff})_{JNW}>0$. On the other hand, for $0.5<n<1$, one can get the innermost stable circular orbit (ISCO) at $r_{(ISCO)}$. Now, employing the general method to derive an orbit equation for the JNW spacetime, one can get the second-order non-linear differential equation as,
\begin{equation}
    \frac{d^2u}{d\phi^2} + u + \frac{b\gamma^2}{2h^2}(2-2n)(1-bu)^{1-2n} - \frac{b(2-n)}{2h^2}(1-bu)^{1-n} - \frac{3bu^2}{2}=0\,\, .
   \label{orbiteqJNW}
\end{equation}

\section{Periastron precession of the timelike bound orbits}
\begin{center}
    \begin{figure}[H]
\centering
    \includegraphics[scale=0.7]{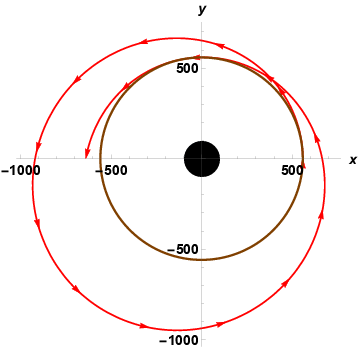}
    \caption{Positive orbital precession in Schwarzschild spacetime}
    \label{SCHprece1}
\end{figure}
\end{center}

Since the fully relativistic orbit equations are second-order non-linear differential equations, it is solved numerically. The solutions of these equations are used to get parametric orbital plots. In figure (\ref{SCHprece1}), the orbital precession in the equatorial plane of the Schwarzschild spacetime is shown for the parameters $M_0=0.09, h=200, E=-0.0268$ and $R_b=1000$.
The dark black region indicates the position of the Schwarzschild black hole. The red line represents the orbit of the freely falling test particle. The brown circle shows the test particle's minimum approach (periastron position) towards the center. Due to the spacetime curvature of the Schwarzschild geometry, a test particle's orbit would precess in a particular direction. In the Schwarzschild black hole case, the orbital precession of a test particle occurs in the direction of the particle's angular motion. One can see from the figure (\ref{SCHprece1}) that the arrow indicates the direction of the particle's angular movement, which is in the anti-clockwise order. Whenever the particle's orbit precesses in the direction of the particle's orbiting path, the angular distance travelled by a particle around the center is always found greater than $2\pi$, known as the positive precession. As seen from the above figure, the angular distance travelled by a particle between two successive periastron points remains greater than $2\pi$ for any parametric values. Therefore, the Schwarzschild black hole spacetime always admits positive precession. the proof of this statement will discuss in the next section. 
\begin{figure*}
\centering
\subfigure[]
{\includegraphics[scale=0.55]{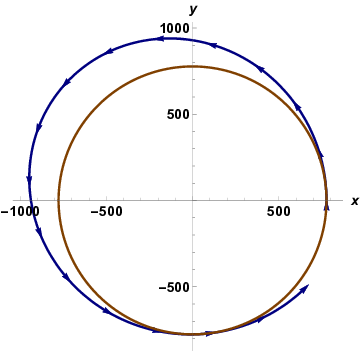}}
\hspace{0.2cm}
\subfigure[]
{\includegraphics[scale=0.55]{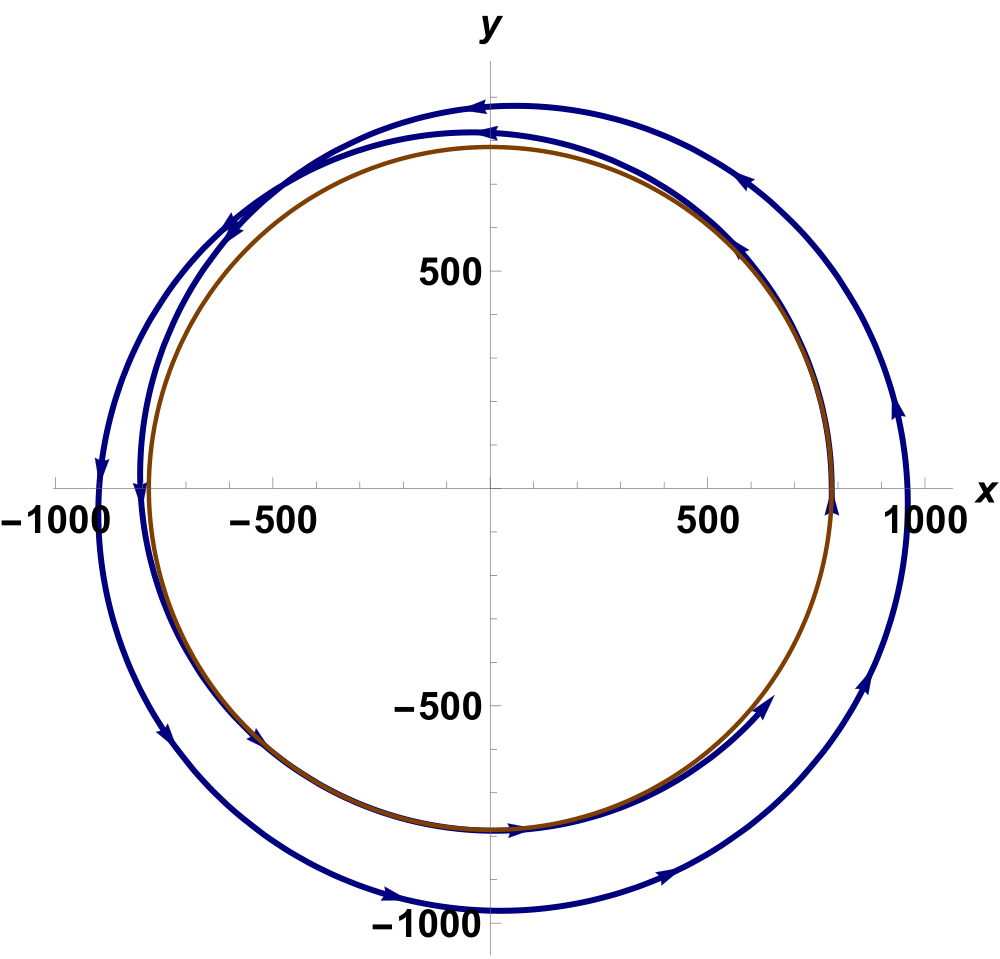}}
\hspace{0.2cm}
\caption
{Negative and Positive orbital precession in the JMN-1 spacetime}\label{JMN1prece1}
\end{figure*}

Figure \ref{JMN1prece1} shows the particle orbits in the JMN-1 spacetime, where the blue line represents the particle's orbit in the JMN-1 spacetime. The brown circle indicates the periastron position. For the comparison, the same parameters is used in the JMN-1 case as it is used in the Schwarzschild case, where $M_0=0.09, h=200, E=-0.0268$ and $R_b=1000$. An interesting result emerged from the orbital precession of a test particle in the JMN-1 spacetime. In this spacetime, the angular distance travelled by a test particle in between two successive periastron points can have less than $2\pi$. In other words, the precession of the orbits will occur in the opposite direction of the particle's angular motion, known as the negative precession. One can see, Figure \ref{JMN1prece1}(a) shows the negative precession of the orbits in the JMN-1 spacetime. On the other hand, JMN-1 spacetime admits the same nature of the precession as Schwarzschild spacetime does. Figure \ref{JMN1prece1}(b), shows the positive precession of the orbits in JMN-1 spacetime for the parameters $M_0=0.55, h=1100, E=-0.004$ and $R_b=1000$. This figure shows that the angular distance travelled by a particle between two successive periastron points is greater than $2\pi$. 
\begin{figure*}
\centering
\subfigure[]
{\includegraphics[scale=0.28]{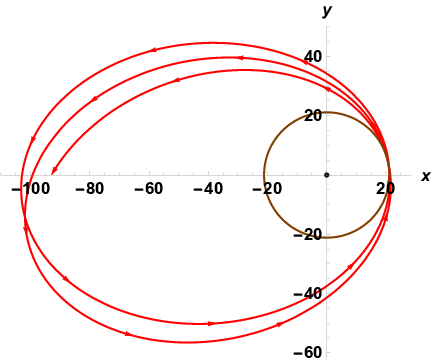}}
\hspace{0.2cm}
\subfigure[]
{\includegraphics[scale=0.28]{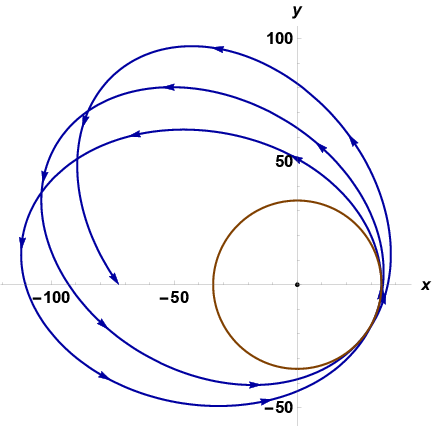}}
\hspace{0.2cm}
\subfigure[]
{\includegraphics[scale=0.28]{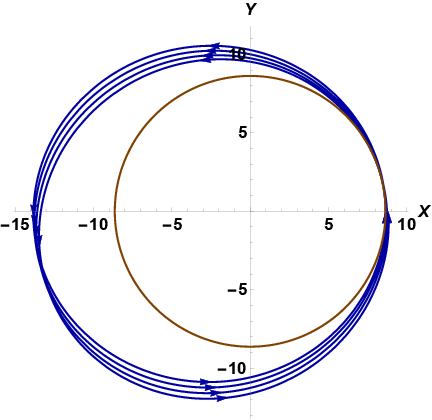}}
    \hspace{0.2cm}
\caption
{Positive and negative orbital precession in the JNW spacetime}\label{JNWprece1}
\end{figure*}

Similarly, the orbital precession in the JNW spacetime have been analysed. As discussed above in the JMN-1 case, JNW spacetime can also allow the negative precession of the orbits. Figure \ref{JNWprece1}(a) shows the particle orbits in the Schwarzschild spacetime corresponding to the JNW case. As shown in figure \ref{JNWprece1}(b), the JNW spacetime admits negative precession of the orbits. Figures \ref{JNWprece1}(a) and \ref{JNWprece1}(b) are shows the comparison of the orbital precession in between the Schwarzschild and JNW spacetime for the same parameters $M=0.25, q=10, h=3$ and $ E=-0.002$. The positive precession of the orbits in JNW spacetime is shown in figure \ref{JNWprece1}(c) for $M=0.25, q=3, h=3$ and $E=-0.002$.

\section{Approximate solutions of the orbit equations}
In the above section (\ref{relativisticorbiteqs}), The fully relativistic second-order non-linear differential orbit equations are derived for Schwarzschild, JMN-1, and JNW spacetimes. The numerical solution of those orbit equations can give the information about the shape of different orbits of freely falling particles in the given spacetime. However, the approximate solutions of those orbit equations are helpful in understanding the nature of orbital precession.

\subsection{Approximate solution of the Schwarzschild orbit equation}
In weak field limits, the fully relativistic Schwarzschild orbit equation becomes the Newtonian orbit equation since the Schwarzschild metric is asymptotically flat. The Schwarzschild orbit equation in the asymptotic limit can be derived by neglecting the $u^3$ term as,
\begin{equation}
    \frac{d^2u}{d\phi^2}+u=\frac{M_0R_b}{2h^2}\,\, ,
\end{equation}
the well-known solution of the above Newtonian orbit equation is given as,
\begin{equation}
    u=\frac{M_0R_b}{2h^2}\left[1+e \cos(\phi)\right], 
\end{equation}
This equation implies that the orbiting particle will reach its previous position after the $2\pi$ rotation. Nevertheless, it is not true for strong field limits since one cannot neglect $u^3$ term from the Schwarzschild orbit equation. We have considered the case in which a small eccentricity approximation is valid in both the strong and weak field regimes. An approximate solution for the orbit equation of Schwarzschild spacetime can be write as,
\begin{equation}
\tilde{u}=\frac{1}{p}\left[1+e\cos(m\phi)+O(e^2)\right]\,\, ,
\label{orbitsch1}
\end{equation}
where $\tilde{u}=uR_b$, and $m$ and $p$ are positive real numbers. Now, by substituting the above solution of the orbit equation (\ref{orbitsch1}) into the second-order non-linear differential Schwarzschild orbit equation (\ref{orbiteqsch}), then one can get the following expressions of $p$ and $m$ from the zeroth and first order of eccentricity terms respectively,
\begin{equation}
    p_\pm=\frac{1\pm \sqrt{1-\frac{3M_0^2R_b^2}{h^2}}}{\frac{M_0R_b^2}{h^2}}\,\, ,\label{psch}\\
\end{equation}
\begin{equation}
    m=\sqrt{1-\frac{3M_0}{p}}\label{msch}\,\, .
\end{equation}
the second and higher-order eccentricity terms are neglected, considering a small eccentricity approximation. We also neglect another solution of $p_-$, as $m$ becomes imaginary when $h>\sqrt3M_0R_b$, which is a necessary condition for real values of $p$.
\begin{equation}
    p=\frac{1-\sqrt{1-\frac{3M_0^2R_b^2}{h^2}}}{\frac{M_0R_b^2}{h^2}}
\end{equation} 
We always get $p_+>3M_0$ and hence $0<m<1$ for $h>\sqrt3M_0R_b$. These expressions of $p_+$ and $m$ imply that the orbiting test particle would not reach its previous minimum approach towards the center (periastron point) after having a full $2\pi$ rotation. A test particle needs to travel some extra angular distance to reach the periastron point again, known as the periastron precession or shift of the orbits. The value of this extra angular distance depends upon the value of $m$. In the Schwarzschild case, $m$ should be within $0<m<1$ as seen from equation (\ref{msch}). It implies that the angular distance travelled by a particle between two successive periastron points is always greater than $2\pi$. In other words, the precession of the orbit will occur in the direction of the particle's rotation. 
   
\subsection{Approximate solution of the JMN-1 orbit equation}
The JMN-1 orbit equation (\ref{orbiteqJMN-1}) can define by transformation $u\rightarrow uR_b$,
\begin{equation}
\tilde{u}\frac{d^2\tilde{u}}{d\phi^2}+\frac{1}{1+2\delta}\tilde{u}^2=C_{\delta}\tilde{u}^{2\delta}\,\, ,
\label{orbiteq2}
\end{equation}
where $\tilde{u}=uR_b$ and
\begin{equation}
    \delta=\frac{M_0}{2(1-M_0)},
\end{equation}
\begin{equation}
    C_{\delta}=\frac{\gamma^2 R_b^2}{h^2}\delta.
\end{equation} 
In \cite{Struck:2005hi,Struck:2005oi}, 
the approximate solution of a small eccentricity is introduced for this differential equation. The solution can be written as,
\begin{equation}
\tilde{u}=\frac1p\left[1+e\cos(m\phi)+O(e^2)\right]^{\frac12+\delta}\,\, ,
\label{approxorbit}
\end{equation}
by substituting equation (\ref{approxorbit}) into (\ref{orbiteq2}), the following expressions of $p$ and $m$ for JMN-1 spacetime can be find from the zeroth and first-order eccentricity terms, respectively,
\begin{equation}
    p=\left[C_{\delta}(1+2\delta)\right]^{-\frac{1}{2(1-\delta)}}\,\, ,
\label{pjmn1}
\end{equation}
\begin{equation}
    m=\sqrt{\frac{2(1-\delta)}{2\delta+1}}=\sqrt{2-3M_0}\,\, ,
\label{mjmn1}
\end{equation}
from the above equation (\ref{mjmn1}), one can clearly see that $m$ is greater than 1 for $0<M_0<1/3$, whereas $m$ is less than 1 for $1/3<M_0<2/3$. Hence, in JMN-1 spacetime, the angular distance travelled by a particle in between two successive periastron points could be greater than (for $1/3<M_0<2/3$) or less than (for $0<M_0<1/3$). Consequently, the nature of the orbits also changes across $M_0=1/3$. The precession of orbits will occur in the opposite direction of the particle's motion for $M_0<1/3$, whereas, for $M_0>1/3$, one can get Schwarzschild-like orbital precession.

\subsection{Approximate solution of the JNW orbit equation} 
Like Schwarzschild and JMN-1 orbit equations, the JNW orbit equation (\ref{orbiteqJNW}) can be write in the following form \cite{parth1},
\begin{figure*}
\centering
\subfigure[$h=3$,$E=-0.002, R_b=1$]
{\includegraphics[width=67mm]{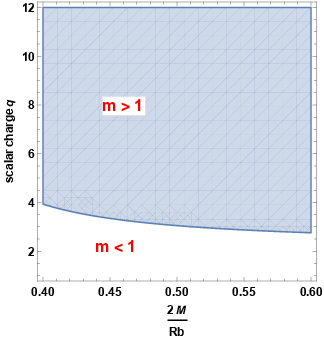}\label{re1}}
\hspace{0.2cm}
\subfigure[$h=0.1$,$E=-0.02, R_b=1$]
{\includegraphics[width=70mm]{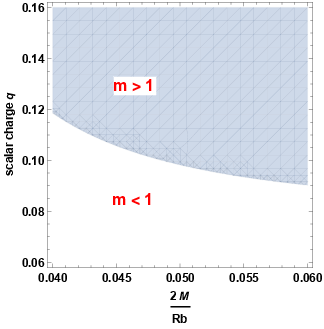}\label{re2}}
 \caption{Regions plots for $m>1$ and $m<1$ in JNW spacetime.}
\end{figure*}
\begin{equation}\label{approxJNW}
\tilde{u}\frac{d^2\tilde{u}}{d\phi^2} + Q\tilde{u} = R\tilde{u}^2 + S\tilde{u}^3\,\, ,
\end{equation}
where, 
\begin{equation}
    Q = \left[\frac{b^2\gamma^2(1-n)}{h^2} - \frac{b^2(2-n)}{2h^2}\right]\,\, ,\\
\end{equation}
\begin{equation}
    R = \left[\frac{b^2\gamma^2(1-n)(1-2n)}{h^2} - \frac{b^2(2-n)(1-n)}{2h^2} - 1\right]\,\, ,\\
\end{equation}
\begin{equation}
    S = \left[\frac{3}{2} - \frac{b^2(2-n)(1-n)n}{4h^2} + \frac{b^2\gamma^2(1-2n)(1-n)n}{h^2}\right]\,\, .\\
\end{equation}
The small eccentricity approximation solution of an orbit equation (\ref{approxJNW}) can be written as,
\begin{equation}
\tilde{u}=\frac{1}{p}\left[1+e\cos(m\phi)+O(e^2)\right]\,\, ,
\label{solapprox}
\end{equation}
where p and m are the positive real number. Using eq.~(\ref{approxJNW}) and (\ref{solapprox}) the expressions of $p$ and $m$ can be found by neglecting higher order terms of eccentricity $e$,
\begin{equation}
   p_{\pm}=\frac{R\pm\sqrt{R^2+4QS}}{2Q}\,\, ,\label{pJNW}\\
\end{equation}
\begin{equation}
    m=\sqrt{Qp-2R-\frac{3S}{p_+}}\label{mJNW}\,\, ,
\end{equation}
now, using equation (\ref{mJNW}), it is shown that in JNW spacetime, one can have two types of orbital precessions. In JNW spacetime, a test particle can travel less than or greater than $2\pi$ angular distance between two successive periastron points.
In Figure \ref{re1} and \ref{re2}, It is shown that the regions of $m>1$ and $m<1$ in  the parameter space of the scalar charge $q$ and ADM mass $M$ of the JNW spacetime.
\section{Discussions and conclusions}
 \label{sec_discussion}
 
 In this chapter, the timelike bound orbits in the JMN-1, JNW and Schwarzschild geometries are examined to explore the causal structure of these spacetimes and to understand the main characteristic differences between the two. The study of these orbits brings out several interesting differences in the causal structure of these black hole and naked singularity spacetimes, some of these are summarized below: 

\begin{itemize}
    \item In Schwarzschild spacetime, the bound trajectories of a massive particle always precess in the direction of the particle motion. However, in JMN-1 and JNW naked singularity spacetimes, for a range of parameter values, the bound orbits precess in the opposite or reverse the direction of particle motion, which is a very novel distinct feature. In JMN-1 spacetime, for  $0<M_0<\frac13$, the orbits of a particle precess in the opposite direction of particle motion and for $\frac13<M_0<\frac23$, the orbits precess in the same direction of particle motion. While for JNW naked singularity spacetimes, $m$ can be greater or less than one, which is shown in fig.~\ref{re1} and \ref{re2}. Therefore, unlike Schwarzschild spacetime, in JNW spacetime, the particle bound trajectories can precess opposite to the direction of particle motion.

    \item In Schwarzschild spacetime, the perihelion point of a timelike orbit always lies in $r> M_0R_b$. On the other hand, in JMN-1 naked singularity spacetime, the perihelion point of a timelike orbit lies in $r>0$ region. Therefore, in JMN-1 spacetime, a massive particle can go very close to the central naked singularity. In Schwarzschild spacetime, an innermost stable circular orbit (ISCO) exists at $r=3M_0R_b$. However, in JMN spacetimes, stable circular orbits for massive 
particles can exist at any radius. On the other hand, in JNW spacetime, for $0.5<n<1$, there exists an ISCO. However, for $0.447<n<0.5$, other than $r_{-}<r<r_{+}$ (eq.~(\ref{iscojnw}) and (\ref{oscojnw})), stable circular orbits of any radius are possible. For $0<n<0.447$, there are no constraints on the radius of the stable circular orbits of particles in JNW spacetime. This difference can create distinguishable accretion disk properties, which can be detectable \cite{Joshi:2013dva, shaikh1}. 
 
\end{itemize}

To better understand the galactic centre's causal structure, mass, and dynamics, one has to study the timelike and null geodesics behaviour around the galactic center. The GRAVITY and SINFONI are continuously eyeing up the Milky Way galactic center to get important observational data of stellar motion around the center. Within such a  context, it is shown in this chapter that the timelike geodesics of a freely falling particle in naked singularity spacetimes can be significantly different from the timelike geodesics in Schwarzschild spacetime. 
There are many `S' stars (e.g. S2, S102, S38, and others) orbiting around the center of the Milky Way galaxy Sgr A*. The minimum approach of those stars lies inside $0.0001-0.0006$ parsec. They are very close to the center, so their dynamics can give the essential information about the central object Sgr A*. One can use the orbit equations of the Schwarzschild, JMN-1 and JNW spacetimes to predict the future trajectories of those stars after best fitting the data of past trajectories for them. 

\chapter{Precession of timelike bound orbits in rotating spacetimes}
Every celestial body has its own intrinsic spin angular momentum. Therefore, including a non-zero spin in a non-rotating spacetime makes the modified spacetime physically more realistic. There are no restrictions on the value of the spin angular momentum of a compact object as long as it is not a black hole. A dimensionless spin parameter typically represents the spin of a celestial object $\tilde{a}$. This spin parameter can be defined as $\tilde{a}=\frac{c J}{GM^2}$, where $c, J, G, M$ are the velocity of light, the intrinsic spin angular momentum of the body, Newton's gravitational constant, and the mass of the celestial body respectively. Earth, with its spin angular momentum $J\sim 7.2\times 10^{33} kg~ m^2 s^{-1}$ and mass $M= 5.972\times 10^{24} kg$ has the spin parameter $\tilde{a}=907$, whereas the sun has spin parameter $\tilde{a}=0.216$. On the other hand, a rapidly spinning massive star VFTS102 has $\tilde{a}=75$ \cite{Nielsen:2016kyw}.
Therefore, one can see that the value of the spin parameter of a celestial object can be much greater than one. However, if one consider the Kerr black hole, which is a rotating generalization of the Schwarzschild black hole, the spin parameter $\tilde{a}$ cannot be allowed to be greater than unity in order to ensure the existence of an event horizon. The Kerr black hole is a vacuum axisymmetric solution of Einstein field equations. Two parameters characterize it: the black hole's total mass $M$ and total angular momentum $J$. Kerr spacetime describes a rotating black hole, if $\tilde{a} \leq 1$, whereas it describes a rotating naked singularity spacetime if $\tilde{a} > 1$.

In this chapter, the prime focus is to discuss the nature of periastron precession of timelike bound orbits in the rotating spacetimes, e.g. Kerr and JNW spacetimes. The previous chapter shows that Schwarzschild spacetime does not admit any negative precession. In this work, it is shown that the precession of the timelike bound orbits in Kerr spacetime is always positive when one consider only low eccentric, equatorial, bound orbits. With the small eccentricity approximation, one can analyze the nature of the timelike bound orbits in the strong field region of Kerr spacetime, where the relativistic effects on the bound orbits are likely to be observed. The timelike bound orbits are studied in rotating JNW spacetime and compared with the orbits in Kerr spacetime.

\section{Timelike geodesics in Kerr spacetime}
\label{two}
In this section, the geodesics of a test particle in the Kerr spacetime are studied. The stationary, axisymmetric and rotating Kerr spacetime is already introduced in the third chapter (\ref{Kerr metric}). Studying test particle motion in Kerr spacetime is important to understand the physical processes occurring in these spacetimes and their observational consequences. The Kerr spacetime is independent of $t$ and $\phi$. Therefore, the conserved energy ($e$) and the angular momentum ($L$) per unit rest mass are given by,
\begin{equation}
    e = g_{tt} U^t + g_{t\phi} U^{\phi}\,\, ,
    \label{conserved1}
\end{equation}
\begin{equation}
     L = -g_{t\phi} U^t + g_{\phi\phi} U^{\phi}\,\, ,
     \label{conserved2}
\end{equation}
where $U^\mu$ are the components of the four velocity of a test particle and $g_{tt} = \left(1-\frac{r_s r}{\Sigma}\right) $, $g_{rr} = \frac{\Sigma}{\Delta}$, $g_{\theta\theta} = \Sigma$, $g_{\phi\phi} = \left(r^2 + a^2 + \frac{r_s r a^2 \sin^2\theta}{\Sigma}\right)$, and $g_{t\phi} = \frac{r_s r a \sin^2\theta}{\Sigma} $. In a physically realistic situation, the orbital angular momentum of a test particle and the spin angular momentum of a central rotating body need not necessarily be aligned. Here, for simplicity, the orbits are considered in the equatorial plane ($\theta=\frac\pi2$).
Eqs. (\ref{conserved1}) and (\ref{conserved2}) can be solved for $U^t$ and $U^{\phi}$ as, 
\begin{equation}
    U^t =  \frac{1}{\Delta} \left[\left(r^2 + a^2 + \frac{r_s  a^2 }{r}\right) e - \left(\frac{r_s  a }{r}\right) L\right] \,\, ,
    \label{four0}
\end{equation}
\begin{equation}
    U^{\phi} = \frac{1}{\Delta} \left[\left(\frac{r_s a }{r}\right) e + \left(1 - \frac{r_s }{r}\right) L\right]\,\, .
    \label{four1}
\end{equation}
Now, using normalization condition $U^{\alpha}U_{\alpha}=-1$ of timelike geodesics and also using the Eqs. (\ref{four0}), (\ref{four1}), one can derive $r$-component of the four velocity $U^r$ as,
\begin{equation}
   U^r = \pm \sqrt{(e^2 - 1) + \frac{r_s}{r} - \frac{L^2 - a^2 (e^2-1)}{r^2} + \frac{r_s (L - a e)^2}{r^3}}\,\, .
   \label{four2}
\end{equation}
Here, $\pm$ signatures correspond to the radially outgoing and incoming timelike geodesics, respectively. The expression, in Eq.~(\ref{four2}), is equivalent to the kinetic energy of a test particle. The total relativistic energy is defined as,
\begin{equation}
    E = \frac{1}{2} (e^2 - 1) = \frac{1}{2} (U^r)^2 + V_{eff}(r)\,\, .\label{E1}
\end{equation}
Using the expression of $U^r$ (Eq.~(\ref{four2})) and the expression of total relativistic energy (Eq.~(\ref{E1})), the following expression of effective potential is derived,
\begin{equation}
    V_{eff}(r) = -\frac{r_s}{2r} + \frac{L^2 - a^2 (e^2-1)}{2r^2} - \frac{r_s (L-a e)^2}{2 r^3}\,\, .
    \label{veff1}
\end{equation}
The above expression of the effective potential is only applicable for equatorial timelike geodesics. For bound orbits, the particle's total energy is greater than or equal to the minimum effective potential. The minimum effective potential is determined as,
\begin{equation}
    \frac{dV_{eff}}{dr}|_{r_b} = 0 ;        \frac{d^2 V_{eff}}{dr^2}|_{r_b}>0\,\, ,
    \label{con1}
\end{equation}
where the effective potential has a minimum at $r=r_b$.
Using Eq.~(\ref{veff1}) and Eq.~(\ref{con1}), following expression of $r_b$ is obtained as, 
\begin{equation}
    r_b = \frac{1}{2M}\big(L^2 + a^2 (1 - e^2) + \sqrt{(L^2 + a^2 (1 - e^2))^2 - 12M^2 (L - ae)^2}\big)\,\, .
    \label{rb1}
\end{equation}
The minimum effective potential at $r = r_b$ is,
\begin{multline}
    V_{eff}(r_b) = \frac{-1}{(L^2 + a^2 (1 - e^2) + \sqrt{(L^2 + a^2(1 - e^2))^2 - 12M^2 (L - a  e)^2})^3}(2M^2 (a^4 (1 - e^2)^2\\ + 16a L e M^2 + L^2(L^2 - 8M^2 + \sqrt{(L^2 + a^2(1 - e^2))^2 - 12M^2 (L - a e)^2}) + a^2 (2L^2 (1 - e^2) 
    \\
    - 8M^2 + (1 - e^2) \sqrt{(L^2 + a^2(1 - e^2))^2 - 12M^2 (L - a e)^2})))\,\, .
\end{multline}
The bound orbits exist for $V_{min}\leq E<0$. Using the bound orbit conditions, one can determine the shape of the orbits, which gives how $r$ changes in the equatorial plane with respect to $\phi$,
\begin{equation}
    \frac{dr}{d\phi} = \frac{\pm \sqrt{(e^2 - 1) + \frac{2 M}{r} - \frac{L^2 - a^2 (e^2-1)}{r^2} + \frac{2 M (L - a e)^2}{r^3}}}{\frac{1}{\Delta} \left[\left(\frac{r_s a }{r}\right) e + \left(1 - \frac{r_s }{r}\right) L\right]}\,\, .
    \label{shapeorbits}
\end{equation}
Using Eq.~(\ref{shapeorbits}), one can derive second order differential orbit equation of a massive test particle in Kerr spacetime,

\begin{multline}
    \frac{d^2u}{d\phi^2}= \frac{1 - 2Mu + a^2 u^2}{(L - 2Mu (L - a e))^3}[M (L - 2a e (e^2 - 1)) + u (L (-L^2 + 3a^2 (e^2 - 1))\\ - 2M^2 (2L + 2a e) + u B(\phi))]\,\, ,
    \label{orbiteq1}
\end{multline}
where $u = \frac{1}{r}$ and 
\begin{multline}
    B(\phi) = M (7L^3 - 6a e L^2 + a^2L (11 - 3e^2) + 2a^3e (e^2-1)) + 4M^3 (L - a e) + u [-3a^2L^3 \\+ 3a^4L (e^2 - 1)
    + 2M^2 (L - a e)(-8L^2 + a (7Le + a (e^2-5))) \\- Mu (L - a e) (a^2(-11L^2 + a (7Le + 4a (e^2 - 1)))
    - 12M^2(L - a e)^2 + 10a^2Mu(L-a e)^2)  ]\,\,\nonumber .
\end{multline}

\begin{figure*}
\centering
	\includegraphics[scale=0.35]{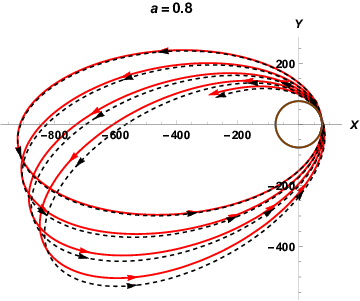}
	\includegraphics[scale=0.35]{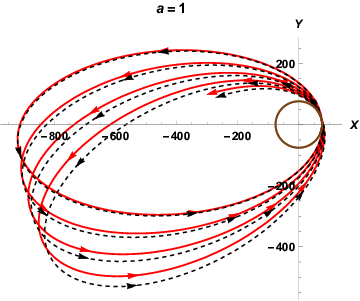}
	\includegraphics[scale=0.35]{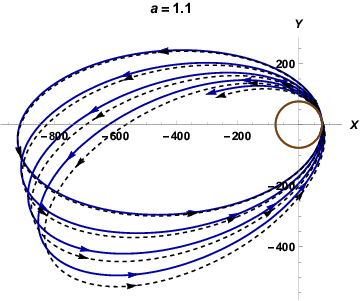}
	\includegraphics[scale=0.35]{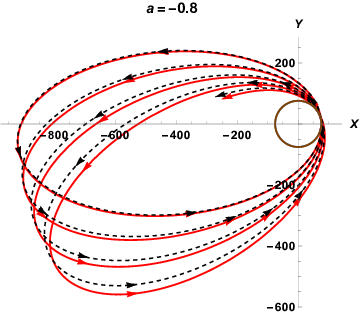}
	\includegraphics[scale=0.35]{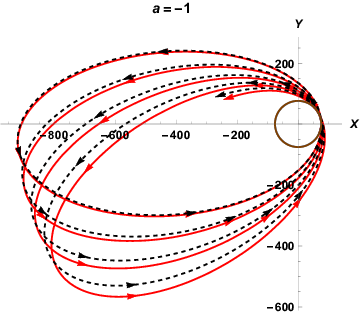}
	\includegraphics[scale=0.35]{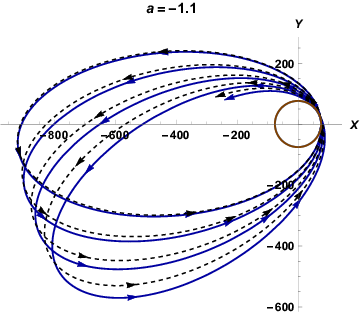}
		\caption{ Orbits of a test particle in the Kerr spacetime for $L=12$.}
	\label{orbit1}
    \end{figure*}
Numerical solution is found for the above orbit equation (Eq.~(\ref{orbiteq1})) to investigate the nature and shape of bound orbits of a test particle that is freely falling in Kerr spacetime. Fig.~(\ref{orbit1}) shows the bound orbits of a test particle in the Kerr black hole spacetime and Kerr naked singularity spacetime. In that figure, the timelike bound orbits are shown for spin parameters $a=\pm 0.8, \pm 1, \pm 1.1$. The values of spin parameter $a=\pm 0.8, \pm 1, \pm 1.1$ correspond to the Kerr black hole, extreme Kerr black hole and Kerr naked singularity, respectively. In Fig.~(\ref{orbit1}), the particle's total energy $E=-0.001$, angular momentum $L=12$ and the mass of the black hole $M=1$ are considered. The orbit shown by black dotted lines represents the timelike orbits in Schwarzschild spacetime (i.e. $a=0$). In Fig.(\ref{orbit1}), the timelike bound orbits in Kerr black hole spacetime (i.e. $a<1$) and in Kerr naked singularity spacetime (i.e. $a>1$) are shown by solid red lines and solid blue lines respectively. It can be seen that the orbital precession in Kerr spacetime is distinguishable from the orbital precession in Schwarzschild spacetime. All the orbits in Fig.~(\ref{orbit1}) show a positive precession, and the non-zero spin parameter changes the minimum approach ($r_{min}$) and perihelion shift of those orbits. One can see that for $a>0$, the minimum approach of the particle (Periastron point) increases as the value of the spin parameter increases.
 \begin{figure*}
 \centering
	\includegraphics[scale=0.35]{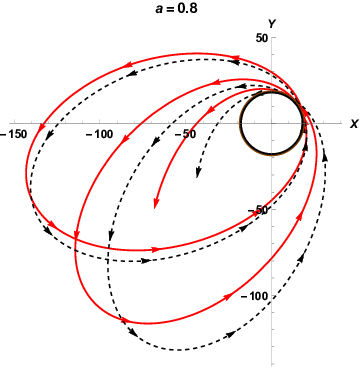}
	\includegraphics[scale=0.35]{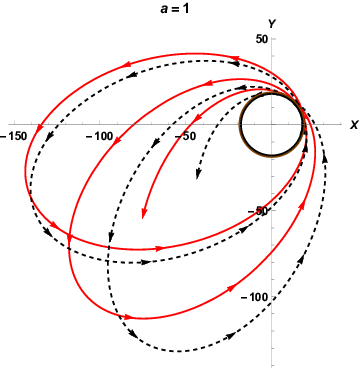}
	\includegraphics[scale=0.35]{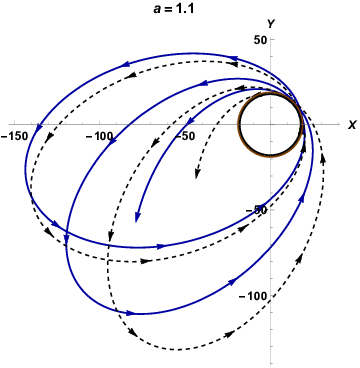}
	\includegraphics[scale=0.35]{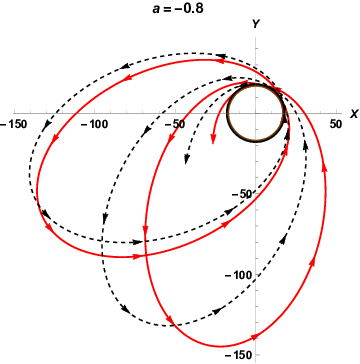}
	\includegraphics[scale=0.35]{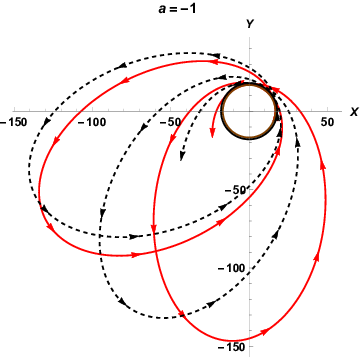}
	\includegraphics[scale=0.35]{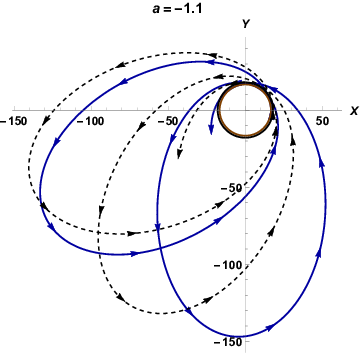}
\caption{Orbits of a test particle in the Kerr spacetime for $L=6$.}
	\label{orbit2}
    \end{figure*}

On the other hand, for $a<0$, the minimum approach of the particle decreases as the value of the spin parameter decreases. This effect of a spin parameter can also be seen in Fig.~(\ref{orbit2}), where the particle's total energy $E=-0.006$ and angular momentum $L=6$. Since the angular momentum ($L$) considered in Fig.(\ref{orbit2}) is smaller than that of Fig.(\ref{orbit1}), the minimum approaches of the orbits in the Fig.~(\ref{orbit2}) are much smaller than the minimum approaches of the orbits in the Fig.~(\ref{orbit1}). Therefore, the frame-dragging effect of Kerr black hole geometry is much higher in the second case (i.e. for $L=6$). However, it can be verified that all the orbits in Fig.~(\ref{orbit2}) always have a positive precession.

There exists a radial limit under which no stable bound orbit is possible in Kerr spacetime. In Schwarzschild spacetime, there exists a minimum value of the radius of the stable circular orbit, and it is known as the radius ($r_{ISCO}$) of the innermost stable circular orbit (ISCO). For Schwarzschild spacetime, the ISCO is at $r_{ISCO}=6M$. If one can put a condition that the $r_b$ in  Eq.~(\ref{rb1}) should always be real, then the $r_{ISCO}$ for Kerr metric can be written as,
\begin{eqnarray}
    r_{ISCO} &=& 3M - \sqrt{3}ae + \sqrt{9M^2-6\sqrt{3}aeM-3a^2(1-e^2)}\nonumber\,\, ,\\
    &\approx& 6M - 2\sqrt{3}a\bigg(e+\dfrac{a}{4\sqrt{3}M} \bigg)\,\, ,
    \label{rISCO}
\end{eqnarray}
where the first expression is the exact expression of $r_{ISCO}$ and the second expression is the approximate one, where small values of the spin parameter is considered. Using the first expression $r_{ISCO}$, one can verify that the value of $r_{ISCO}$ is real and finite when $a<M$ (i.e.  Kerr black hole). However, when $a>M$ (i.e.  Kerr naked singularity), there exists no real value of $r_{ISCO}$, which implies that stable circular orbits can extend up to the singularity.
The second expression is helpful in understanding how much the ISCO radius differs from $6M$ (i.e. the ISCO radius in Schwarzschild spacetime) due to the non-zero value of the spin parameter. Any bound orbits with a minimum approach ($r_{min}$) close to $r_{ISCO}$ can have a very large perihelion shift.
One can derive the expression for the smallest possible value of $r_{min}$ ($r_{min0}$) of a bound timelike orbit in Kerr spacetime by finding the solution of $V_{eff}|_{r_{min0}}=0$, $\frac{dV_{eff}}{dr}|_{r_{min0}}=0$ and $\frac{d^2 V_{eff}}{dr^2}|_{r_{min0}}<0$. The expression of $r_{min0}$ can be written as,
\begin{equation}
r_{min0}\approx 4M-a\left(2e+\frac{a}{4M}\right)\,\, .
\end{equation}
Timelike bound orbits with the above value of minimum approach have a large amount of perihelion shift. One can verify that though the perihelion shift is large near $r_{min0}$, the precession is always positive.
However, this significant perihelion shift of timelike bound orbits cannot be seen or verified in the stellar motions of `S' stars. For example, the S2 star has a minimum approach which is $r_{min}= 2800M$; therefore, if the central body (Sgr-A*) is considered as a Kerr black hole, then the perihelion shift of the orbit of S2 would have a very small positive value, and that would be very close to the expected value of the perihelion shift of the S2 star in a Schwarzschild background. Hence, the orbits of the S2 star should always have a positive precession if Schwarzschild or Kerr black hole is assumed at the center of the Milky way galaxy (Sgr-A*).  

In Fig.~(\ref{orbit1}) and Fig.~(\ref{orbit2}), the positive precession of timelike bound orbits (on $\theta=\frac\pi2$ plane) is shown in Kerr spacetime for some particular values of $L$, $E$ and $a$. However, by those figures, one cannot or do not claim to prove that the phenomenon of negative precession is always forbidden or absent in Kerr spacetimes. To prove that for timelike bound orbits in Kerr spacetime, one need to analytically solve the orbit equation (Eq.~(\ref{orbiteq1})). However, as Eq.~(\ref{orbiteq1}) is a fairly complicated non-linear differential equation, it is solved numerically or using a suitable approximation. Therefore, in the next section, the nature of the perihelion shift of the orbits are studied in Kerr spacetime using an approximation technique. 

\section{An approximate solution of orbit equation}\label{three}
In this section, it is investigated that whether a negative precession of timelike orbits of particles is possible for any value of $L,e$ and $a$. For this purpose, an approximate method is used where only low eccentric orbits are considered. Therefore, for the approximate solution of the Eq.~(\ref{orbiteq1}), the first order expression in eccentricity $\epsilon$ is considered \cite{Struck:2005hi,Struck:2005oi}. The approximate solution can be written as, 
\begin{equation}
u(\phi)=\frac{1}{M p}(1+\epsilon\cos(m\phi)+\mathcal{O}(\epsilon^2))\,\, .
\label{18}
\end{equation}
where $p$ and $m$ are real positive constants, and $m>1$ represents the precession of the timelike bound orbits in the negative direction, while $m<1$ represents the precession in the positive direction. 
When the above expression of $u(\phi)$ is substitute in the orbit equation (Eq.~(\ref{orbiteq1})) and separate the zeroth order terms and the first order terms of $\epsilon$, one can get an expression of $m$ in terms of $p$. From the zeroth order terms, equation of fifth order polynomial of $p$ is obtain as,
\begin{eqnarray}
    &g_5&(L,a,e,M) p^5 + g_4(L,a,e,M) p^4 + g_3(L,a,e,M) p^3\nonumber \\&+& g_2(L,a,e,M) p^2 + g_1(L,a,e,M) p + g_0(L,a,e,M) = 0\,\, ,\nonumber\\ 
    \label{polynomial}
\end{eqnarray}
where
\begin{eqnarray}
    g_5(L,a,e,M) &=& -(2ae(1-e^2)+L)M^4\,\, ,\nonumber\\
    g_4(L,a,e,M) &=& (3a^2(1-e^2)L+L^3+2aeM^2+4LM^2)M^2\,\, ,\nonumber\\
    g_3(L,a,e,M) &=& (2a^3e(1-e^2)-a^2(11-3e^2)L\nonumber\\&-&7L^3-4LM^2+2ae(3L^2+2M^2) ) M^2\,\, ,\nonumber\\
    g_2(L,a,e,M) &=& 3a^4L(1-e^2)-2ea^3M^2(5-e^2)-30aeL^2M^2\nonumber\\&+&16L^3M^2+a^2L(3L^2+2M^2(5+6e^2))\,\, ,\nonumber\\
    g_1(L,a,e,M) &=& -a^2(ae-L) (-4a^2(1-e^2)+7aeL-11L^2)\nonumber\\&+&12M^2(ae-L)^3\,\, ,\nonumber\\
    g_0(L,a,e,M) &=& -10a^2 (ae-L)^3\,\, .\nonumber
\end{eqnarray}
It is not very easy to get an analytical solution to the above fifth-order polynomial equation (Eq.~(\ref{polynomial})). Therefore, one can use the numerical technique and get five solutions of $p$. One can verify that among those five solutions, only one solution has a real and positive value. Now, the following expression of $m$ can be obtain from the first order term of $\epsilon$, 

\begin{multline}
    m^2 = \dfrac{-1}{M^4 (2ae+L(p-2))^4 p^3} \big(f_7(L,a,e,M) p^7 + f_6(L,a,e,M) p^6 + f_5(L,a,e,M) p^5 \\+ f_4(L,a,e,M) p^4
    + f_3(L,a,e,M) p^3 + f_2(L,a,e,M) p^2 + f_1(L,a,e,M) p + f_0(L,a,e,M)\big)\,\, ,
    \label{msquare}
\end{multline}
where,

\begin{eqnarray}
 f_7(L,a,e,M) &=&  M^4 (-L^4 - 8 a e^3 L M^2 + 3 a^2 (-1 + e^2) (L^2 + 4 e^2 M^2)),\\
   f_6(L,a,e,M) &=& -2 M^4 (6 a^3 e (-1 + e^2) L - 7 L^4 + 4 a e L (L^2 - 2 e^2 M^2) \nonumber\\&+& a^2 (3 (-4 + e^2) L^2 + 4 e^2 (-3 + 2 e^2) M^2)),\\
  f_5(L,a,e,M) &=& -6 L^2 M^2 (-3 a^4 (-1 + e^2) - 16 a e L M^2 + 12 L^2 M^2 \nonumber\\&+& 2 a^2 (L^2 + 2 (3 + e^2) M^2)),\\
   f_4(L,a,e,M) &=& 4 L M^2 (6 a^5 e (-1 + e^2) + 3 a^4 (7 - 2 e^2) L - 96 a e L^2 M^2 + 44 L^3 M^2 \nonumber \\&-& 8 a^3 e (3 L^2 + (3 + e^2) M^2)
   + 12 a^2 L (2 L^2 + (2 + 5 e^2) M^2)),\\
f_3(L,a,e,M) &=& 15 a^6 (-1 + e^2) L^2 - 15 a^4 L^4 + 12 a^2 (a e - L) (a^3 e (-1 + e^2) \nonumber\\&-& a^2 (-11 + e^2) L
 - 24 a e L^2 + 24 L^3) M^2 - 16 (-a e + L)^2 (a^2 (3 + e^2)\nonumber\\ &-& 14 a e L + 13 L^2) M^4,\\
  f_2(L,a,e,M) &=& 6 a^4 (a e - L) L (6 a^2 (-1 + e^2) + 7 a e L - 13 L^2)\nonumber \\&+& 24 a^2 (-a e + L)^2 (3 a^2- 16 a e L + 16 L^2) M^2
  + 96 (-a e + L)^4 M^4,\\
   f_1(L,a,e,M) &=& 8 a^4 (-a e + L)^2 (3 a^2 (-1 + e^2) + 14 a e L - 17 L^2)\nonumber\\ &-& 192 a^2 (-a e + L)^4 M^2,\\
  f_0(L,a,e,M) &=& 80 a^4 (-a e + L)^4.
\end{eqnarray}

The numerical solutions of $p$ is substituted in the expression of $m$ given in Eq.~(\ref{msquare}) and get the numerical values of $m$. As mentioned before, $m>1$ implies negative precession of the timelike orbits in Kerr spacetime. Therefore, it is verified whether any parameter space regions exist where $m$ is greater than one. To get a numerical solution of $p$, the specific physically realistic parameter spaces for the parameters $a, L$ and $e$ are considered. In this numerical analysis, $M=1$ is considered, while $a$ varies from $-40$ to $40$, $L$ varies from $3.8$ to $35$ and $e$ varies from $0.895$ to $0.999$. The spin parameters $a<-40$ and $a>40$ are not physically realistic as they may represent somewhat extreme spin situations. When specific energy $e<0.8$, solutions for bound stable orbits become hard to find, specific energy cannot have values beyond one since the particle's total energy becomes positive, which implies unbound orbits. Approximate solution mentioned in Eq.~(\ref{18}) is not a good approximation for the orbits of high eccentricity. For $L$ less than $3.8$, the orbits' eccentricity becomes very high; therefore, using this approximation method, one cannot do numerical analysis in $L<3.8$ region.

\begin{figure*}
\centering
\subfigure[$M=1$, $~3.8\leq L\leq 5$, $~0.895\leq e\leq 0.999$]
{\includegraphics[scale=0.37]{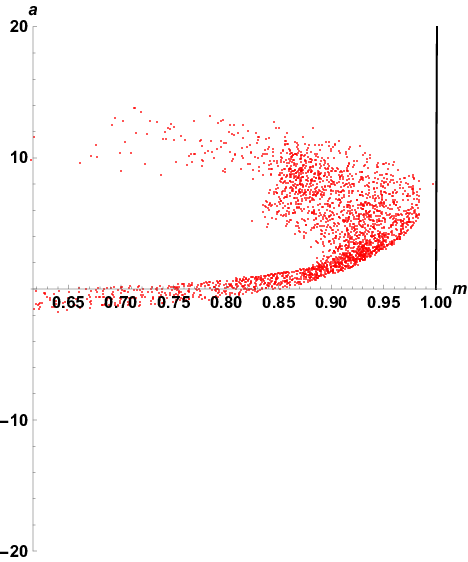}\label{re11}}
\hspace{0.1cm}
\subfigure[$M=1$, $~5\leq L\leq 10$, $~0.895\leq e\leq 0.999$]
{\includegraphics[scale=0.37]{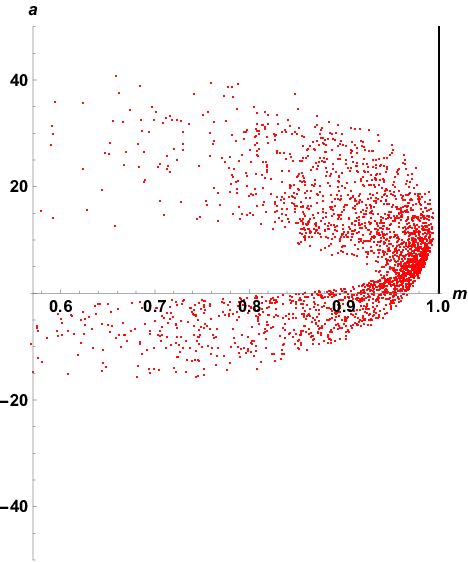}\label{re22}}
\hspace{0.1cm}
\subfigure[$M=1$, $~10\leq L\leq 20$, $~0.895\leq e\leq 0.999$]
{\includegraphics[scale=0.37]{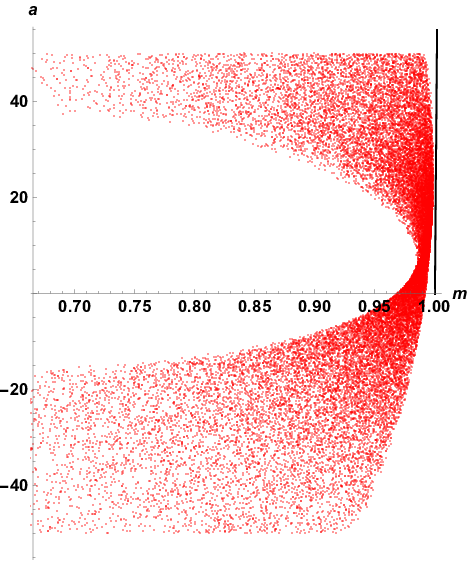}\label{re3}}
\hspace{0.1cm}
\subfigure[$M=1$, $~20\leq L\leq 35$, $~0.895\leq e\leq 0.999$]
{\includegraphics[scale=0.37]{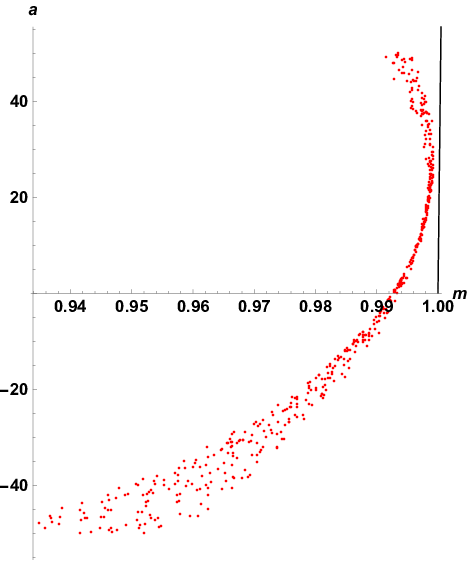}\label{re4}}
 \caption{The solution points for the timelike bound orbits in $m$ and $a$ coordinates.}
 \label{mvsa}
\end{figure*}

In Fig.~(\ref{mvsa}), the solution points for which bound stable orbits are possible are shown. Those points are given in the $m$ and $a$ coordinates, where $m$ is a function of $a$, $L$ and $e$. Therefore, for a particular value of $a$, one can get many values for $m$ since, at every point, the other two variables ($L$ and $e$) vary. The ranges of angular momentum $3.8\leq L\leq 35$ are separated into four sections to get more data points in every section. From Figs.~(\ref{re1}, \ref{re2}, \ref{re3}, \ref{re4}), one can see that all the solution points are inside the region $0\leq m\leq 1$ which indicates that the precession of orbits in Kerr spacetime is always positive for the given classes. One can see from Fig.~(\ref{mvsa}) that for any absolute value of the spin parameter, the positive perihelion shift becomes larger when the sign of the spin changes from positive to negative. This phenomenon also can be seen from Figs.~(\ref{orbit1},\ref{orbit2}). Therefore, if negative precession is not possible for any positive value of the spin parameter, it would not be possible for any negative value of the spin parameter either. One can consider larger interval of $a$ to verify whether negative precession of orbits is possible for larger values of $a$ (i.e. $a>50$), or smaller values of $a$ (i.e. $a<-50$). However, it can be verified from Fig.~(\ref{mvsa}) that the solution points are very close to $m=1$ in the interval $0\leq a\leq 30$, and after that range, those points are diverging away from $m=1$. Therefore, from the above analysis, it can be stated that the negative precession is forbidden under the given approximation in the Kerr spacetime. 

In this section, all the analysis have been done considering only the low eccentricity approximation. It is shown that the negative precession of the orbits of a test particle is not possible in Kerr spacetime \cite{Bambhaniya:2020zno}. Any weak field approximation have been not considered for this numerical analysis.
Since the eccentricity of an orbit is not directly related to the perihelion distance, a highly eccentric orbit can be far from the center. In contrast, a low eccentric orbit can be very close to the center. Therefore, in this analysis, the orbit of the test particle can be very close or far away from the singularity.

One can consider weak field approximation along with the small eccentricity approximation to get analytical solutions of $p$ and $m$. From that expression of $m$, one can get the expression of $m$ in the Schwarzschild limit (i.e. $a\rightarrow 0$). In weak field approximation, one can neglect third and higher order powers of $u(\phi)$ in the expression of the orbit equation (Eq.~(\ref{orbiteq1})). The approximate orbit equation up to the second order power of $ u(\phi)$ is given by,

\begin{multline}
    \frac{d^2u}{d\phi^2} = \frac{1}{L^5}[M L^2(L + 2a e (1-e^2)) - L (L^4 + 8a Le^3M^2 + 3a^2(1-e^2)(L^2 + 4e^2M^2))u\\ + 3M (L^2(L^3+3a^2Le^2 
    + 6a^3e (1-e^2)) - 4a e^2(2e L^2 + 3a L(1-2e^2)\\-4a^2e(1-e^2))M^2)u^2].
    \label{weak}
\end{multline}
Substituting the expression of $u(\phi)$ (Eq.~(\ref{18})) in the above orbit equation (\ref{weak}), the following quadratic equation of $p$ from the coefficient of zeroth order power of eccentricity ($\epsilon$) can be derived as,
\begin{eqnarray}
\tilde{g}_2(L,a,e,M) p^2 +\tilde{g}_1(L,a,e,M) p + \tilde{g}_0(L,a,e,M) = 0\,\, ,\nonumber\\ 
    \label{polynomial2}
\end{eqnarray}
where,
\begin{eqnarray}
\tilde{g}_2(L,a,e,M)&=& 2a L^2 M^2 e (e^2 - 1) - L^3 M^2,\,\\
\tilde{g}_1(L,a,e,M)&=& L\left[ L^4 + 8a L e^3 M^2 - 3a^2 (e^2 - 1)(L^2 + 4e^2 M^2)\right],\,\\
\tilde{g}_0(L,a,e,M)&=&[-3L^2(L^3 + 3L e^2 a^2 
    +6e a^3(1 - e^2)) \nonumber\\&+& 12a e^2 M^2(2e L^2 + 3a L (1 - 2e^2) - 4e a^2(1-e^2))]\,\, .
\end{eqnarray}
Now, the following expression of $m$ from the first-order term of eccentricity ($\epsilon$) can be obtain as, 
\begin{multline}
  m^2 = \frac{1}{p L^5}[96a^3 e^3 M^2 (e^2 - 1) + L^5 (p - 6) + 3a^2 L^3 (p - e^2 (p + 6))
\\+ 4a e L^2 (9a^2 (e^2 - 1) + 2e^2 M^2 (p + 6))
- 12L a^2 e^2 M^2 (-p - 6 + e^2 (p + 12))]\,\, .
\end{multline}
One can solve the above quadratic equation analytically and get the following two roots of $p$,
\begin{multline}
    p_1 = \frac{1}{2L^2 M^2 (L + 2a e (1 - e^2))}[L^5 + 3a^2 L^3 (1 - e^2) + 8a L^2 M^2 e^3 + 12L a^2 e^2 M^2 (1 - e^2) 
    \\
    + \{L^2 (-12M^2 (L + 2a e (1 - e^2)) 
    (L^2 (L^3 + 3L a^2 e^2 + 6e a^3 (1 - e^2)) - 4a e^2 (2e L^2 + 3a L(1 - 2e^2)
    \\
    - 4e a^2(1 - e^2)) M^2) + (L^4 + 8a L e^3 M^2 + 3a^2 (1 - e^2)(L^2 + 4e^2 M^2))^2) \}^{1/2}],\\\\
     p_2 = \frac{1}{2L^2 M^2 (L + 2a e (1 - e^2))} [L^5 + 3a^2 L^3 (1 - e^2) + 8a L^2 M^2 e^3 + 12L a^2 e^2 M^2 (1 - e^2)
    \\
    - \{L^2 (-12M^2 (L + 2a e (1 - e^2)) (L^2 (L^3 + 3L a^2 e^2 + 6e a^3 (1 - e^2)) - 4a e^2 (2e L^2 + 3a L(1 - 2e^2)
    \\
    - 4e a^2 (1 - e^2)) M^2) + (L^4 + 8a L e^3 M^2 + 3a^2 (1 - e^2)(L^2 + 4e^2 M^2))^2) \}^{1/2}].
\end{multline}
Now, it can be verified that for $p=p_2$, $m$ becomes imaginary and therefore, the real solution of the quadratic equation of $p$ is $p=p_1$. The following expression of $m$ can be determined after substituting $p=p_1$ in the expression of $m$,
\begin{equation}
m = (1 + f(a,M,L,e))^{1/4}\,\, ,
\label{m}
\end{equation}
where,
\begin{multline}
    f(a,M,L,e) = \frac{1}{L^8}(-12M^2 L^6 + 8a e M^2 L^3 (L^2 (5e^2 - 3) + 12e^2 M^2) \\+ 24 L e a^3 M^2 (e^2 - 1)(L^2 (e^2 + 3)
    + 4e^2 M^2(4e^2 - 1)) - 3a^4 (e^2 - 1)^2 (-3L^4 + 24L^2 e^2 M^2 + 80e^4 M^4)\\ - 2a^2 L^2(3L^4 (e^2 - 1)
    + 6L^2 e^2 M^2(1 + 2e^2) + 8e^2 M^4(8e^4 + 6e^2 - 9)))\,\, .
    \label{f}
\end{multline}
In the above expressions of $m$, $f>0$ implies $m>1$, which is the necessary condition for the negative precession and $f<0$ implies $m<1$, which is the necessary condition for the positive precession. As $m$ must be a real and positive number,  $f\ge -1$. Therefore, for the positive precession, one must have satisfied $-1<f<0$. In the Schwarzschild limit or with the approximation of a negligibly small value of $a$, the expression of $m$ written above reduces to,
\begin{equation}
m = \left(1 -\dfrac{12 M^2}{L^2}\right)^{1/4}\,\, ,
\label{SCHm}
\end{equation}
where $f=\dfrac{-12 M^2}{L^2}$.
 Therefore, to satisfy the condition $-1<f<0$, one must need,
\begin{equation}
\dfrac{L}{M}>\sqrt{12}\,\, .
\label{ratioLM}
\end{equation}
From the expression of $m$ in Eq.~(\ref{SCHm}), one can verify that in Schwarzschild spacetime, $m$ is always less than one, which implies positive precession of timelike bound orbits \cite{parth1}. If $L>>M$ is considered then one can write down an approximate expression of $m$ for the Schwarzschild spacetime,
\begin{equation}
m = \left(1 -\dfrac{3 M^2}{L^2}\right)\,\, .
\end{equation}
From the above expression of $m$, one can get the following positive perihelion shift of timelike bound orbits in Schwarzschild spacetime \cite{Hartle:2021pel},  
\begin{equation}
\delta\phi_{prec}=\frac{6\pi M^2}{L^2}\,\, .
\end{equation}
In Kerr spacetime, the minimum value of $\frac{L}{M} $ can be obtained for bound orbits,
\begin{equation}
    \dfrac{L}{M} \approx \sqrt{12}-\dfrac{a}{M}\bigg(e+\dfrac{a}{\sqrt{12}M} \bigg)\,\, ,
    \label{KerrLMratio}
\end{equation}
where the second order power of $a$ is considered.
\begin{figure*}
\centering
\subfigure[$~M=1$, $L=6$]
{\includegraphics[scale=0.38]{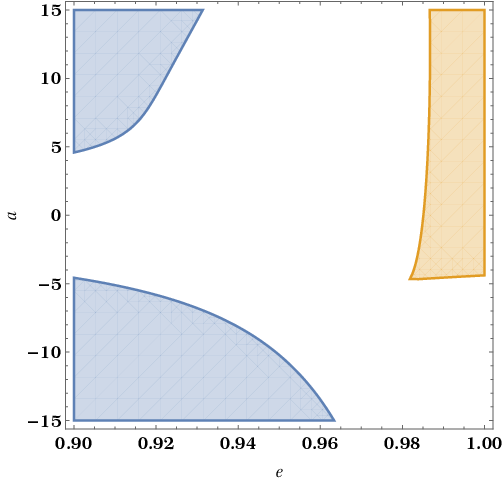}\label{figh6}}
\hspace{0.1cm}
\subfigure[$~M=1$, $L=12$]
{\includegraphics[scale=0.38]{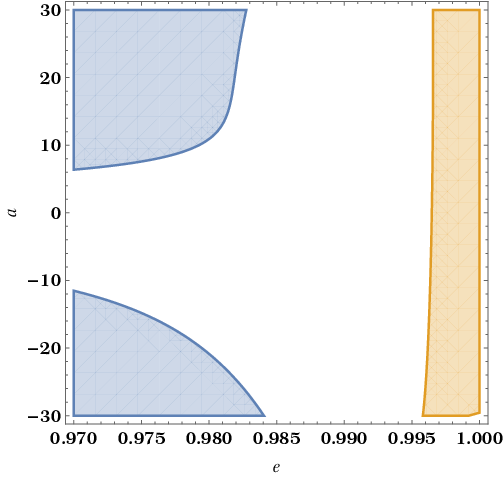}\label{figh12}}
	\caption{The region plots of $f>0$ and $V_{min}\leq E<0$ for M=1 and $L=6$~(left) and $L=12$ (right). }
	\label{fig:3}
\end{figure*}
Previously, it is shown that the negative precession does not occur in Kerr spacetime, where weak field approximation is not considered. Now, with the weak field approximation, it should be evident that one can get the same result. To verify that, in Fig.~(\ref{fig:3}), the two region plots for fixed mass (M) and angular momentum (L) are shown, where the regions of the negative precession $f>0$ (i.e. the blue region) and the bound orbits $V_{min}\leq E<0$ (i.e. the orange region) are shown for different values of $a$ and $e$. The existence of any common region between these two regions implies that in Kerr spacetime, negative precession of bound timelike orbit is possible. However, such overlapping regions are not found. Therefore, the same result is verified previously without considering any weak field approximation. 

\section{Timelike geodesics in rotating JNW spacetime}
\label{Sec_Orbit}
 In this section, an effective potential and orbit equation are derived in the rotating JNW spacetime. Solving the orbit equation numerically, the nature of the orbital precession is determined. The general form of the spacetime metric, i.e.,
 \begin{equation}
     ds^2 = -g_{tt} dt^2 + g_{rr} dr^2 - 2g_{t\phi} dtd\phi + g_{\theta\theta} d\theta^2 + g_{\phi\phi} d\phi^2,
     \label{general rotating metric}
 \end{equation}
 which is written in the Boyer-Lindquist Coordinates (BLC). For the above spacetime metric, the effective potential takes the form
\begin{equation}
    V_{eff}=\frac{1}{2 g_{rr}} + \frac{e^2-1}{2} + \frac{L^2 g_{tt}+2eL g_{t\phi}-e^2 g_{\phi\phi}}{2g_{rr}(g_{t\phi}^2+g_{tt}g_{\phi\phi})},
\label{Veff general}
\end{equation}
Substituting the metric components and taking $\theta=\frac{\pi}{2}$ into eq. (\ref{Veff general}), an effective potential can be obtain for the rotating JNW spacetime,
 \begin{eqnarray}
     V_{eff} = \frac{(L-ae)^2}{2r^2} \left(1-\frac{2M}{r\nu} \right)^{2\nu-1} - \frac{1}{2} + \frac{\Delta+2ae\left(L-ae \right)}{2r^2} \left(1-\frac{2M}{r\nu} \right)^{\nu-1}.
     \label{Veff rotating JNW}
 \end{eqnarray}
For the general rotating spacetime metric (\ref{general rotating metric}), orbit equation can be obtained as \cite{solanki},
\begin{eqnarray}
\left(\frac{dr}{d\phi}\right)^2=-\frac{(g_{t\phi})^2+g_{tt}g_{\phi\phi}}{g_{rr}(e g_{t\phi} + L g_{tt})^2} [(g_{t\phi})^2-e^2 g_{\phi\phi}+2eL g_{t\phi}+g_{tt}&(L^2+g_{\phi\phi})] \,\,\,
\label{general orbit}
\end{eqnarray}
 where, $\theta=\frac{\pi}{2}$. Now, using the differential relation,
 \begin{equation}
     \frac{d^2r}{d\phi^2}=\frac{1}{2\frac{dr}{d\phi}}\frac{d}{d\phi}\left(\frac{dr}{d\phi}\right)^2,
 \end{equation}
the orbit equation can be derived as,
\begin{equation}
    \frac{d^2u(\phi)}{d\phi^2} = -\frac{1}{r^2} \frac{d^2r}{d\phi^2} + \frac{2}{r^3} \left(\frac{dr}{d\phi}\right)^2,
    \label{orbiteq}
\end{equation} 
 where, $u=\frac{1}{r}$. The above eq. (\ref{orbiteq}) is the non-linear second-order differential equation. Thus it is tough to find the complete analytic solution to this equation. Hence, one can solve this orbit equation numerically for the rotating JNW metric. 

\begin{figure*}
\centering
\subfigure[Bound orbit for $q =0.4$ and $a = 0.3$.]
{\includegraphics[width=60mm]{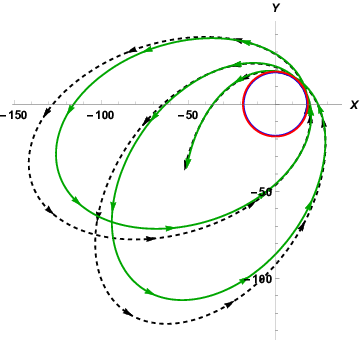}\label{orbit11}}
\hspace{0.2cm}
\subfigure[Bound orbit for $q =0.4$ and $a = 0.5$.]
{\includegraphics[width=60mm]{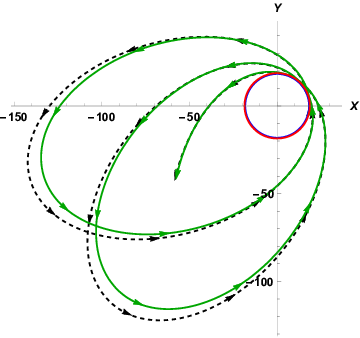}\label{orbit12}}
\subfigure[Bound orbit for $q =12$ and $a = 0.3$.]
{\includegraphics[width=60mm]{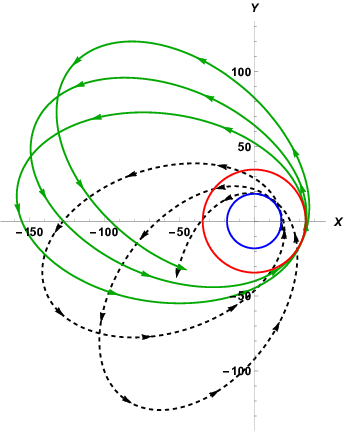}\label{orbit21}}
\subfigure[Bound orbit for $q =12$ and $a = 0.5$.]
{\includegraphics[width=60mm]{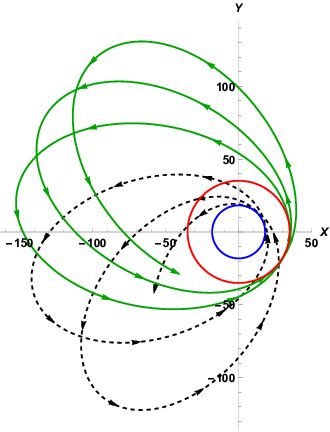}\label{orbit22}}

 \caption{The timelike bound orbits in the rotating JNW naked singularity spacetime.}
\label{figorbit}
\end{figure*}

In figure (\ref{figorbit}), the timelike bound orbits in rotating JNW spacetime are shown and compared with the orbits in Kerr spacetime. In fig. (\ref{orbit11}),(\ref{orbit12}), (\ref{orbit21}) and (\ref{orbit22}), the bound orbits for scalar field charge $q=0.4, 12$ are shown corresponding to the spin parameter $a=0.3, 0.5$. The green and black dotted lines represent the orbits of the particles in the rotating JNW and Kerr spacetimes, respectively. The red and blue circles denote the periastron points (minimum approach towards the center) for rotating JNW and Kerr spacetime, respectively. The mass (ADM mass for JNW) $M=1$, conserved angular momentum $L=6$, total energy $E=-0.006$ are taken.

One can see from the figure (\ref{figorbit}) that the timelike bound orbital motion in the JNW spacetime is significantly different from the orbital motion in Kerr spacetime. The timelike bound orbits can have negative (or opposite) precession in rotating JNW spacetime. In other words, if the angular distance travelled by a particle between two successive periastron points is less than $2\pi$, then the next orbit would shift in the opposite direction of the particle's orbiting direction. Hence, it is known as negative (or opposite) precession. On the other hand, in \cite{Bambhaniya:2020zno}, it is shown that the negative precession is not present in the Kerr spacetime. The Kerr spacetime exhibits positive precession since the angular distance travelled by a particle between two successive periastron points is always greater than $2\pi$.

\section{Discussion and Conclusion}\label{four}
In this work, the nature of timelike bound orbits of a test particle have been studied in Kerr and rotating JNW spacetimes. In order to find the particle trajectories in those spacetimes, an orbit equation is derived and solve the same numerically. In Fig.~(\ref{orbit1}), it is shown that the trajectories of a test particle in Kerr spacetime for the spin parameter values $a=\pm 0.8, \pm 1, \pm 1.1$, where black hole mass $M=1$, specific angular momentum of the test particle $L=12$ and the total energy of the test particle $E=-0.001$ are taken. In Fig.~(\ref{orbit2}), the particle trajectories are shown for the same spin parameter values where $M=1, L=6$ and $ E=-0.006$. In both cases, positive precession of the bound orbits is identified. 

Then in section (\ref{three}), an approximation solution of the orbit equation (Eq.~(\ref{orbiteq1})) is considered in order to understand the nature of perihelion shift of timelike bound orbits in Kerr spacetime. In that approximation, a small eccentricity values ($\epsilon$) are considered, and therefore, the second and higher-order terms of $\epsilon$ are neglected. With this approximation, it is shown that the negative precession of the timelike bound orbits is forbidden in Kerr spacetime, no matter how far or close the orbit is from the center.

Finally, using a weak field approximation, shown that the Kerr spacetime solutions reduce to the solution of Schwarzschild with the approximation $a\rightarrow 0$. It is shown that negative precession of timelike orbits is not possible in Kerr and Schwarzschild spacetimes, respectively. In the previous chapter, it is showed that naked singularity models, such as JMN-1 and JNW spacetimes, admit both negative and positive precession of timelike orbits. Moreover, the precession of timelike bound orbits in rotating JNW spacetime have been examined. It is found that the rotating JNW spacetime can have positive and negative (or opposite) precession of the timelike bound orbits. Hence, any evidence of negative precession of any `S' star can raise a big question on the existence of Kerr black hole at the Milky Way galactic center.  

Of course, this analysis cannot scanned the entire space of bound orbits around the Kerr black hole or the Kerr naked singularity. However, clearly points out that in the classes of bound orbits are analysed, using the approximation and numerical techniques are stated, there have been not found any negative precession for the bound orbits in both of these cases. In particular, the case of the Kerr black hole needs to be analyzed in more detail to ensure it forbids the negative precession for the bound timelike trajectories. If that turns out to be the case, that will support the conjecture that the vacuum solutions of black holes never allow for negative precession. However, naked singularities allow the same as shown by some naked singularity spacetimes investigated, as pointed out here.  

\chapter{Relativistic orbits of S-stars}
It is a general belief that the spacetime of the central supermassive object (Sgr A*) is a vacuum solution of Einstein field equations (e.g., Schwarzschild and Kerr solutions). However, from the observational results, it can be understood that the center of a galaxy is surrounded by highly concentrated matter. Therefore, a vacuum solution is unlikely to be present around a galactic center. Hence, one could consider a matter distribution near the central region and investigate the corresponding physical signature. 

In this chapter, the scalar field effect on the orbital dynamics of the `S2' star is investigated. A scalar field is the simplest constitute of matter, and at the beginning, one may model the matter distribution around the galactic center using the scalar field. This is not related to anything about the particle physics model of the scalar field. It may be interpreted by some beyond the standard model (BSM) of particle physics, which is not the scope of the present study. We are interested in constraining the scalar charge by best-fitting the astrometric data of `S2' with the theoretical prediction using the MCMC technique. In order to do that, the JNW spacetime which is the minimally coupled, mass-less scalar field solution of Einstein equations is considered. The JNW spacetime possesses a naked singularity at the center. One can retain the Schwarzschild spacetime from the JNW spacetime by considering zero scalar charge. However, a small but non-zero scalar field charge can drastically change the causal structure of the Schwarzschild spacetime. The best-fitted value of the scalar charge comes out to be non-zero, which may be interpreted as the existence of non-vacuum spacetime seeded by a scalar field around the galactic center.

\section{S-stars near Milky Way galactic center}
Near the center of the Milky Way galaxy, many stars are hovering around with very high speeds. Due to the presence of the massive central object of mass around $4\times 10^6 M_{\odot}$, these stars can move at $\frac{1}{60}$ the speed of light and they can have highly eccentric orbits. These stars are known as `S'-stars. Since they are very close to the galactic center, there exists a possibility that they can show up some general relativistic effects. However, it is challenging to follow the dynamics of those stars since they are far from us ($\approx 25,000$ light years). The highly sensitive infra-red instruments, namely GRAVITY, SINFONI, and NACO in the European Southern Observatory (ESO), are capable of tracking the trajectory of the `S'-stars. They recently released 23 years of astrometric data of the `S2' star \cite{Do:2019txf,GRAVITY:2018ofz,GRAVITY:2020gka}, which is one of the important star of the `S' star family. As it is mentioned above, general relativistic effects can be seen in the dynamics of the `S2' star. Therefore, its trajectory can give us information about the spacetime around the Sgr A*.
\section{Orbital parameters of the real and apparent orbits}

In the forth chapter, the properties of timelike orbits in the JNW spacetime are discussed and compared with the timelike orbits in Schwarzschild spacetime. The vital difference which is coming out from the analysis is that the perihelion precession of bound timelike orbits in the JNW spacetime can be negative, i.e., the direction of particle motion is opposite to the direction of precession. This unique characteristic of the timelike orbit is forbidden in Schwarzschild spacetime. In \cite{Igata:2022rcm,Igata:2022nkt}, it is shown that the presence of matter may be responsible for the negative precession. 
Therefore, negative precession is very important in the context of the trajectories of `S' stars around the Sgr A*. To predict the possible trajectory of the `S2' star, JNW orbit equation is used. Furthermore, using the astrometric data of that star, the parameters' space of the JNW spacetime can be constrained.

\begin{figure}
    \centering
    \includegraphics[width=12cm]{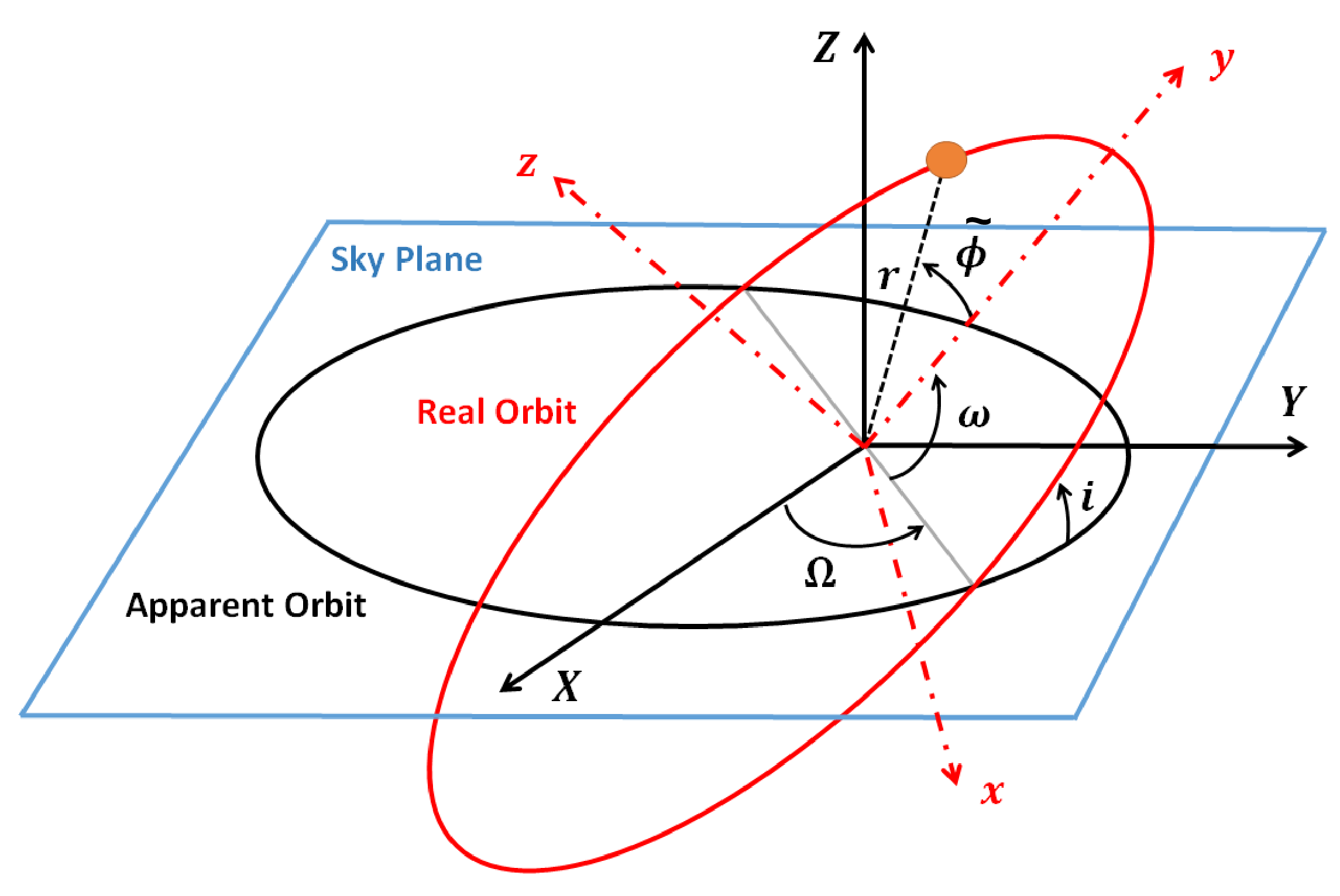}
    \caption{Real orbit projection into the sky plane \cite{Becerra-Vergara:2020xoj}.}
    \label{orbitproject}
\end{figure}
\begin{figure}
    \centering
    \includegraphics[width=13cm]{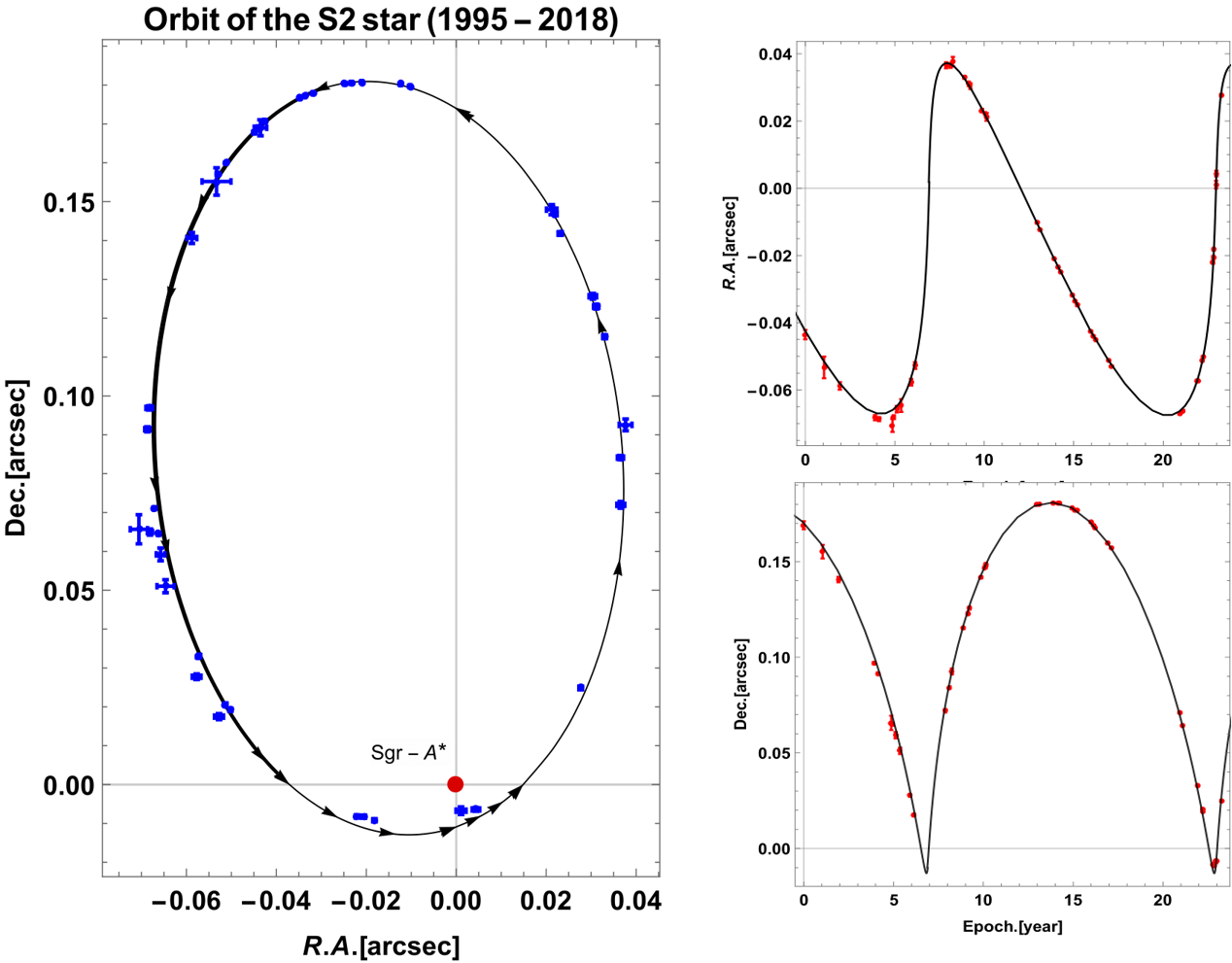}
    \caption{Left: The best fitting orbit of the JNW model (black) along with the observed astrometric position (Blue) of the S2 star from 1995 to 2018. The red dot shows the position of the Sgr A* at (0,0). Right: R.A. (right ascension $\alpha$ on top) and Dec. (declination $\delta$ on bottom) offset of S2 star with the orbital period.}
    \label{fig:orbit_JNW}
\end{figure}

It is necessary to transform the real orbit into the apparent orbit since the observed astrometric data is given onto the plane of the sky. The position and velocity components of the real orbit are defined in Cartesian coordinates as $x, y, z$ and $v_x, v_y, v_z$, respectively. As in this case, $\theta=\pi/2$, the position and velocity of the real orbit can be obtained by the transformation from spherical coordinates to Cartesian coordinates as
\begin{equation}
    x=r\cos \phi, \; \; \;
    y=r\sin\phi, \; \; \;
    z=0,
    \label{position1}
\end{equation}
and the corresponding three velocities transform as,
\begin{equation}
    v_x=v_r\cos \phi-rv_{\phi}\sin\phi, \; \; \;
    v_y=v_r\sin\phi+rv_{\phi}\cos\phi, \; \; \;
    v_z=0,
    \label{velocity1}
\end{equation}
where, $v_r=dr/dt$ and $v_{\phi}=d\phi/dt$ are three velocity components and the corresponding four velocity can be written as $u^r=v_r u^0$ and $u^{\phi}=v_{\phi} u^0$. Now, to fit the real orbit with the astrometric observational data, one must project the real orbit on the apparent plane of the sky, as shown in fig. (\ref{orbitproject}). In that figure, the axes begin at Sgr A* (the focus of the ellipse) and which depicts the orbital parameters: $\tilde{\phi}=\phi-\pi/2$, where $\phi$ is the azimuth angle of the spherical coordinate system associated with the $x, y, z$ Cartesian coordinates, i.e. the true anomaly for an elliptic motion in the $x-y$ plane, $i$ is the angle of inclination between the real orbit and the sky plane, $\Omega$ is the ascending node angle, and $\omega$ is the pericenter argument. It is worth mentioning that the vector going from the solar system to the galactic centre defines the coordinate system's Z-axis. The observed astrometric positions $X_{obs}$ and $Y_{obs}$ of the star in Cartesian coordinates are defined by the observed angular positions, right ascension $(\alpha)$ and declination $(\delta)$ \cite{Becerra-Vergara:2020xoj}.

\begin{equation}
    X_{obs}=r_d (\alpha-\alpha_{SgrA^*}),\; \; \;
    Y_{obs}=r_d (\delta-\delta_{SgrA^*}),\; \; 
\end{equation}
where $r_d$ is the distance between the Sgr A* and the earth. Note that the center of the coordinate system represents the position of Sgr A*. The positions $X, Y, Z$ of the apparent orbit can be obtained from the real orbit positions $x$ and $y$ by using classic Thiele-Innes constants with the same notation given in \cite{Becerra-Vergara:2020xoj} as
\begin{equation}
    X=xB+yG, \; \; \;
    Y=xA+yF, \; \; \;
    Z=xC+yH, \; \;
    \label{position2}
\end{equation}
 and the corresponding velocity components of the apparent orbit are,
 \begin{equation}
    V_X=v_xB+v_yG, \; \; \;
    V_Y=v_xA+v_yF, \; \; \;
    V_Z=v_xC+v_yH, \; \;
    \label{velocity2}
\end{equation}
where,
\begin{equation}
    A=\cos\Omega\cos\omega-\sin\Omega\sin\omega\cos i,
\end{equation}
\begin{equation}
    B=\sin\Omega\cos\omega+\cos\Omega\sin\omega\cos i,
\end{equation}
\begin{equation}
    C=\sin\omega\sin i,
\end{equation}
\begin{equation}
    F=-\cos\Omega\sin\omega-\sin\Omega\cos\omega\cos i,
\end{equation}
\begin{equation}
    G=-\sin\Omega\sin\omega+\cos\Omega\cos\omega\cos i,
\end{equation}
\begin{equation}
    H=\cos\omega\sin i,
\end{equation}
where the osculating orbital elements $\Omega$, $i$, and $\omega$ are the ascending node angle, inclination angle, and the argument of pericenter, respectively. The fully general relativistic orbit equation of the JNW spacetime is given as
\begin{eqnarray}
\frac{d^2u}{d\phi^2}+u-\frac{3bu^2}{2}+\frac{e^2b}{c^2L^2}(1-n)(1-bu)^{1-2n}-\frac{c^2b}{2L^2}(2-n)(1-bu)^{1-n}=0\,\, ,
\label{orbitgen}
\end{eqnarray}
and the solution of this above equation gives the $r(\phi)$, which one has to transform into Cartesian coordinates using the relation given in (\ref{position1}). The apparent orbital coordinates $X, Y$ can be obtained from the real orbital positions $x, y$ using the transformation (\ref{position2}). Now, one can fit the observed apparent orbital data with the apparent orbital plane, which will give us the actual nature of the orbital shape as shown in fig. (\ref{fig:orbit_JNW}). The best fitting parameters are estimated for the JNW metric using the MCMC technique (see table \ref{table2}).

\section{Numerical techniques}
\begin{table}
    \centering
    \vspace{0.05cm}
    \begin{tabular}{l c c}
        \hline\hline
    &       &         \\
{\bf Parameter}      &                                 {\bf JNW (95\% limits)}   \\
\hline
\hline
\\
{\boldmath$L^2         $} $(pc^2(km/s)^2)$ &  $4.44216^{+0.00075}_{-0.00075}$ \\
\\
{\boldmath$\log{E_n}          $} $(km/s)^2$ &  $10.95422391^{+0.00000034}_{-0.00000033}$ \\
\\
{\boldmath$t_{\rm ini}          $} (year) &  $1.199^{+0.038}_{-0.040}   $ \\
\\
{\boldmath$\log{M}         $} $(M\odot)$ & $6.666^{+0.010}_{-0.012}   $ \\
\\
{\boldmath$\log{q}          $} $(M\odot)$ & $-7.46^{+0.58}_{-0.57}     $ \\
\\
{\boldmath$\theta_{\rm inc}        $} (radian) & $2.316^{+0.025}_{-0.025}   $ \\
\\
{\boldmath$\Omega         $}  (radian)& $4.017^{+0.035}_{-0.033}   $ \\
\\
{\boldmath$\omega         $} (radian) & $1.199^{+0.029}_{-0.029}   $ \\
\\
{Distance (parsec), \boldmath$r_d          $} & $8169^{+34}_{-39}          $ \\
\\
{Time period (yr), \boldmath$T         $} & $16.1379          $ \\\\
{Minimal \boldmath$\chi^2         $} & $4.71          $ \\
\hline
    \end{tabular}
    \caption{Estimated best-fit values of the parameters for the JNW metric.}
    \label{table2}
\end{table}
\begin{figure}
    \centering
    \includegraphics[width=\linewidth]{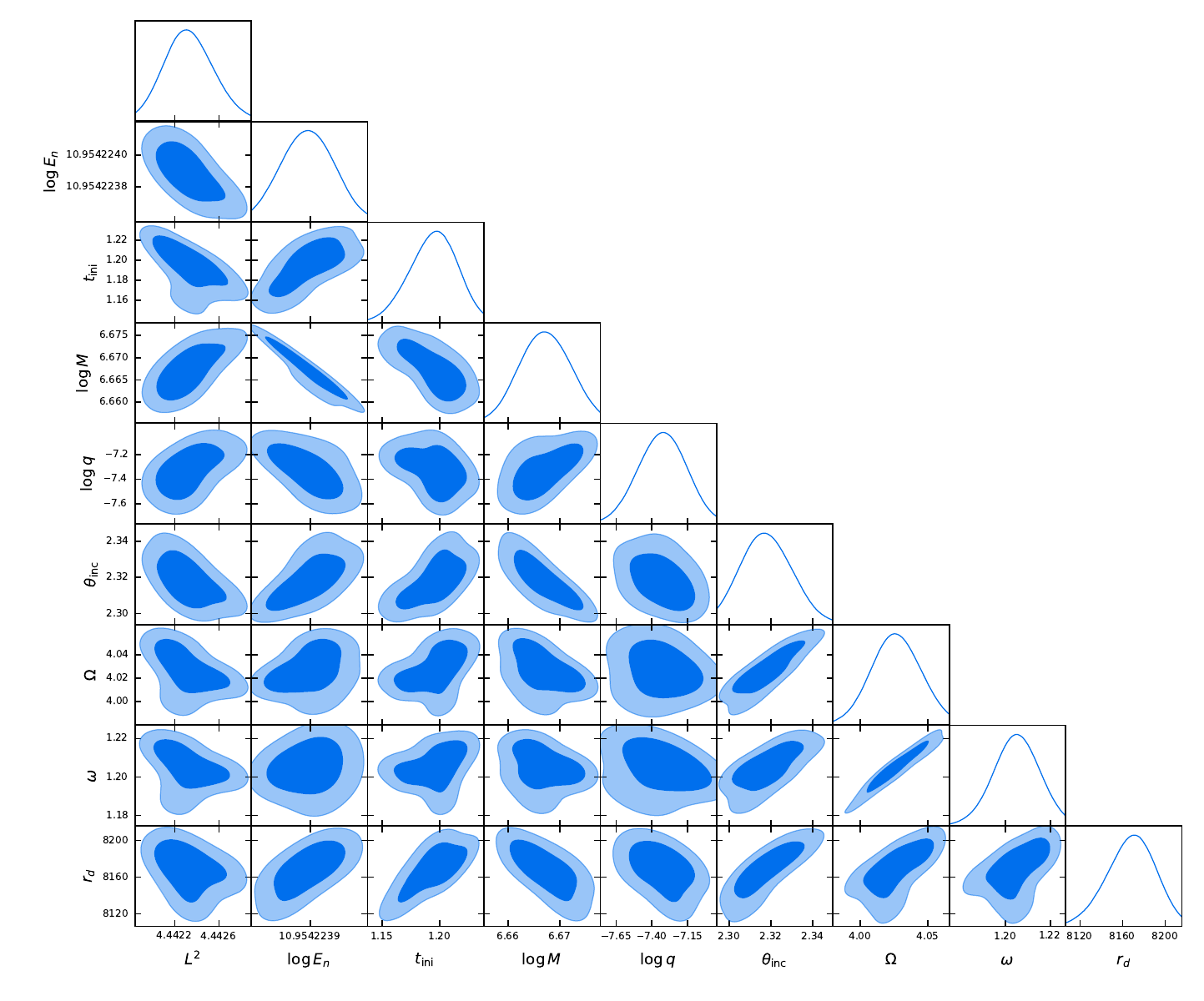}
    \caption{1-$\sigma$ and 2-$\sigma$ best-fit regions and the posterior distributions of the parameters for the JNW metric and Sgr A* derived using MCMC. The lowest $\chi^2$ value obtained is 4.71.}
    \label{fig:param_JNW}
\end{figure}
The MCMC analysis performed in the paper for the astrometric data of S2 star is based upon the Metropolis-Hastings algorithm \cite{Hastings:1970aa}. The likelihood function used in the analysis for symmetric error is as follows,
\begin{equation}
-\log {\mathcal L} \propto \sum_i \left[\left(X_i - \bar{X}_i\over \sigma^X_i\right)^2 + \left(Y_i - \bar{Y}_i\over \sigma^Y_i\right)^2\right].
\end{equation}
Here $X_i$ and $Y_i$ represent the observed coordinates of the S2 star in its orbit, and $\bar{X}_i$ and $\bar{Y}_i$ are theoretically calculated values. The errors of the observation in the $X$ and $Y$ direction are $\sigma^X$ and $\sigma^Y$ correspondingly. To
implement the MCMC analysis, the value of $q$ is always
put within the open bounded interval $(0, q_0)$, where $q_0$ is
taken to some large positive value. This means that the
present analysis excludes the possibility of $q = 0$, i.e., the
Schwarzschild spacetime, all along.

Details about the data set: 46 data points are considered here from the astrometric positions of the S2 star. The Source of the orbital data is adopted from the supplement material of the paper \cite{Do:2019txf}. For the MCMC analysis, the Gaussian priors are used. The details of the priors are given in the table~(\ref{tab2}).

\begin{table}
\centering
    \vspace{0.2cm}
    \begin{tabular}{|l | c |c|}
\hline
    &       &\\        Parameter & Mean & 1-$\sigma$ \\
     \hline
    &       &\\      {\boldmath$L^2          $} ($pc^2(km/s)^2$) & $4.44$ & $0.04$

\\

{\boldmath$\log{E_n}          $} & $10.90$ & $0.01$\\

{\boldmath$t_{\rm ini}          $} (yr) & $1.22$  & $0.01$ \\

{\boldmath$\log{M}          $} & $6.56$ & $0.05$ \\

{\boldmath$\log{q}          $} & $-8.00$ & $0.005$ \\

{\boldmath$\theta_{\rm inc}          $} (radian) & $2.30$ & $0.02$ \\

{\boldmath$\Omega          $} (radian) & $4.00$ & $0.04$ \\

{\boldmath$\omega          $} (radian) & $1.20   $ & $0.01$\\

{Distance, \boldmath$r_d          $} (parsec) & $8200$ & $10$\\
\hline
\end{tabular}
\caption{Details of the Gaussian priors of different parameters.}
\label{tab2}
\end{table}

\section{Discussion and Conclusion}
In this work, the fully relativistic orbit equation is derived for the JNW spacetime, which is a second-order non-linear differential equation.  This equation is solved numerically since it is difficult to solve analytically. To obtain the best fitting orbital parameters of the JNW model, the astrometric data of the S2 star are used from \cite{Do:2019txf}. The best fitting parameters for the JNW metric (see table \ref{table2}) is estimated using the MCMC technique under the assumption of the positive scalar charge and obtain the lowest $\chi^2$ value is 4.71 (see the figures (\ref{fig:orbit_JNW}) and (\ref{fig:param_JNW})). The nature of the Sgr A* is probed using the available observed astrometric data of the S2 star. The results show that, like Schwarzschild black hole, the JNW naked singularity could be a possible candidate for the compact object Sgr A* at Milky Way galaxy center.  

\chapter{Concluding remarks and future scopes}
The way mankind has witnessed a series of scientific
breakthroughs in astrophysics, including the detection of
gravitational waves from the merger of two black holes,
the shadow images of M87* and Milky Way galactic center
(Sgr- A*), it has drawn attention of not just scientific
community but also general public. In the universe, an ultra-dense region can be anticipated by modeling various astrophysical compact objects, such as black holes, naked singularities, worm holes and other specific types of these objects. These compact objects, together with gravitational waves and shadows, are considered one of the most efficient sources of energy in the Universe. As a result, they are assumed to be responsible for a gigantic electromagnetic environment in their near vicinity, as well
as high-energy jet emission outbursts destroying nearby stars and galaxies.
These observations by the EHT group have opened up the way for gravitational theories to be verified in strong gravity regimes.
The EHT group’s findings have been used to constrain
and study various aspects of gravity theories, extending
from general relativity to its alternatives.

\section{Concluding remarks from shadow properties}
In this thesis, the second and third chapters deal with the properties of the accretion disks and the shadows cast by various compact objects. The concluding remarks from these chapters are as follows: Although the shadow is observed in the image of Sgr A*, it cannot be conclusively commented that whether it contains a SMBH, as it is evident from the naked singularity models and a black hole. Moreover if in future observations, shadow is absent in the image, it may hint towards the presence of a naked singularity, and not a black hole. Concerning the general theory of relativity (and not the other modified gravities), these naked singularity solutions present an excellent alternative to the conventionally used black hole. The shape of the shadows can be find in rotating JNW spacetime and it becomes a prolate arc shape from the prolate contour shape as with increasing the scalar field charge $q$ and/or spin parameter ‘a’. This novel signature of the shadow shapes in rotating JNW spacetime (prolate arc) could be observationally significant to differentiate among the two presented models.

\section{Concluding remarks from relativistic orbits}
The fourth and fifth chapters are concerned with the relativistic orbits of a test particle and its periastron precession. The sixth chapter deals with the relativistic orbits of the S2 star. One need to study the timelike and null geodesics behaviour around the galactic centre to better understand its causal structure, mass, and dynamics. GRAVITY and SINFONI are actively monitoring the Milky Way galactic centre for important observations of stellar motion around the centre. Many 'S' stars (for example, S2, S102, S38, and others) orbit the central area of the Milky Way galaxy, Sgr A*. The stars' closest approach is within $0.0001-0.0006$ parsec. Since they are so close to the centre, their dynamics can provide vital information about the central object Sgr A*.

The concluding remarks from these chapters are as follows: In Schwarzschild spacetime, the bound trajectories of a massive particle always precess in the direction of the particle motion. However, in JMN-1 and JNW naked singularity spacetimes, for an allowed range of parameter values, the bound orbits precess in the opposite or reverse the direction of particle motion, presenting us a novel distinguishing feature. We show that negative precession of timelike orbits is not possible in the Schwarzschild and Kerr black hole spacetimes. Hence, any evidence of negative precession of any ‘S’ star can raise a big question on the existence of Schwarzschild and Kerr black holes (vacuum solutions) at the Milky Way galaxy center and might hint towards the existence of a naked singularity. The nature of Sgr A* is probed using the available observed astrometric data of the S2 star. The best fitting parameters for the JNW metric are estimated using the MCMC technique, and obtain the $\chi^2$ value of 4.71. Therefore, the JNW naked singularity could be a possible candidate which might represent the spacetime structure of the Sgr-A*.

\section{Future scopes}
In future, one could devote time to the problems of the Janis-Newman Algorithm (JNA). JNA is used to get the rotating generalization of static spacetimes. However, JNA is not applicable in general for any arbitrary spherically symmetric and static spacetimes. It is only applicable to those spacetimes which at least satisfy the given condition $f(r)*g(r)=-1$. The JNA can be generalised using the null tetrad formalism and time would be dedicated to the problem of rotating compact objects. Afterwards the spin of the Sgr A* can be estimate using the astrometric and spectroscopic data of the S2 star. This would be one of the major look-outs and area of research in this work.

Current observations from different telescopes and instruments are presenting us with an opportunity with unparalleled access to data-sets of various astrophysical phenomenon.
In future projects, the shadow images of the rotating JNW naked singularity (arc shape) in the visual domain can be constructed using the available data of the Sgr A* shadow image. Near the Milky Way galactic center, there are other observed data sets of the S-stars and G-cluster objects, which can be useful to precisely predict the nature of Sgr-A*. In outcome, one can constrain the models of the various compact objects by estimating best fitting parameters of the observed S-stars orbital data (Astrometric as well as spectroscopic). One can constrain the values of several parameters in given models and identify the causal structure of the Sgr-A* galactic centre by comparing the observed shadow with the theoretically generated shadow of the compact object.

 There are other astrophysical phenomena which requires further study to determine observational signatures of these compact objects. This includes tidal force effects as well. Using these effects, one can study the tidal disintegration of stars near ultra-high dense compact objects such as horizon-less compact objects (naked singularities).

For now, researchers have various theoretical predictions and results, and what they require is a conclusive observational confirmation. Investigations in these directions can open future avenues for such and many other unanswered questions and mysteries. In the coming years, we will be moving a step further in the journey of understanding our simple yet complex universe.





\label{Bibliography}
\bibliographystyle{agsm}

\clearpage
\pagestyle{plain}
\begin{center}
\textbf {\Large \textbf{Appendix A}}
\end{center}
\begin{enumerate}
\item \textbf{Energy conditions of rotating JNW spacetime:} We discuss about the energy conditions of the metric (\ref{Rotating_JNW_NJA}) obtained using the original NJA. Consider the inverse form of the spacetime metric.

\begin{eqnarray}
    \partial_s^2 = -\left(1-\frac{2Mr}{\nu\rho^2} \right)^{-\nu} \left\{1 - \frac{a^2\sin^2\theta}{\Delta} \left[1 - \left(1-\frac{2Mr}{\nu\rho^2} \right)^{\nu} \right]^2 \right\} \partial_t^2 \nonumber\\+\, \frac{1}{\rho^2} \left(1-\frac{2Mr}{\nu\rho^2} \right)^{\nu-1} \big\{\Delta\partial_r^2 + \partial_\theta^2 \big\} -\frac{2a}{\Delta} \left[1 - \left(1-\frac{2Mr}{\nu\rho^2} \right)^{\nu} \right] \partial_t \partial_\phi \nonumber\\+\, \frac{1}{\Delta\sin^2\theta} \left(1-\frac{2Mr}{\nu\rho^2} \right)^{\nu} \partial_\phi^2\nonumber.
\end{eqnarray}
It can be written as
\begin{eqnarray}
    \partial_s^2 = -\left(1-\frac{2Mr}{\nu\rho^2} \right)^{-\nu} \partial_t^2 + \frac{1}{\rho^2} \left(1-\frac{2Mr}{\nu\rho^2} \right)^{\nu-1} \big[\Delta\partial_r^2 + \partial_\theta^2 \big] \nonumber\\+\, \left(1-\frac{2Mr}{\nu\rho^2} \right)^{\nu} \bigg[\frac{a\sin\theta}{\sqrt{\Delta}} \left\{1-\left(1-\frac{2Mr}{\nu\rho^2} \right)^{-\nu} \right\} \partial_t + \frac{1}{\sqrt{\Delta}\sin\theta} \partial_\phi \bigg]^2\nonumber. \,\,\,
    \label{Inverse JNW NJA}
\end{eqnarray}
Now, we choose a set of orthonormal basis $\{u^\mu, e_r^\mu, e_\theta^\mu, e_\phi^\mu \}$, which satisfies $u^\mu u_\mu = -1$, $e_{i}^{\mu} (e_i)_\mu = 1$, $u^\mu (e_i)_\mu = 0$; where, $(i \to r,\theta,\phi)$,
\begin{eqnarray}
    && u^\mu = \left(\left(1-\frac{2Mr}{\nu\rho^2} \right)^{-\nu/2}, 0, 0, 0 \right)\nonumber, \\
    && e_r^\mu = \left(0, \frac{\sqrt{\Delta}}{\rho}\left(1-\frac{2Mr}{\nu\rho^2} \right)^{\frac{\nu-1}{2}}, 0, 0 \right)\nonumber,\\
    && e_\theta^\mu = \left(0, 0, \frac{1}{\rho}\left(1-\frac{2Mr}{\nu\rho^2} \right)^{\frac{\nu-1}{2}}, 0 \right)\nonumber, \\
    && e_\phi^\mu = \left(1-\frac{2Mr}{\nu\rho^2} \right)^{\nu/2} \left(\frac{a\sin\theta}{\sqrt{\Delta}} \left\{1-\left(1-\frac{2Mr}{\nu\rho^2} \right)^{-\nu} \right\}, 0, 0,\frac{1}{\sqrt{\Delta}\sin\theta} \right)\nonumber.
\end{eqnarray}
Using eq. (\ref{rho general}), (\ref{Pr general}), (\ref{Ptheta general}), (\ref{Pphi general}), we determine the components of the energy momentum tensor. For mathematical simplicity we take $\theta=\pi/2$.
\begin{eqnarray}
    && \rho_e = P_r + \frac{2a^2M^2(1-\nu)^2}{r^6\nu^2} \left(1-\frac{2M}{r\nu} \right)^{-3+\nu}\nonumber, \\
    && P_r = - P_\theta = \frac{M^2 (1-\nu^2)}{r^4 \nu^2} \left(1 - \frac{2M}{r\nu} \right)^{-2+\nu} \label{Pr Ptheta}\nonumber,\\
    && P_\phi = P_\theta - \frac{2a^2M^2(1-\nu^2)}{r^6\nu^2} \left(1-\frac{2M}{r\nu} \right)^{-3+\nu}\nonumber.
\end{eqnarray}
We see that
\begin{equation}
    \rho_e + P_r + P_\theta + P_\phi = \rho_e + P_\phi = -\frac{4a^2M^2(1-\nu)}{r^6\nu} \left(1-\frac{2M}{r\nu} \right)^{-3+\nu} < 0\nonumber,
\end{equation}
which implies that the spacetime metric (\ref{Rotating_JNW_NJA}) does not obey the weak energy condition, null energy condition, and the strong energy condition. We note that the dominant energy condition is also violated. Thus, the rotating spacetime metric obtained using the Newman-Janis algorithm with complexification is not physically a valid solution of the Einstein field equation.

\end{enumerate}

\clearpage
\pagestyle{plain}
\begin{center}
\textbf {\Large \textbf{List of Publications}}
\end{center}
\begin{enumerate}

\item Bambhaniya, P., Joshi, A., Dey, D. and Joshi, P. S. (2019) \lq Timelike geodesics in Naked Singularity and Black Hole Spacetimes\rq, \textit{Phys. Rev. D}, vol. 100, no. 12, pp. 124020. \url{https://doi.org/10.1103/PhysRevD.100.124020}. (JCR (2019) Impact Factor: 4.833)\par

\item Dey, D., Joshi, P. S., Joshi, A. and Bambhaniya, P. (2019) \lq Towards an observational test of black hole versus naked singularity at the galactic center\rq, \textit{International Journal of Modern Physics D}, vol. 28, no. 14, pp. 1930024. \url{https://doi.org/10.1142/S0218271819300246}. (JCR (2019) Impact Factor: 2.154)\par

\item Bambhaniya, P., Solanki, D. N., Dey, D., Joshi, A. and Joshi, P. S. (2021) \lq Precession of timelike bound orbits in Kerr spacetime\rq, \textit{Eur. Phys. J. C}, vol. 81, no. 3, pp. 205.  \url{https://doi.org/10.1140/epjc/s10052-021-08997-x}. (JCR (2021) Impact Factor: 4.590)\par

\item Bambhaniya, P., Dey, D., Joshi, A., Joshi, P. S., Solanki, D. N. and Mehta, A. (2021) \lq Shadows and negative precession in non-Kerr spacetime\rq, \textit{Phys. Rev. D}, vol. 103, no. 8, pp. 084005.  \url{https://doi.org/10.1103/PhysRevD.103.084005}. (JCR (2021) Impact Factor: 4.833)\par

\item Solanki, D. N., Bambhaniya, P., Dey, D., Joshi, P. S. and Pathak, K. N. (2022) \lq Shadows and precession of orbits in rotating Janis–Newman–Winicour spacetime\rq, \textit{Eur. Phys. J. C}, vol. 82, no. 1, pp. 77.  \url{https://doi.org/10.1140/epjc/s10052-022-10045-1}. (JCR (2022) Impact Factor: 4.590)\par

\item Bambhaniya, P., Saurabh, Jusufi, K. and Joshi, P. S. (2022) \lq Thin accretion disk in the Simpson-Visser black-bounce and wormhole spacetimes\rq, \textit{Phys. Rev. D}, vol. 105, no. 2, pp. 023021. \url{https://doi.org/10.1103/PhysRevD.105.023021}. (JCR (2022) Impact Factor: 5.296)\par

\item Bambhaniya, P., Verma, J., Dey, D., Joshi, P. S. and Joshi, A. (2021) \lq Lense-Thirring effect and precession of timelike geodesics in slowly rotating black hole and naked singularity spacetimes\rq, \textit{Physics of the Dark Universe}, vol. 40, pp. 101215. \url{https://doi.org/10.1016/j.dark.2023.101215}. (JCR (2023) Impact Factor: 5.09)\par

\item Vagnozzi, S., Roy, R., Tsai, Y. D., Visinelli, L., Lee, D., Afrin, M., Allahyari, A., Bambhaniya, P., Dey, D., Ghosh, S. G., Joshi, P. S., Jusufi, K., Khodadi, M., Walia, R. K., Avgun, A. and Bambi, C. (2023)
\lq Horizon-scale tests of gravity theories and fundamental physics from the Event Horizon Telescope image of Sagittarius A$^*$\rq,
 \textit{Class.Quant.Grav.}, vol. 40, pp. 165007.  \url{https://doi.org/10.1088/1361-6382/acd97b}. (JCR (2023) Impact Factor: 3.5)\par

\item Joshi, A. B., Bambhaniya, P., Dey, D. and Joshi, P. S. (2019)
\lq Timelike Geodesics in Naked Singularity and Black Hole Spacetimes II\rq,
\textit{arXiv:1909.08873 [gr-qc].}

\item Saurabh, Bambhaniya, P., Joshi, P. S. (2022) \lq Probing the Shadow Image of the Sagittarius A* with Event Horizon Telescope\rq, \textit{arXiv:2202.00588 [gr-qc]}.\par

\item Bambhaniya, P., Joshi, A. B., Dey, D., Joshi, P. S., Mazumdar, A., Harada, T. and Nakao, K. I. (2022)
\lq Relativistic orbits of S2 star in the presence of scalar field\rq,
\textit{arXiv:2209.12610 [gr-qc]}.\par


\end{enumerate}

\clearpage
\pagestyle{plain}
\begin{center}
\textbf {\Large \textbf{Other Publications}}
\end{center}
\begin{enumerate}

\item Joshi, A., Dey, D., Joshi, P. S. and Bambhaniya, P. (2020) \lq Shadow of a naked singularity without photon sphere\rq, \textit{Phys. Rev. D}, vol. 102, no. 2, pp. 024022. \url{https://doi.org/10.1103/PhysRevD.102.024022}. (JCR (2020) Impact Factor: 4.833)\par

\item Patel, V, Acharya, K., Bambhaniya, P. and Joshi, P. S. (2022)
\lq Rotational energy extraction from the Kerr black hole's mimickers\rq, \textit{Universe}, vol. 8, no. 11, pp. 571.  \url{https://doi.org/10.3390/universe8110571}. (JCR (2022) Impact Factor: 2.813)\par

\item Patel, V, Acharya, K., Bambhaniya, P. and Joshi, P. S. (2023)
\lq Energy extraction from Janis-Newman-Winicour naked singularity\rq, \textit{Phys. Rev. D}, vol. 107, no. 6, pp. 064036. \url{https://doi.org/10.1103/PhysRevD.107.064036}. (JCR (2023) Impact Factor: 5.296)\par

\item Madan, S. and Bambhaniya, P. (2022) \lq Tidal force effects and periodic orbits in null naked singularity spacetime\rq, \textit{arXiv:2201.13163 [gr-qc]}.\par

\item Arora, D., Bambhaniya, P., Dey, D. and Joshi, P. S. (2023) \lq Tidal forces in the Simpson-Visser black-bounce and wormhole spacetimes\rq, \textit{arXiv:2305.08082 [gr-qc]}.\par

\item Acharya, K., Pandey, K., Bambhaniya, P., Joshi, P. S. and Patel, V. (2023) \lq Naked Singularity as a Possible Source of Ultra-High Energy Cosmic Rays\rq, \textit{arXiv:2303.16590 [gr-qc]}.\par

\item Saurabh, Bambhaniya, P., Joshi, P. S. (2023) \lq Imaging ultra-compact objects with radiative inefficient accretion flows\rq, \textit{arXiv:2308.14519 [astro-ph.HE]}.\par


\end{enumerate}

\clearpage
\pagestyle{plain}
\begin{center}
\textbf {\Large \textbf{Details of the Work Presented in Conferences}}
\end{center}
\begin{enumerate}
\item Oral presentation at, \textit{International Workshop on Astrophysics and Cosmology} organized by International Center for Cosmology, Charotar University of Science and Technology, 20th -24th December, 2019. 

\item Participated in \textit{Virtual Conference of the Polish Society on Relativity 2020} organized by Polish Society on Relativity – Poland, 24-26th September, 2020. 

\item Participated in \textit{31st meeting of the Indian Association for General Relativity and Gravitation (IAGRG)} organized by IIT-Gandhinagar, 19-20th  December, 2020. 

\item Oral presentation at \textit{11th Central European Relativity Seminar (CERS11)}, Vienna, Austria, February 11-13th 2021. 

\item Oral presentation at \textit{21st BritGrav meeting (BritGrav21)} organized by University of college Dublin, Ireland, 12-16th April 2021. 

\item Participated in \textit{International online Workshop on Relativistic Astrophysics and Gravitation (IWRAG-2021)} organized by Astronomical Institute of the Uzbekistan Academy of Sciences, 12-14th  May, 2021.

\end{enumerate}


\end{document}